\documentclass {article}
\usepackage{graphicx}

\begin{document}

\title{The holographic solution - Why general relativity must be understood in terms of strings}

\author{Michael Petri\thanks{email: mpetri@bfs.de} \\Bundesamt f\"{u}r Strahlenschutz (BfS), Salzgitter, Germany}

\date{{May 1, 2004}}

\maketitle

\begin{abstract}

This paper discusses the so called holographic solution, in short
"holostar". The holostar is the simplest exact, spherically
symmetric solution of the original Einstein field equations with
zero cosmological constant, including matter. The interior
matter-distribution follows an inverse square law $\rho = 1 / (8
\pi r^2)$. The interior principal pressures are $P_r = - \rho$ and
$P_\perp = 0$, which is the equation of state for a collection of
radial strings with string tension $\mu = -P_r = 1/(8 \pi r^2)$.
The interior space-time is separated from the exterior vacuum
space-time by a spherical two-dimensional boundary membrane,
consisting out of (tangential) pressure. The membrane has zero
mass-energy. Its stress-energy content is equal to the holostar's
gravitational mass. The total gravitational mass of a holostar can
be determined by a proper integral over the Lorentz-invariant
trace of the stress-energy tensor.

The holostar exhibits properties similar to the properties of
black holes. The exterior space-time of the holostar is identical
to that of a Schwarzschild black hole, due to Birkhoff's theorem.
The membrane has the same properties (i.e. the same pressure) as
the fictitious membrane attributed to a black hole according to
the membrane paradigm. This guarantees that the dynamical action
of the holostar on the exterior space-time is identical to that of
a black hole. The holostar possesses an internal temperature
proportional to $1/\sqrt{r}$ and a surface redshift proportional
to $\sqrt{r}$, from which the Hawking temperature and entropy
formula for a spherically symmetric black hole are derived up to a
constant factor.

The holostar's interior matter-state is singularity-free. It can
be interpreted as the most compact spherically symmetric (i.e.
radial) arrangement of classical strings: The radially outlayed
strings are densely packed, each string occupies exactly one
Planck area in its transverse direction. This maximally dense
arrangement is the reason why the holostar does not collapse to a
singularity and why the number of interior degrees of freedom
scales with area. Although the holostar's {\em total} interior
matter state has an overall string equation of state, {\em part}
of the matter can be interpreted in terms of particles. The number
of zero rest-mass particles within any concentric region of the
holostar's interior is shown to be proportional to the proper area
of its boundary, implying that the holostar is compatible with the
holographic principle and the Bekenstein entropy-area bound, not
only from a string but also from a particle perspective.

In contrast to a black hole, the holostar-metric is static
throughout the whole space-time. There are no trapped surfaces and
no event horizon. Information is not lost. The weak and strong
energy conditions are fulfilled everywhere, except for a
Planck-size region at the center. Therefore the holostar can serve
as an alternative model for a compact self-gravitating object of
any conceivable size.

The holostar is the prototype of a closed system in thermal
equilibrium. Its lifetime is several orders of magnitude higher
than its interior relaxation time. The thermodynamic properties of
the interior space-time can be derived exclusively from the
geometry. The local entropy-density $s$ in the interior space-time
is exactly equal to the proper geodesic acceleration of a
stationary observer, $s = g / \hbar$. It is related to the local
energy-density via $s T = \rho$. The free energy in the interior
space-time is minimized to zero, i.e. $F = E - ST = 0$.
Disregarding the slow process of Hawking evaporation, energy and
entropy are conserved locally and globally.

Although the holostar is a static solution, it behaves dynamically
with respect to the interior motion of its constituent particles.
Geodesic motion of massive particles in a large holostar is nearly
radial and exhibits properties very similar to what is found in
the observable universe: Any material observer moving geodesically
outward will observe an isotropic outward directed Hubble-flow of
massive particles from his local frame of reference. An inward
moving observer experiences an inward directed Hubble-flow. The
outward motion is associated with an increase in entropy, the
inward motion with a decrease. The radial motion of the
geodesically moving observer is accelerated, with the proper
acceleration falling off over time (for an outward moving
observer). The acceleration is due to the interior metric, there
is no cosmological constant.

Geodesic motion of massive particles is highly relativistic, as
viewed from the stationary coordinate system. The $\gamma$-factor
of an inward or outward moving observer is given by $\gamma
\approx \sqrt{r/r_0}$, where $r_0$ is a fundamental length
parameter which can be determined experimentally and
theoretically. $r \approx 2 r_{Pl}$. Inmoving and outmoving matter
is essentially decoupled, as the collision cross-sections of
ordinary matter must be divided by $\gamma^2$, which evaluates to
$\approx 10^{60}$ at the radial position of an observer
corresponding to the current radius of the universe.

The total matter-density $\rho$, viewed from the extended
Lorentz-frame of a geodesically moving observer, decreases over
proper time $\tau$ with $\rho \propto 1 / \tau^2$. The radial
coordinate position $r$ of the observer evolves proportional to
$\tau$. The local Hubble value is given by $H = 1/ \tau$. Although
the relation $\rho \propto 1/r^2$ seems to imply a highly
non-homogeneous matter-distribution, the large-scale
matter-distribution seen by a geodesically moving observer within
his observable local Hubble-volume is homogeneous by all practical
purposes. The large-scale matter-density within the Hubble-volume
differs at most by $\delta \rho / \rho \approx 10^{-60}$ at radial
position $r \approx 10^{61}$, corresponding to the current
Hubble-radius of the universe. The high degree of homogeneity in
the frame of the co-moving observer arises from the combined
effect of radial ruler distance shrinkage due the radial metric
coefficient ($\sqrt{g_{rr}} \propto \sqrt{r/r_{Pl}}$ and
Lorentz-contraction in the radial direction because of the highly
relativistic geodesic motion, which is nearly radial with $\gamma
\propto \sqrt{r/r_{Pl}}$ .

The geodesically moving observer is immersed in a bath of zero
rest-mass particles (photons), whose temperature decreases with $T
\propto 1 / \sqrt{\tau}$, i.e. $\rho \propto T^4$. Geodesic motion
of photons within the holostar preserves the Planck-distribution.
The radial position of an observer can be determined via the local
mass-density, the local radiation-temperature, the local
entropy-density or the local Hubble-flow. Using the experimentally
determined values for the matter-density of the universe, the
Hubble-value and the CMBR-temperature, values of $r$ between
$8.06$ and $9.18 \cdot 10^{60} r_{Pl}$ are calculated, i.e values
very close to the current radius and age of the universe.
Therefore the holographic solution might serve as an alternative
model for a universe with an overall negative (string type)
equation of state, without need for a cosmological constant.

The holostar as a model for a black hole or the universe contains
no free parameters: The holostar metric and fields, as well as the
initial conditions for geodesic motion are completely fixed and
arise naturally from the solution. An analysis of the
characteristic properties of geodesic motion in the interior
space-time points to the possibility, that the holostar-solution
might contribute substantially to the understanding of the
phenomena in our universe. The holostar model of the universe is
free of most of the problems of the standard cosmological models,
such as the "cosmic-coincidence-problem", the "flatness-problem",
the "horizon-problem", the "cosmological-constant problem" etc. .
The cosmological constant is exactly zero. There is no
horizon-problem, as the co-moving distance $r$ evolves exactly
proportional to the Hubble-radius ($r \propto \tau$ for $r \gg
r_{Pl}$). Inflation is not needed. There is no initial
singularity. The expansion (= radially outward directed geodesic
motion) starts out from a Plank-size region at the
Planck-temperature, which contains at most one "particle" with
roughly 1/8 to 1/5 of the Planck-mass. The maximum angular
separation of particles starting out from the center is limited to
roughly $60^\circ$, which could explain the low quadrupole and
octupole-moments in the CMBR-power spectrum. The relation $r
\propto \tau$ for $r \gg r_{Pl}$, whose fundamental origin can be
traced to the zero active gravitational mass-density of the
string-type matter in the interior space-time, can be interpreted
in terms of a permanently unaccelerated expansion, from which $H
\tau = 1$ follows. This genuine prediction of the holostar model
is very close to the observations, which give values between
$0.98$ and $1.04$. Permanently undecelerated expansion is also
compatible with the luminosity-redshift relation derived from the
most recent supernova-measurements, although the experimental
results favor the concordance $\Lambda$CDM-model over the
holostar-model at roughly $1 \sigma$ confidence level. The Hubble
value in the holostar-model of the universe turns out lower than
in the concordance model. $H \simeq 1/r = 62.35 \, (km/s) / MPc$
is predicted. This puts $H$ into the range of the other absolute
measurements, which consistently give values $H \approx 60 \,
(km/s) / MPc$ and is compatible with the recent supernova-data,
which favor values in the range $H \approx 60-65$. Geodesic motion
of particles in the holostar space-time preserves the relative
energy-densities of the different particle species (not their
number-densities!), from which a baryon-to photon ratio of roughly
$\eta \approx 10^{-9}$ at $T_{CMBR} = 2.725 \, K \approx 10^{-9}
(m_e c^2 / k)$ is predicted.

The holographic solution also admits microscopic self-gravitating
objects with a surface area of roughly the Planck-area and zero
gravitating mass.

\end{abstract}

\section{\label{sec:intro}Introduction:}

In a series of recent papers new interest has grown in the problem
of finding the most general solution to the spherically symmetric
equations of general relativity, including matter. Many of these
papers deal with anisotropic matter states.\footnote{Relevant
contributions in the recent past (most likely not a complete list
of relevant references) can be found in the papers of
\cite{Barve/Witten, Burinskii, Dev/00, Dev/03, Dymnikova,
Elizalde, Gair, Goncalves, Giambo, Hernandez, Herrera,
Herrera/2002, Ivanov, Mak, McManus/Coley, Mielke/BosonStars,
petri/bh, Salgado, Singh/Witten} and the references given
therein.} Anisotropic matter - in a spherically symmetric context
- is a (new) state of matter, for which the principal pressure
components in the radial and tangential directions differ. Note
that an anisotropic pressure is fully compatible with spherical
symmetry, a fact that appears to have been overlooked by some of
the old papers. One of the causes for this newly awakened interest
could be the realization, that models with anisotropic pressure
appear to have the potential to soften up the conditions under
which spherically symmetric collapse necessarily proceeds to a
singularity.\footnote{See for example \cite{Singh/Witten}, who
noted that under certain conditions a finite region near the
center necessarily expands outward, if collapse begins from rest.}
Another motivation for the renewed interest might be the prospect
of the new physics that will have to be developed in order to
understand the peculiar properties of matter in a state of highly
anisotropic pressure and to determine the conditions according to
which such matter-states develop.

The study of anisotropic matter states on a large scale might also
become relevant from recent cosmological observations. It is well
known, that the cosmic microwave background radiation (CMBR)
contains a dipole with a direction pointing roughly to the Virgo
cluster. The common explanation for this anisotropy is the
relative motion of the earth with respect to the preferred
rest-frame of the universe, which is identified with the frame in
which the CMBR appears spherically symmetric. Yet despite years of
research the physical cause for this peculiar motion has remained
unclear. The observations have failed to deliver independent
conclusive evidence for a sufficiently large mass concentration in
the direction of Coma/Virgo. In a recent paper the question was
raised, whether the universe might exhibit an intrinsic anisotropy
at large scales \cite{Ralston}. Not only the CMBR-dipole, but also
the quadrupole and the octupole coefficients of the CMBR-multipole
expansion single out preferred axes, which all point in the same
direction as the dipole, i.e. roughly to the Virgo
direction.\footnote{In order to get directional information from
the higher multipoles, the authors analyzed not only the
(directionless) $l$-terms of the multipole expansion of the
CBMR-powerspectrum in spherical harmonics, but also the
$m$-terms.} In an earlier paper the same authors found an
alignment of optical and radio polarization data with respect to
Virgo \cite{Jain/polarization, Jain}.\footnote{In
\cite{Jain/polarization} the authors analyzed the statistics of
offset angles of radio galaxy symmetry axes relative to their
average polarization angles, indicating an anisotropy for the
propagation of radiation on cosmological scales, which lies in the
direction of Virgo. Another preferred axis, parallel to the
CMBR-dipole within the measurement errors, was identified in
\cite{Jain}. Here the optical polarization data from
cosmologically distant and widely separated quasars showed an
improbable degree of coherence. A significant clustering of
polarization coherence in large patches in the sky was identified,
with the axis of correlation again lying in the direction of
Virgo.} In \cite{Ralston} a Bayesian analysis was performed with
the result, that the common alignment of five different axes
appears very unlikely in the context of the Standard Cosmological
Model of a homogeneously expanding Friedman Robertson-Walker
universe. The authors argue, that within the standard big bang
hypothesis one rather expects an isotropic distribution of the
different multipole alignment axes for the post-inflationary
state. As the experimental evidence rather points to the contrary
the authors conclude, that we might be living in a universe that
exhibits an intrinsic anisotropic on very large scales. In a very
recent paper another team of authors pointed out, that the
quadrupole and (three) octupole planes are correlated with 99.97
\% confidence level and that the alignment of the normals of these
planes with the cosmological dipole and the equinoxes is
inconsistent with Gaussian random skies at 99.8 \% confidence
level \cite{Schwarz2004}.

Anisotropic matter states are also predicted by string theory. The
equation of state for a cosmic string or for a 2D-membrane is
naturally and necessarily anisotropic. The interior matter-state
of the solution discussed in this paper has a definite string
interpretation: It is that of a collection of radially outlayed
strings, attached to a spherical 2D-boundary membrane.
The string tension falls off as $1/r^2$. The strings are "densely"
packed in the sense, that the transverse extension of the strings
amounts to exactly one Planck area \cite{petri/string}.

The absence of singularities and trapped surfaces in the holostar
space-time is compatible with a very recent result in string
theory. According to \cite{Senovilla} the so called VSI
space-times\footnote{VSI = vanishing scalar invariants.
VSI-space-times are solutions where all scalar invariants vanish}
- exact solutions of string theory - are incompatible with trapped
surfaces. The proof is quite general. It is based on geometric
arguments and doesn't require the field equations. If this result
is confirmed, string theory might turn out to be incompatible with
trapped surfaces and - consequentially - classical space-time
singularities, such as can be found in the classical black hole
solutions.\footnote{In \cite{Chinea2004} it was shown, that
trapped surfaces (or more generally: the occurrence of any causal
trapped set in the space-time) are an {\em essential} requirement
for the singularity theorems: Neither the energy-conditions nor
the causality conditions alone or in combination lead to a
singularity. Whereas the positivity of the energy and the
principle of causality are fundamental requirements for a
self-consistent physical theory, trapped surfaces are not
necessary for a self-consistent physical description. The {\em
assumption} that a {\em physically realistic} space-time should
develop a trapped surface (or a trapped causal set) at some
particular time is the most questionable of the assumptions
underlying the singularity theorems. One can regard it as the key
assumption on which the common belief is based, that a physically
realistic space-time must contain singularities. But this
assumption has not been proven. It was formulated when the only
solutions of physical relevance that were known at that time {\em
all} contained trapped surfaces: the black hole solutions. Now we
know of physically relevant solutions {\em without} trapped
surfaces and singularities. The claim, that a physically realistic
space-time must necessarily contain trapped surfaces (and
therefore singularities), must be viewed in the proper historical
context. If history had taken a different route, chances are, that
today's claims would have been quite different: If realistic
singularity free solutions to the field equations had been known
at the time the singularity theorems were formulated, one would
rather have argued, that singularities are unphysical and
therefore trapped surfaces cannot be elements of a physically
realistic space-time. In view of the new singularity free
solutions it is appropriate to exercise some caution in raising an
assumption to the status of unquestionable physical truth, as can
be found occasionally in today's scientific literature. If we are
honest, we must admit that despite decades of research we still
don't know, what properties a "physically realistic" space-time
should actually have. In fact, our preconceptions about how a
physically realistic space-time should "look" like have changed
dramatically in the course of history. A most radical change
occurred in the recent past: The luminosity dependence of distant
supernovae on red-shift makes it quite clear, that we are not
living in a dust-type universe with equation of state $P \approx
0$, as had been thought decades before, but rather in a universe
which consists mostly out of (cold) "dark matter" and "dark
energy". Although we neither know what the "dark matter" is, and
even less what the "dark energy" could be, the equation of state
for the large scale distribution of mass-energy in the universe
has been measured fairly accurately: It is of the form $\rho +P
\approx 0$, which is compatible with vacuum-type, string-type or
domainwall-type matter.}

Therefore the result reported in \cite{Senovilla} has two possible
outcomes: Either string theory is the - essentially - correct
description for a physically realistic space-time in the high {\em
and} the low energy limit. Then trapped surfaces, the classical
vacuum black hole solutions and - most likely - space-time
singularities (of "macroscopic scale") cannot be part of a
physically realistic description of the world we live in. We will
have to search for other solutions of the classical field
equations without singularities and trapped surfaces,
preferentially with a strong string character. The holostar is the
simplest member of such a class of solutions.

The other outcome - which might be the preferred scenario by the
1000 or more researchers who invested much time and effort in the
study of singularities, event horizons, apparent horizons and
trapped surfaces - might be, that despite their highly undesirable
properties event horizons, trapped surfaces and singularities are
real. Then string theory cannot fulfill its main objective to
provide a unified description for all forces, including gravity,
on all energy scales. We will have to look for another approach.

Whatever the final outcome is going to be, for the present time we
have to work with the theories and solutions we know.

In this paper the geometric properties of the so called
holographic solution, an exact solution of the Einstein field
equations, are studied in somewhat greater detail. The holographic
solution turns out to be a particularly simple model for a
spherically symmetric, self gravitating system with a highly - in
fact maximally - anisotropic, string-type pressure. Despite its
simplicity and its lack of adjustable parameters, the holostar
appears very well suited to explain many of the phenomena
encountered in various self gravitating systems, such as black
holes, the universe and - possibly - even elementary particles.

The paper is divided into three sections. In the following
principal section some characteristic properties of the
holographic solution are derived. In the next section these
properties are compared to the properties of the most fundamental
objects of nature that are known so far, i.e. elementary
particles, black holes and the universe. The question, whether the
holostar can serve as an alternative, unified model for these
fundamental objects (within the limitations of a classical
description!) will be discussed. The paper closes with a discussion
and outlook.

\section{Some characteristic properties of the holographic solution}

In this section I derive some characteristic properties of the
holostar solution. As the exterior space-time of the holostar is identical
to the known Schwarzschild vacuum solution, only the interior
space-time will be covered in detail.

Despite the mathematically simple form of the interior metric, the
holostar's interior structure and dynamics turns out to be far
from trivial. A remarkable list of properties can be deduced from
the interior metric, indicating that the holostar has much in
common with a spherically symmetric black hole and with the
observable universe.

\subsection{The holostar metric and fields}

For any spherically symmetric problem the metric is characterized
by two functions, which only depend on the radial coordinate value
$r$. In units $c=G=1$ and with the (+ - - -) sign convention the
most general spherically symmetric metric can be expressed as:

\begin{equation} ds^2 = B(r) dt^2 - A(r) dr^2 - r^2 d\Omega
\end{equation}

The holostar solution is the simplest spherically symmetric
solution of the Einstein field equations containing matter. It's
remarkable properties arise from a delicate cancelation of terms
in the Einstein field equations, which only arises for a
matter-distribution equal to

\begin{equation} \rho(r)= \frac{1}{8 \pi r^2}
\end{equation}

and a string type equation of state

\begin{equation} P_r(r)= - \rho(r)
\end{equation}

The equation of state $\rho + P_r = 0$ implies $A B =
1$, which "linearizes" the field equations (in a spherically
symmetric problem) and decouples the set of two non-linear
differential equations for the metric coefficients $A$ and $B$ to
a single linear first order differential equation:

\begin{equation} \label{eq:rB}
\left(r B \right)' = 1 - 8 \pi r^2 \rho
\end{equation}

For $AB = 1$ the tangential pressure is a linear function of the
metric's first and second derivatives:

\begin{equation} \label{eq:Pt}
8 \pi P_\theta = \frac{B''}{2} + \frac{B'}{r}
\end{equation}

For the linearized equations a (weighted) superposition principle
holds: Any solution can be expressed as the superposition of the
weighted sum of already known solutions, as long as the weights
$w_i$ satisfy the norm condition $\sum{w_i} = 1$, i.e. add up to
one. The metric coefficient $B$ and the fields $\rho$, ${P_r}$,
${P_\theta}$ of any conceivable solution can be derived from the
weighted sum of the metric coefficients $B_i$ and fields $\rho_i$
etc. of the already known solutions. See \cite{petri/bh} for a
somewhat more detailed discussion and derivation.

Independent from the simplifications arising from an equation of
state of the form $\rho + P_r = 0$, the "special"
matter-distribution $\rho = 1 / (8 \pi r^2)$ renders the generally
non-homogeneous differential equation for the {\em radial} metric
coefficient $A$ to a {\em homogeneous} differential equation. For
any spherically symmetric problem the radial metric coefficient
$A$ can be derived exclusively from the (total) matter-density
$\rho$ by a simple integration:

\begin{equation} \label{eq:r/A}
\left(\frac{r}{A}\right)' = 1 - 8 \pi r^2 \rho
\end{equation}

It is important to keep in mind that equation (\ref{eq:r/A}) is
completely general and does not depend on the equation of state.
The only requirement is spherical symmetry and the Einstein field
equations with zero cosmological constant. Therefore in any
spherically symmetric problem the spatial part of the metric
$A(r)$ is independent of the pressure.

It is easy to see from equation (\ref{eq:r/A}) that a
matter-density of the form $\rho = 1 / (8 \pi r^2)$ is special.
For this matter-density the spatial part of the metric attains its
simplest possible form: $A \propto r$.

With the additional assumption, that the holostar resides in an
exterior vacuum space-time one arrives at the following form for
the matter-distribution

\begin{equation} \rho(r)= \frac{1}{8 \pi r^2}  \left(1 - \theta(r-r_h)\right)
\end{equation}

$\theta(r)$ denotes the Heavyside step functional and $\delta(r)$
the Dirac delta-distribution.

With the additional simplification due to the (string) equation of
state $\rho + P_r = 0$ we get:

\begin{equation} B(r) = \frac{1}{A(r)} = \frac{r_0}{r} (1-\theta(r-r_h))
+ \left( 1 - \frac{r_+}{r}\right) \theta(r-r_h)
\end{equation}

\begin{equation} P_\theta = \frac{1}{16 \pi r_h}
\delta(r-r_h)
\end{equation}

$r_h = r_+ + r_0$ is the position of the holostar's boundary
membrane, which separates the interior matter distribution from
the exterior vacuum space-time. $r_+ = 2 M$ is the gravitational
radius of the holostar, $M$ its gravitating mass and $r_0$ is a
fundamental length. There is some significant theoretical evidence
\cite{petri/charge, petri/thermo} that $r_0$ is comparable to the
Planck-length up to a numerical factor of order unity ($r_0^2
\approx 4 \sqrt{3/4} \, A_{Pl} \simeq 3.5 \, \hbar$). An
experimental determination of the fundamental area $r_0^2$ is
given in section \ref{sec:r0} of this paper. The experimental
result $r_0^2 \approx 3.5 \,  \hbar$ agrees quite well with the
theoretical expectation.\footnote{With the convention ($c=G=k=1$)
used throughout this paper, Planck-units for the fundamental
physical quantities such as the Planck time $t_{Pl}$, the Planck
mass $m_{Pl}$, etc. are given as powers of $\hbar$. For instance
$A_{Pl} = \hbar$, $m_{Pl} = E_{Pl} = \sqrt{\hbar}$,$ r_{Pl} =
t_{Pl} = \sqrt{\hbar}$, $\rho_{Pl} = 1 / \hbar$, etc. For the
temperature units k=1 are used throughout this paper. The choice
$k=1$ implies that temperature and energy are regarded as
equivalent (or dual) quantities, so that the Planck temperature
$T_{Pl} = E_{Pl} = \sqrt{\hbar}$. Another natural choice for $k$
would be to set $k = 4 \pi$, which has the effect that some
factors of $4 \pi$ are eliminated in the equations.}

At the holostar's center a negative point mass $M_0 = -r_0/2$ of
roughly Planck mass is situated.\footnote{At least this is true in
a {\em formal} sense: As can be seen from the metric, a holostar
with $r_h=r_0$ has an exterior metric which is exactly Minkowski,
i.e. $A = B = 1$ for $r \geq r_h$. This means, that the total
gravitational mass of such a (minimal) holostar is zero. Yet there
is a non-zero matter-distribution in the interior region $r \leq
r_0$. The integral over the matter-distribution of the interior
region $[0, r_0)$ formally gives a mass of $M = r_0/2$. In order
to get a mass of zero for the total space-time one has to
introduce a negative point-mass $M_0 = -r_0/2$ at the origin. In
\cite{petri/charge} it was shown, that $r_0 \approx 2 \, r_{Pl}$,
so $M_0 \approx m_{Pl}$.} From a purely classical point of view
the holostar space-time therefore contains a "naked" singularity,
which - however - is "covered up" by the surrounding matter. The
question is, whether this (formal) negative point mass is
observable. In a spherically symmetric context the only relevant
observable for a spherically symmetric object is its gravitating
mass, which can be calculated from the (improper) integral over
the mass-energy density. For the holographic solution the mass
"visible" for an observer situated at radial position $r$ is given
by

$$M(r) = \frac{r-r_0}{2}$$

$M(r)$ is zero at $r=r_0$ and negative only for $r < r_0$. We find
that the observable effect of the negative mass singularity is
"shielded" from the outside world  by the surrounding matter such,
that the presence of a negative mass is "noticeable" only within the
Planck-sized region $r < r_0$. But for distances smaller than the
Planck distance, we expect the classical description to break
down. In fact, according to recent results in loop quantum
gravity, measuring the geometry with sub-Planck precision makes no
sense, as the geometric operators are quantized. With this in mind
the negative point mass singularity at the center rather appears
as an artifact of the purely classical description: For $r \gg
r_0$, i.e. for an energy/distance range where classical general
relativity can be expected to be a fairly good approximation, the
quantum singularity at $r < r_0$ has no noticeable adverse effects
for any classical observer.

\subsection{Proper volume and radial distance}

The proper radius, i.e. the proper length of a radial trajectory
from the center to radial coordinate position $r$, of the holostar
scales with $r^{3/2}$:

\begin{equation}
l(0, r) =  \int_{0}^{r}{\sqrt{A}dr}=\frac{2}{3} r \sqrt{\frac{r}{r_0}}
\end{equation}

The proper radial distance between the membrane at $r_h$ and the
gravitational radius at $r_+$ is given by:

\begin{equation}
l(r_+, r_h) =  \frac{2}{3} r_+ \sqrt{\frac{r_+}{r_0}} \big(
(1+\frac{r_0}{r_+})^{\frac{3}{2}}-1\big)\cong r_0
\sqrt{\frac{r_+}{r_0}} = \sqrt{r_+ r_0}
\end{equation}

The proper volume of the region enclosed by a sphere with proper
area $4 \pi r^2$ scales with $r^{7/2}$:

\begin{equation}
V(r) = \int_0^{r}{4 \pi r^2 \sqrt{\frac{r}{r_0}} dr} =\frac{6}{7}
\sqrt{\frac{r}{r_0}} V_{flat}
\end{equation}

$V_{flat} = (4 \pi /3) r^3$ is the volume of the respective sphere
in flat space.

Therefore both volume and radial distance in the interior
space-time region of the holostar are enhanced over the respective
volume or radius of a sphere in flat space by the square-root of
the ratio between $r_h = r_+ + r_0$ and the fundamental distance
defined by $r_0$.

The proper integral over the mass-density, i.e. the sum over the
total constituent matter, scales as the proper radius, i.e. as
$r_h^{3/2}$.

The enormous shrinkage of radial ruler distances due to the radial
metric coefficient $\sqrt{A} = \sqrt{r/r_0}$ implies, that the
matter-density in any extended volume is nearly homogeneous. A
stationary observer at radial position $r$ will find, that a
concentric sphere with proper radius $r_p \approx r$ will have
negligible extension in terms of the radial coordinate parameter
$r$ (Keep in mind, that $r$ {\em not} a proper distance measure!):
With $r_p = \sqrt{A} \delta r$ we find that $\delta r = r_p /
\sqrt{A}$. The matter-density within any sphere of {\em proper}
radius $r_p$ will differ at most by:

$$ \frac{\delta \rho}{\rho} \leq \frac{\rho(r+\delta r) - \rho(r-\delta r)}{\rho(r)} \simeq 4 \frac{\delta r}{r} = 4 \frac{r_p}{r} \sqrt{r_0/r}$$

For $r_p \approx r$ the maximum relative difference in density
will be given by $\delta \rho / \rho \approx 4 \sqrt{r_0/r}$. As
$r_0$ is comparable to the Planck length, the matter-density in
the holostar space-time can be regarded as nearly homogeneous for
any stationary observer further than a million Planck-distances
from the center. We will see later in section
\ref{sec:massive:acc} that the situation is even more favorable
for a {\em geodesically} moving observer. For such an observer the
local Hubble distance $r_H = 1/H$ is equal in magnitude to the
radial coordinate value, i.e. $r_H \approx r$. Far away from its
turning point of the motion such an observer moves highly
relativistically on a nearly radial trajectory with a
$\gamma$-factor proportional to $\sqrt{A}$. Due to
Lorentz-contraction in the radial direction a geodesically moving
observer will find that the large scale matter-distribution is
homogeneous with a relative deviation from homogeneity that is
unmeasurable by all practical purposes: $\delta \rho / \rho
\approx 1 / A = r_0 / r$ within the local Hubble volume of such an
observer. For example, at radial position $r \approx 10^{61} \,
r_{Pl}$, corresponding to the current radius of the universe, the
density within the a sphere of proper radius $r_p \approx 10^{61}$
varies at most by $\delta \rho / \rho \approx 10^{-60}$. Keep in
mind that in practice the matter-distribution will show much
larger fluctuations due to the magnification of quantum
fluctuations during the expansion and by the various processes of
structure formation.

\subsection{\label{sec:energy:conditions}Energy-conditions}

For a space-time with a diagonal stress-energy tensor, $T_\mu^\nu
= diag(\rho, -P_1, -P_2, -P_3)$, the energy conditions can be
stated in the following form:

\begin{itemize}
\item {weak energy condition: $\rho \geq 0$ and $\rho + P_i \geq 0$ }
\item {strong energy condition: $\rho + \sum{P_i} \geq 0$ and $\rho + P_i \geq 0$}
\item {dominant energy condition: $\rho \geq | P_i |$}
\end{itemize}

It is easy to see that the holostar fulfills the weak and strong
energy conditions at all space-time points, except at $r=0$, where
- formally - a negative point mass of roughly Planck-size is situated.
The dominant energy condition is fulfilled everywhere except at
$r=0$ and at the position of the membrane $r=r_h$.

According to the recent results of loop quantum gravity, the
notion of space-time points is ill defined. Space-time looses its
smooth manifold structure at small distances. At its fundamental
level the geometry of space time should be regarded as discrete
\cite{Penrose/spin-networks, Penrose/twistor}. The minimum
(quantized) area in loop quantum gravity is non-zero and roughly equal
to the Planck area \cite{Ashtekar/area, Rovelli/Smolin}. Therefore
the smallest physically meaningful space-time region will be
bounded by a surface of roughly Planck-area. Measurements probing
the interior of a minimal space-time region make no sense. A
minimal space-time region should be regarded as devoid of any
physical (sub-)structure.

A similar result follows from string theory. Although string
theory treats space-time as a smooth manifold structure up to the
highest energies and smallest scales, all observables, such as the
frequency of a vibrating string, are quantized in terms of the
string scale. Probing the high - trans-Planckian - energy regime,
i.e. at distances far smaller than the string scale, just leads to
the dual description of an equivalent string theory at low
energies. For instance a string state at low energies with N
"winding modes" and M "longitudinal modes" cannot be distinguished
- with respect to its total vibrational energy - from a string
state at high energies, where the numbers N and M of the "winding"
and "longitudinal" are interchanged.

If the energy conditions are evaluated with respect to physically
meaningful space-time regions\footnote{With "Planck-sized region"
a compact space-time region of non-zero, roughly Planck-sized
volume bounded by an area of roughly the Planck-area is meant.},
i.e. by integrals over at least Planck-sized regions, the
following picture emerges: Due to the negative point mass at the
center of the holostar the weak energy condition is violated in
the sub Planck size region $r < r_0$. However, from the viewpoint
of loop quantum gravity this region should be regarded as inaccessible
for any meaningful physical measurement.

I therefore propose to discard the region $r < r_0$ from the
physical picture. This deliberate exclusion of a classically well
defined region might appear somewhat conceived. From the viewpoint
of loop quantum gravity it is quite natural. Furthermore, disregarding
the region $r < r_0$ is not inconsistent. In fact, the holographic
solution in itself very strongly suggests, that whatever is
"located" in the region $r<r_0$ is irrelevant to the (classical)
physics outside of this region, not only from a quantum, but also
from a purely classical perspective: If we "cut out" the region $r
< r_0$ from the holostar space-time and identify all space-time
points on the sphere $r = r_0$, we arrive at exactly the same
space-time in the physically relevant region $r \geq r_0$, as if
the region $r < r_0$ had been included. The reason for this is,
that the gravitational mass of the region $r \leq r_0$ - evaluated
classically - is exactly zero. This result is due to the
(unphysical) negative point mass at the (unphysical) position $r =
0$ in the classical solution, which cancels the integral over the
mass-density $\propto 1/r^2$ in the interval $(0, r_0]$. Although
neither the negative point mass nor the infinite mass-density and
pressure at the center of the holostar are acceptable, they don't
have any classical effect outside the physically meaningful region
$r > r_0$.

Thus the holographic solution satisfies the weak (positive) energy
condition in any physically meaningful space-time region
throughout the whole space-time manifold. The same is true for the
strong energy condition.

The two space-times with the region $r \leq r_0$ "cut out" or with
$m_0 = -r_0/2$ at $r=0$ can only be distinguished from each other
by measurements in the sub-Planck size region $r<r_0$. Whether
these two space-times are different therefore appears rather as a
philosophical question with no practical consequence. Only if our
measurement capabilities increase dramatically, so that
Planck-precision measurements become possible, one might be able
to answer this question. However, if the results of loop quantum
gravity (quantization of the geometric operators) or of string
theory (dualities) are essentially correct, chances are, that it
will be impossible in principle to distinguish both space-times.

Can one "mend" the violation of the dominant energy condition in
the membrane by a similar argument? Due to the considerable
surface pressure of the membrane the dominant energy condition is
clearly not satisfied within the membrane. Unfortunately there is
no way to fulfill the dominant energy condition by "smoothing"
over a Planck-size region\footnote{If the term "Planck-sized
region" would only refer to volume, allowing an arbitrary large
boundary area, we could construct a cone-shaped region extending
from the center of the holostar to the membrane, with arbitrary
small solid angle, but huge area and radial extension. In such a
cone-shaped region the integrated dominant energy condition would
not be violated, if the region extends from the membrane at least
half-way to the center.}, as is possible in case of the weak and
strong energy conditions. Therefore the violation of the dominant
energy condition by the membrane must be considered as a real
physical effect, i.e. a genuine property of the holostar space-time.

Is the violation of the dominant energy condition within the
membrane incompatible with the most basic physical laws? The
dominant energy condition can be interpreted as saying, that the
speed of energy flow of matter is always less than the speed of
light. As the dominant energy condition is violated in the
membrane, one must expect some "non-local" behavior of the
membrane. Non-locality, however, is a well known property of
quantum phenomena. Non-local behavior of quantum systems has been
verified experimentally up to macroscopic dimensions.\footnote{See
for example the spin "entanglement" experiments, "quantum
teleportation" and quantum cryptography, just to name a few
phenomena, that depend on the non-local behavior of quantum
systems. Some of these phenomena, such as quantum-cryptography,
are even being put into technical use.} This suggests that the
membrane might be a macroscopic quantum phenomenon. In
\cite{petri/thermo} it is proposed, that the membrane should
consist of a condensed boson gas at a temperature far below the
Bose-temperature of the membrane. In such a case the membrane
could be characterized as a single macroscopic quantum state of
bosons. One would expect collective non-local behavior from such a
state.

\subsection{"Stress-energy content" of the membrane}

One of the outstanding characteristics of the holographic solution
is the property, that the "stress-energy content" of the membrane
is equal to the gravitating mass $M$ of the holostar.

For any spherically symmetric solution to the equations of general
relativity the total gravitational mass $M(r)$ within a concentric
space-time region bounded by $r$ is given by the following mass
function:

\begin{equation} \label{eq:massfunction}
M(r) = \int_0^r{\rho \, \widetilde{dV}} = \int_0^r{\rho \, \, 4 \pi r^2 dr}
\end{equation}

$M(r)$ is the integral of the mass-density $\rho$ over the
(improper) volume element $\widetilde{dV} = 4 \pi r^2 dr$, i.e.
over a spherical shell with radial coordinate extension $dr$
situated at radial coordinate value $r$. Note, that in a
space-time with $A B = 1$ the so called "improper" integral over
the interior mass-density appears just the right way to evaluate
the asymptotic gravitational mass $M$: The gravitational mass can
be thought to be the genuine sum (i.e. the proper integral) over
the constituent matter, corrected by the gravitational red-shift.
The proper volume element is given by $dV = 4 \pi r^2 \sqrt{A}
dr$. The red-shift factor for an asymptotic observer situated at
infinity, with $B(\infty) = 1$, is given by $\sqrt{B}$. Therefore
the proper integral over the constituent matter, red-shift
corrected with respect to an observer at spatial infinity, is
given by: $M = \int{\rho \, \sqrt{A B} \, 4 \pi r^2 dr}$. This is
equal to the improper integral in equation
(\ref{eq:massfunction}), whenever $A B = 1$.

If the energy-content of the membrane is calculated by the same
procedure, replacing $\rho$ with the two principal non-zero
pressure components $P_\theta = P_\varphi$ in the membrane, one
gets:

\begin{equation} \label{eq:Energy:Pt}
\int_0^\infty{(2 P_{\theta}) 4 \pi r^2 dr} = \frac{r_h}{2} = M +
\frac{r_0}{2}\cong M
\end{equation}

The holostar's membrane is nothing else than its - real,
physically localizable - boundary. The single main characteristic
of the holostar solution with respect to the exterior space-time,
its gravitating mass $M$, can be derived exclusively from the
properties of its boundary - a result which is quite compatible
with the holographic principle. Note also, that the  tangential
pressure of the holostar's membrane, $P_{\theta} = 1 / (32 \pi (M
+ r_0/2))$, is (almost) exactly equal to the tangential pressure
attributed to the event horizon of a spherically symmetric black
hole by the membrane paradigm \cite{Price/Thorne/mem, Thorne/mem},
$P_{\theta} = 1 / (32 \pi M)$. According to the membrane paradigm
all properties of an (uncharged, non-rotating) black hole - with
respect to the {\em exterior} space-time - can explained in terms
of a - fictitious - membrane with $\rho = 0$ and $P_\theta = 1 /
(32 \pi M)$. For charged / rotating black holes a similar result
holds. The equivalence of the - fictitious - black hole membrane
with the - real - holostar membrane therefore guarantees, that the
holostar's dynamical action on the exterior space-time is
practically indistinguishable from that of a black hole with the
same gravitating mass.

\subsection{On the physical interpretation of the trace of the stress-energy tensor in the holographic space-time}

The integral over the mass-density (or over some particular
pressure components) might not be considered as a very
satisfactory means to determine the total gravitational mass of a
self gravitating system. Neither the mass-density nor the principal
pressures have a coordinate independent meaning: They transform
like the components of a tensor, not as scalars.

In this respect it is quite remarkable, that the gravitating mass
of the holostar can be derived from the integral over the {\em
trace} of the stress-energy tensor $T = T_\mu^\mu$. In fact, the
integral over $T$ is exactly equal to $M$ if the negative point
mass $M_0 = -r_0/2$ at the center is included, or - which is the
preferred procedure - if the integration is performed from $r \in
[r_0, \infty]$, as was suggested in the previous section.

\begin{equation} \label{eq:Energy:T}
\int_0^\infty{T \, \sqrt{-g} d^3x} = \int_0^\infty{T \, 4 \pi r^2
dr} = \frac{r_h}{2} + M_0 = M
\end{equation}

with

$$T = \frac{1}{4 \pi r^2}\left(1-\theta(r-r_h)\right) - \frac{1}{8 \pi r_h} \delta(r-r_h)$$

Contrary to the mass-density or the pressure, $T$ is a
Lorentz-scalar and therefore has a definite coordinate-independent
meaning at any space-time point. Note, that the scalar $T(x^\mu)$
gives quite an accurate account for the "strength" of the local
source-distribution at $x^\mu$, which creates the gravitational
field of the holostar.

The gravitating mass of the holostar can also be derived from the
so called Tolman-mass $M_{Tol}$ of the space-time:

\begin{equation} \label{eq:M:Tolman}
M_{Tol} = \int_0^\infty{(\rho + P_r + 2 P_{\theta}) \sqrt{-g}
d^3x} = \frac{r_h}{2} + M_0 = M
\end{equation}

The Tolman mass is often referred to as the "active gravitational
mass" of a self-gravitating system. The motion of particles in a
wide class of exact solutions to the equations of general
relativity indicate that the sum of the matter density and the
three principle pressures can be interpreted as the "true source"
of the gravitational field.\footnote{Poisson's equation for the
local relative gravitational acceleration $g$ is given by $\nabla
\cdot \textbf{g} = -R_{00} = - 4 \pi G (\rho + \sum{P_i})$ in
local Minkowski-coordinates. Therefore the sum of $\rho$ and the
three principal pressures appears as source-term in Poisson's
equation for the local gravitational acceleration $g$, i.e. can be
considered as the "true" gravitational mass density.}

The holostar space-time therefore has three identical measures for
it's total gravitating mass $M$, the "normal" mass $M_\rho$,
derived by an integral over the energy-density $\rho$, the
"invariant" mass $M_T$, derived by an integral over the
(coordinate independent) trace $T$ of the stress-energy tensor,
and the "Tolman" mass $M_{Tol}$, derived by an integral over the
active gravitational mass-density $\rho + P_r + 2 P_\theta$.

Also note, that the rr- and tt-components of the Ricci-tensor are
zero everywhere, except for a $\delta$-functional at the position
of the membrane:

$$R_0^0 = R_1^1 = -\frac{1}{2 r_h}\delta = -\frac{1}{4 M + 2 r_0}\delta$$

\subsection{\label{sec:eq:motion}The equations of geodesic motion}

In this section the equations for the geodesic motion of particles
within the holostar are set up. Keep in mind that the results of
the analysis of pure geodesic motion have to be interpreted with
caution, as pure geodesic motion is unrealistic in the interior of
a holostar. In general it is not possible to neglect the
pressure-forces totally. In fact, as will be shown later, it is
quite improbable that the motion of massive particles will be
geodesic throughout the holostar's {\em whole} interior space-time.
Nevertheless, the analysis of pure geodesic motion, especially for
photons, is a valuable tool to discover the properties of any space-time.

Disregarding pressure effects, the interior motion of massive
particles or photons can be described by an effective potential.
The geodesic equations of motion for a general spherically
symmetric space-time, expressed in terms of the "geometric"
constants of the motion $r_i$ and $\beta_i$, were given in the
appendix of \cite{petri/bh}:

\begin{equation} \label{eq:beta:r:2}
\beta_r^2(r) + V_{eff}(r) = 1
\end{equation}

\begin{equation} \label{eq:Veff:beta}
 V_{eff}(r) = \frac{B(r)}{B(r_i)}\left(1 -
\beta_i^2(1-\frac{r_i^2}{r^2})\right)
\end{equation}

\begin{equation} \label{eq:beta:t:2}
\beta_{\perp}^2(r) = \frac{B(r)}{B(r_i)}\frac{r_i^2}{r^2} \,
\beta_i^2
\end{equation}

$r_i$ is the interior turning point of the motion, $\beta_r$ is
the radial velocity, expressed as fraction to the local velocity
of light in the radial direction, $\beta_{\perp}$ is the
tangential velocity, expressed as fraction to the local velocity
of light in the tangential direction. $\partial r, \partial
\varphi$ and $\partial \theta$ are orthogonal for a stationary
observer in the coordinate frame, so $\beta^2 = \beta_r^2 +
\beta_{\perp}^2$. The quantities $\beta^2, \beta_r^2$ and
$\beta_\perp^2$ all lie in the interval $[0,1]$.

One would think, that the turning point $r_i$ of the motion of a
particle in any spherically symmetric space-time can be chosen
arbitrarily. We will see later, that this is not the case in the
holostar space-time. Self-consistency requires, that $r_i \simeq
r_0$. However, in the following discussion we will assume that
$r_i$ can - in principle - take on any value $r_i \leq r_h$.

$\beta_i = \beta(r_i) = \beta_{\perp}(r_i)$ is the velocity of the
particle at the turning point of the motion, $r_i$. By definition
$\beta_r(r_i) = 0$ at any turning point of the motion, therefore
$\beta_i$ is purely tangential at $r_i$. For photons $\beta_i =1$.
Pure radial motion for photons is only possible, when $r_i=0$. In
this case $V_{eff} = 0$. Therefore the purely radial motion of
photons can be considered as "force-free".\footnote{The region $r
< r_0$ should be considered as unaccessible, which means that pure
radial motions for photons is impossible. Aiming the photons
precisely at $r=0$ also conflicts with the quantum mechanical
uncertainty postulate. The photons will therefore always be
subject to an effective potential with $r_i > r_0$, i.e. an
effective potential that is not constant. But whenever the
photon has moved an appreciable distance from its turning point of
the motion, i.e. whenever $r \gg r_i$, the effective potential is
nearly zero. Therefore non-radial motion of photons can be
regarded as nearly "force-free", whenever $r \gg r_i$.}

In order to integrate the geodesic equations of motion, the
following relations are required:

\begin{equation} \label{eq:beta:r}
\beta_r(r) = \frac{dr}{dt} /
\sqrt{\frac{B}{A}}
\end{equation}

\begin{equation} \label{eq:beta:t}
\beta_{\perp}(r) = \frac{r d\varphi}{dt} / \sqrt{B}
\end{equation}

$t$ is the time measured by the asymptotic observer at spatial
infinity. If the equations of motion are to be solved with respect
to the proper time $\tau$ of the particle (this is only reasonable
for massive particles with $\beta <1$), the following relation is
useful:

\begin{equation} dt = d\tau \frac{\gamma_i \sqrt{B(r_i)}}{B}
\end{equation}

$\gamma^2 = 1/(1-\beta^2)$ is the special relativistic
$\gamma$-factor and $\gamma_i = \gamma(r_i)$, the local
$\gamma$-factor of the particle at the turning point of the
motion.

Within the holostar's interior $B = r_0 / r$. Therefore the
equations of motion reduce to the following set of simple
equations:

\begin{equation} \label{eq:beta:r2:int}
\beta_r^2(r) = 1 - \frac{r_i}{r}\left(1 -
\beta_i^2\left(1-\frac{r_i^2}{r^2}\right)\right) =
\left(\frac{dr}{dt} \frac{r}{r_0}\right)^2
\end{equation}

\begin{equation} \label{eq:beta:t2:int}
\beta_\perp^2(r) = \beta_i^2 \left(\frac{r_i}{r}\right)^3 =
\left(\frac{r d\varphi}{dt}\right)^2 \frac{r}{r_0}
\end{equation}

The equation for $\varphi$ can be expressed as a function of
radial position $r$, instead of time:

\begin{equation} \label{eq:phi}
\frac{d\varphi}{dr} = \frac{r_i \beta_i}{r^2 \beta_r(r)}
\sqrt{\frac{r_i}{r_0}}
\end{equation}

Far away from the turning point of the motion one can approximate
equation (\ref{eq:phi}) This approximation will turn out useful
later. We know from equation (\ref{eq:beta:t2:int}) that the
tangential velocity component $\beta_\perp$ along the trajectory
of any geodesically moving particle becomes arbitrarily small,
whereas the radial velocity component $\beta_r$ becomes nearly
unity, when we are far away from the turning point of the motion.
(This shows, that geodesic motion always is nearly radial far away
from the turning point of the motion $r_i$) Whenever we know the
(small) tangential velocity-component $\beta_e = \beta_\perp(r_e)$
of a particle at any arbitrary radial position $r_e$ along its
trajectory, we can express the above equation in terms of
$\beta_e$ and $r_e$:

\begin{equation} \label{eq:phi:j}
\frac{d\varphi}{dr} = \frac{r_e \beta_e}{\beta_r r^2}
\sqrt{\frac{r_e}{r_0}} \simeq \frac{r_e \beta_e}{r^2}
\sqrt{\frac{r_e}{r_0}}
\end{equation}

because

$$\beta_r^2 = 1 - \frac{r_i}{r}(1-\beta_i^2) - \beta_i^2 \frac{r_i^3}{r^3} \simeq 1$$

for $r > r_e \gg r_i$.

and

$$ \beta_i r_i^{\frac{3}{2}} = \beta_\perp(r_e) r_e^{\frac{3}{2}}$$

from equation (\ref{eq:beta:t2:int}).

Equation (\ref{eq:phi}) determines the orbit of the particle in
the spatial geometry. It is not difficult to integrate. For $r \gg
r_i$ the radial velocity $\beta_r(r)$ is nearly unity, independent
of the nature of the particle and of its velocity at the turning
point of the motion, $\beta_i$. In the region $r \gg r_i$ we find
$d\varphi \propto dr/r^2$, so that the angle remains nearly
unchanged. This implies, that the number of revolutions of an
interior particle around the holostar's center is limited. The
radial coordinate position $r_{1}$ from which an interior particle
can at most perform one more revolution is given by
$\varphi(\infty) - \varphi(r_{1}) < 2 \pi$. Expressing $r_1$ in
multiples of $r_i$ yields:

\begin{equation}
\frac{r_1}{r_i} > \beta_i \sqrt{\frac{r_i}{r_0}} \frac{1}{2 \pi}
\end{equation}

Whenever $r_i \approx r_0$ the particle will not be able to
complete even one full revolution. For $r_i \simeq r_0$ and
$\beta_i \simeq 1$, a choice of parameters which will turn out to
be of relevance later, the maximum angle that can be covered is
$\Delta \varphi \simeq 1$, corresponding to roughly $57^\circ$.

The total number of revolutions of an arbitrary particle, emitted
with tangential velocity component $\beta_i$ from radial
coordinate position $r_i$ is very accurately estimated by the
number of revolutions in an infinitely extended holostar:

\begin{equation} \label{eq:Nrev}
N_{rev} \simeq \frac{1}{2 \pi}
\int_{r_i}^\infty{\frac{d\varphi}{dr}} = \beta_i
\sqrt{\frac{r_i}{r_0}} \frac{1}{2 \pi} \int_{0}^1{\frac{dx}{
\sqrt{1 - x \left(1 - \beta_i^2(1-x^2)\right)}}}
\end{equation}

The definite integral in equation (\ref{eq:Nrev}) only depends on
$\beta_i$. It is a monotonically decreasing function of
$\beta_i^2$. Its value lies between $1.4022$ and $2$ (the lowest
value is attained for $\beta_i = 1$, i.e. for photons\footnote{The
exact value of the definite integral for $\beta_i=1$ is given by
$\frac{\sqrt{\pi}}{3}
\frac{\Gamma(\frac{1}{3})}{\Gamma(\frac{5}{6})}$}, and the higher
value for $\beta_i = 0$, i.e. for pure radial motion of massive
particles.\footnote{In this case the number of revolutions is
zero, as $\beta_i = 0$}) From equation (\ref{eq:Nrev}) it is quite
obvious that particles emitted from the central region of the
holostar will cover only a small angular portion of the interior
holostar space-time.

In the following sections we are mainly interested in the radial
part of the motion of the particles.

From equation (\ref{eq:Veff:beta}) it can be seen, that whenever
$B(r)$ is monotonically decreasing, the effective potential is
monotonically decreasing as well, independent of the constants of
the motion, $r_i$ and $\beta_i$. Within the holostar $B(r) =
r_0/r$, therefore the interior effective potential decreases
monotonically from the center at $r=0$ to the boundary at $r =
r_h$, which implies that the radial velocity of an outmoving
particle increases steadily. The motion appears accelerated from
the viewpoint of an exterior asymptotic observer. The perceived
acceleration decreases over time, as the effective potential
becomes very flat for $r \gg r_i$.

For all possible values of $r_i$ and $\beta_i$ the position of the
membrane is a local minimum of the effective potential. Within the
holostar's interior the effective potential is monotonically
decreasing with $r$. Therefore any particle starting out from the
holostar's interior and moving geodesically must cross the
membrane, finding itself in the exterior space-time. Any interior
particle has two options: Either it oscillates between the
interior and the exterior space-time or it passes over the angular
momentum barrier situated in the exterior space-time (or tunnels
through) and escapes to infinity.

In order to discuss the general features of the radial motion it
is not necessary to solve the equations exactly. For most purposes
we can rely on approximations.

For photons the radial equation of motion is relatively simple:

\begin{equation} \label{eq:beta:r2:photon}
\beta_r(r) = \sqrt{1 - \left(\frac{r_i}{r}\right)^3} =
\frac{dr}{dt} \frac{r}{r_0}
\end{equation}

An exact integration requires elliptic functions. For $r \gg r_i$
the term $(r_i/r)^3$ under the root can be neglected, so that the
solution can be expressed in terms of elementary functions.

For massive particles the general equation (\ref{eq:beta:r2:int})
is very much simplified for pure radial motion, i.e. $\beta_i =
0$:

\begin{equation} \label{eq:beta:r2:m:radial}
\beta_r(r) = \sqrt{1 - \frac{r_i}{r}}=\frac{dr}{dt} \frac{r}{r_0}
\end{equation}

This equation can be integrated with elementary mathematical
functions. The general equation of motion (\ref{eq:beta:r2:int})
requires elliptic integrals. However, in the general case of the
motion of a massive particle ($\beta_i \neq 0$ and $\beta_i^2 <
1$) the equation for the radial velocity component
(\ref{eq:beta:r2:int}) can be approximated as follows for large
values of the radial coordinate coordinate, $r \gg r_i$:

\begin{equation} \label{eq:beta:r2:int_approx}
\beta_r^2(r) \simeq 1 - \frac{r_i}{r}\left(1 - \beta_i^2\right) = 1 -
\frac{r_i}{r} \frac{1}{\gamma_i^2}
\end{equation}

Whenever $r \gg r_i$ the solution to the general radial equation
of motion for a massive particle is very well approximated by the
much simpler, analytic solution for pure radial motion of a
massive particle given by equation (\ref{eq:beta:r2:m:radial}).
One only has to replace $r_i$ by $r_i / \gamma_i^2$. The radial
component of the motion of a particle emitted with arbitrary
$\beta_i$ from $r_i$ is nearly indistinguishable from the motion
of a massive particle that started out "at rest" in a purely
radial direction from a somewhat smaller "fictitious" radial
coordinate value $\widetilde{r_i} = r_i/\gamma_i^2$.

\subsection{Bound versus unbound motion}

In this section I discuss some qualitative features of the motion
of massive particles and photons in the holostar's gravitational
field.

An interior particle is bound, if the effective potential at the
interior turning point of the motion, $r_i$, is equal to the
effective exterior potential at an exterior turning point of the
motion, $r_e$. The effective potential has been normalized such,
that at any turning point of the motion $V_{eff}(r) = 1$. A
necessary (and sufficient) condition for bound motion then is,
that the equation

\begin{equation} \label{eq:cond:turningpoint}
V_{eff}(r_e)=1
\end{equation}

has a real solution $r_e \geq r_h$ in the exterior (Schwarzschild)
space-time.

\subsubsection{Bound motion of massive particles}

For pure radial motion and particles with non-zero rest-mass
equation (\ref{eq:cond:turningpoint}) is easy to solve. We find
the following relation between the exterior and interior turning
points of the motion:

\begin{equation} \label{eq:re:0}
r_{e} = \frac{r_+}{1-\frac{r_0}{r_i}}
\end{equation}

Equation (\ref{eq:re:0}) indicates, that for massive particles
bound orbits are only possible if $r_i > r_0$. If the massive
particles have angular momentum, the turning point of the motion,
$r_i$, must be somewhat larger than $r_0$, if the particles are to
be bound. Angular motion therefore increases the central "unbound"
region. The number of massive particles in the "unbound" central
region, however, is very small. In \cite{petri/thermo} it will be
shown, that the number of (ultra-relativistic) constituent
particles within a concentric interior region of the holostar in
thermal equilibrium is proportional to the boundary surface of the
region with $N \approx {(r/(r_0/2))}^2$.

Let us assume that a massive particle has an interior turning
point of the motion far away from the central region, i.e. $r_i
\gg r_0$. Equation (\ref{eq:re:0}) then implies, that for such a
particle the exterior turning point of the motion will lie only a
few Planck-distances outside the membrane. This can be seen as
follows:

If a massive particle is to venture an appreciable distance away
from the membrane, the factor $1- r_0 / r_i$ on the right hand
side of equation (\ref{eq:re:0}) must deviate appreciably from
unity. This is only possible if $r_0 \approx r_i $. In the case
$r_i \gg r_0$ equation (\ref{eq:re:0}) gives a value for the
exterior turning point of the motion, $r_e$, that is very close to
the gravitational radius of the holostar. Any particle emitted
from the region $r_i\gg r_0$ will barely get past the membrane.
Even particles whose turning point of the motion is as close as
two fundamental length units from the center of the holostar, i.e.
$r_i = 2 r_0$, have an exterior turning point of the motion
situated just one gravitational radius outside of the membrane at
$r_e = 2 r_+$.

The above analysis demonstrates, that only very few, if any,
massive particles can escape the holostar's gravitational field on
a classical geodesic trajectory. As has been remarked before, an
escape to infinity is only possible for massive particles (with
zero angular momentum) emanating from a sub Planck-size region of
the center ($r_i < r_0$). It is quite unlikely that this region
will contain more than one particle.

In the case of angular motion the picture becomes more
complicated. In general the equation $V(r_e) = 1$ is a cubic
equation in $r_e$:

\begin{equation} \label{eq:re}
B(r_i) = \frac{r_0}{r_i} = (1 -
\frac{r_+}{r_e})\left(1-\beta_i^2(1-\frac{r_i^2}{r_e^2})\right)
\end{equation}

It is possible to solve this equation by elementary methods. The
formula are quite complicated. The general picture is the
following: For any given $r_i$ the particle becomes "less
bound\footnote{i.e. $r_e$ becomes larger than the value given in
equation (\ref{eq:re:0})}", the higher the value of $\beta_i^2$ at
the interior turning point gets. Particles with interior turning
point of the motion close to the center are "less bound" than
particles with interior turning point close to the boundary. For
sufficiently small $r_i$ bound motion is not possible, whenever a
certain value of $\beta_i^2$ is exceeded.

For particles with $r_i \ll r_h < r_e$ equation (\ref{eq:re}) is
simplified, as $r_i^2/r_e^2$ can be neglected. We find

\begin{equation} \label{eq:re:small}
\frac{r_0}{r_i} \approx (1 - \frac{r_+}{r_e})(1-\beta_i^2) = (1 -
\frac{r_+}{r_e})\frac{1}{\gamma_i^2}
\end{equation}

which allows us to determine $r_e$ to a fairly good approximation:

\begin{equation} \label{eq:re:approx}
r_{e} \approx \frac{r_+}{1-\frac{r_0 \gamma_i^2}{r_i}}
\end{equation}

The motion is bound, as long as $r_i > \gamma_i^2 r_0$. This
inequality seems to indicate, that any massive particle is able to
escape the holostar in principle, as long as $\gamma_i$ is high
enough. This however, is not true:

The geodesic motion of massive particles and photons is described
by an effective potential, which is a function of $r_i$ and
$\beta_i$. From equation (\ref{eq:Veff:beta}) it can be seen, that
the exterior effective potential for fixed $r_i$ is a
monotonically decreasing function of $\beta_i^2$, i.e. $V_{eff}(r,
\beta_i^2) \geq V_{eff}(r, 1)$. This is true for any radial
position $r$. Therefore the exterior effective potential of any
particle with fixed $r_i$ and arbitrary $\beta_i$ is bounded from
below by the effective potential of a photon. This implies that
whenever a photon possesses an exterior turning point of the
motion, any particle with $\beta_i^2 < 1$ must have an exterior
turning point as well, somewhat closer to the holostar's membrane.
Therefore escape to infinity from any interior position $r_i$ for
an arbitrary rest-mass particle is only possible, if a photon can
escape to infinity from $r_i$.

In the following section the conditions for the unbound motion of
photons are analyzed.

\subsubsection{Unbound motion of photons}

The effective potential for photons at spatial infinity is zero.
Therefore in principal any photon has the chance to escape the
holostar permanently. Usually this doesn't happen, due to the
photon's angular momentum.

Disregarding the quantum-mechanical tunnel-effect, the fate of a
photon, i.e. whether it remains bound in the gravitational field
of the holostar or whether it escapes permanently on a classical
trajectory to infinity, is determined at the angular momentum
barrier, which is situated in the exterior space-time at $r = 3
r_+ / 2$. The exterior effective potential for photons possesses a
local maximum at this position. If the effective potential for a
photon at $r = 3 r_+ / 2$ is less than 1, i.e. $V_{eff}(3 r_+/2) <
1$, the photon has a non-zero radial velocity at the maximum of
the angular momentum barrier. Escape is classically inevitable.
The condition for (classical) escape for photons is thus given by:

\begin{equation} \label{eq:Veff:1}
\frac{r_i}{r_0} \leq 3 \left(\frac{r_+}{2
r_0}\right)^{\frac{2}{3}}
\end{equation}

Any photon emitted from the region defined by equation
(\ref{eq:Veff:1}) will escape, if its motion is purely geodesic.

On the other hand, any photon whose internal turning point of the
motion lies outside the "photon-escape region" defined by equation
(\ref{eq:Veff:1}) will be turned back at the angular momentum
barrier and therefore is bound. Likewise, all massive particles
outside the photon escape region must be bound as well, because
the exterior effective potential of a massive particle is always
larger than that of a photon with the same $r_i$, i.e. the angular
momentum barrier for a massive particle is always higher than that
of a photon emitted from the same $r_i$. Therefore every particle
with interior turning point of the motion outside the "photon
escape region" will be turned back by the angular momentum
barrier.

For large holostars, the region of classical escape for photons
becomes arbitrarily small with respect to the holostar's overall
size. A holostar of the size of the sun with $r_+/r_0 \approx
10^{38}$ has an "unbound" interior region of $r_i \leq 4. 10^{25}
r_0 \approx 0.5 \, nm$. The radial extension of the "photon escape
region" is 13 orders of magnitude less than the holostar's
gravitational radius. The gravitational mass of this region is
negligible compared to the total gravitational mass of the
holostar.\footnote{Quite interestingly, for a holostar of the mass
of the universe ($r \approx 10^{61} r_{Pl}$), the temperature at
the radius of the photon-escape region is $T \approx 2.7 \cdot
10^{10} K \approx 2.3 MeV$, which is quite close to the
temperature of nucleosynthesis.}

\subsubsection{Unbound motion of massive particles}

For particles with non-zero rest-mass the analysis is very much
simplified, if the effect of the angular momentum barrier is
neglected.\footnote{For most combinations of $r_i$ with $\beta_i^2
< 1$, the angular momentum barrier hasn't a significant effect on
the question, whether a the particle is bound or unbound.}
Massive particles generally have an effective potential at spatial
infinity larger than zero. A necessary, but not sufficient
condition for a massive particle to be unbound is, that the
effective potential at spatial infinity be less than 1. This
condition translates to:

\begin{equation} \label{eq:re:1}
r_i < \frac{r_0}{1 - \beta_i^2} = r_0 \gamma_i^2
\end{equation}

According to equation (\ref{eq:re:1}) escape on a classical
geodesic trajectory for a massive particle is only possible from a
region a few Planck-lengths around the center, unless the particle
is highly relativistic. For example, massive particles with
$\beta_i^2 = 0.5$ can only be unbound, if they originate from the
region $r_i < 2 r_0$. If the region of escape for massive
particles is to be macroscopic, the proper tangential velocity
$\beta_i^2$ at the turning point of the motion must be
phenomenally close to the local speed of light. Note however that
in such a case there usually is an angular momentum barrier in the
exterior space-time (see the discussion in the previous section).

\subsection{An upper bound for the particle flux to infinity}

The lifetime of a black hole, due to Hawking evaporation, is
proportional to $M^3$. Hawking radiation is independent of the
interior structure of a black hole. It depends solely on the
exterior metric up to the event horizon. As the exterior
space-times of the holostar and a black hole are identical
(disregarding the Planck-size region between gravitational radius
and membrane) the estimated lifetime of the holostar, due to loss
of interior particles, should not significantly deviate from the
Hawking result.

An upper bound for the flux of particles from the interior of the
holostar to infinity can be derived by the following, albeit
very crude argument:

Under the - as we will later see, unrealistic - assumption, that
the effects of the negative radial pressure can be neglected, the
particles move on geodesics and the results derived in the
previous sections can be applied.

For large holostars and ignoring the pressure the particle flux to
infinity will be dominated by photons or other zero rest-mass
particles, such as neutrinos, emanating from the "photon escape
region" $r_i < C r_+^{2/3} r_0^{1/3}$ defined by equation
(\ref{eq:Veff:1}).

The gravitational mass $\Delta M$ of this region, viewed by an
asymptotic observer at infinity, is proportional to $r_+^{2/3}
r_0^{1/3}$.

The exterior asymptotic time $\Delta t$ for a photon to travel
from $r_i$ to the membrane at $r_h \simeq r_+$ is given by:

$$\Delta t = \int_{r_i}^{r_+}{\sqrt{\frac{A}{B}}\frac{dr}{\beta_r}}
= \int_{r_i}^{r_+}{\frac{r}{r_0}
\frac{dr}{\sqrt{1-(\frac{r_i}{r}})^3}}$$

For a large holostar with $r_i \ll r_+$ the integral can be
approximated by:

$$\int_{r_i}^{r_+}{\frac{r}{r_0} \frac{dr}{\sqrt{1-(\frac{r_i}{r}})^3}}
\approx \int_{0}^{r_+}{\frac{r}{r_0} dr} = \frac{r_+^2}{2 r_0}$$

Note, that the time of travel from the membrane, $r_h$, to a
position $r$ well outside the gravitational radius of the holostar
is of order $r - r_h$, i.e. very much shorter than the time of
travel from the center of the holostar to the membrane, which is
proportional to $r_h^2$.

Under the assumption that the continuous particle flux to infinity
is comparable to the time average of the - rather conceived -
process, in which the whole "photon escape region" is moved in one
bunch from the center of the holostar to its surface, one finds
the following estimate for the mass-energy-flux to infinity for a
large holostar:

\begin{equation} \label{eq:flux:1}
\frac{\Delta M}{\Delta t} \propto {\left( \frac{r_0}{r_+}
\right)}^{4/3} \propto {\left( \frac{\sqrt{\hbar}}{M}
\right)}^{4/3}
\end{equation}

This flux is larger than the flux of Hawking radiation, for
which the following relation holds:

$$\frac{dM}{dt} \propto \left(\frac{\sqrt{\hbar}}{M}\right)^2$$

However, the pressure effect has not been taken into account in
equation (\ref{eq:flux:1}). As will be shown in the following
sections, the pressure reduces the photon flux two-fold: First it
reduces the local energy of the outward moving photons, so that
less energy is transported to infinity. Second, if the local
energy of the individual photons is reduced with respect to pure
geodesic motion, the chances of classical escape for a photon are
dramatically reduced, because most photons will not have enough
"energy" to escape when they finally reach the pressure-free
region beyond the membrane.

The first effect reduces the energy of the photon flux by a factor
$r_+^{-1/3}$, as can be derived from the results of the following
section. This tightens the bound given in equation
(\ref{eq:flux:1}) to $dM/dt \propto 1/M^{5/3}$.

The second effect will effectively switch off the flux of photons.
As will be shown later, the energy of an ensemble of photons
moving radially outward or inward changes in such a way, that the
ensemble's local energy density is always proportional to the
local energy density of the interior matter it encounters along
its way. Therefore an ensemble of photons "coming from the
interior", having reached the radial position $r_h$ of the
membrane, will be indistinguishable from the photons present at
the membrane. The majority of the photons at the membrane,
however, will have a turning point of the motion close to $r_h$,
meaning that escape on a classical trajectory is impossible.
Therefore, whenever photons coming from the interior have reached
the surface of the holostar, $r_h$, their energy will be so low,
that the vast majority of the photons are bound and will become
trapped in the membrane. Classically it appears as if no photon
will be able to escape from the holostar.

Massive particles which have high velocities at their interior
turning points of the motion behave like photons. Therefore the
discussion of the previous paragraph applies to those particles as
well. Highly relativistic massive particles will not be able to
carry a significant amount of mass-energy to infinity. For massive
particles escape to infinity is only possible, if the motion
starts out from a region within one (or two) fundamental lengths
of the center (see equation (\ref{eq:re:1})) . But this region
contains only very few particles, if any at all. Furthermore one
has to ensure, that the motion of the massive particles does not
become highly relativistically later on. In such a case the
massive particles would behave similar to photons and would be
subject to the same energy-loss as experienced by the photons.

The holostar therefore must be regarded as classically stable,
just as a black hole. Once in a while, however, a particle
undergoing random thermal motion close to the surface might
acquire sufficient energy in order to escape or tunnel through the
angular momentum barrier. Furthermore there are the tidal
forces in the exterior space-time, giving rise to "normal" Hawking
evaporation.

Taking the pressure-effects into account, the mass-energy flux to
infinity of the holographic solution will be quite comparable to
the mass-energy flux due to Hawking-evaporation of a black hole.
The exponent $x$ in the energy-flux equation $dM/dt \propto 1/M^x$
will lie somewhere between $5/3$ and $2$, presumably quite close,
if not equal to $2$.

Even the very crude upper bound of equation (\ref{eq:flux:1})
yields quite long life-times. For a holostar with the mass of the
sun, the evaporation time due to equation (\ref{eq:flux:1}) is
still much larger than the age of the universe ($T \approx 10^{44}
s$).

\subsection{On the motion of ultra-relativistic particles - pressure effects and self-consistency}

In this section the effect of the pressure on the internal motion
of ultra-relativistic particles within the holostar is studied. I
will demonstrate that the negative, purely radial pressure, equal
in magnitude to the mass-density, is an essential property of the
holostar, if it is to be a self-consistent static solution.

\subsubsection{Geodesic motion of a spherical thin shell of photons}

Let us consider the radial movement (outward or inward) of a
spherical shell of particles with a proper thickness $\delta l =
\sqrt{r/r_0} \delta r$, situated at radial coordinate position $r$
within the holostar. The shell has a proper volume of $\delta V =
4 \pi r^2 \delta l$ and a total local energy content of $\rho
\delta V = \delta l /2$.

In this section the analysis will be restricted to zero rest-mass
particles, referred to as photons in the following discussion. In
the geometric optics approximation photons move along
null-geodesic trajectories. Note that the pressure will have an
effect on the local energy of the photon. However, as the local
speed of light is independent of the photon's energy, the pressure
will not be able to change the geometry of a photon trajectory,
i.e. the values of $r(t), \theta(t), \varphi(t)$ as determined
from the equations for a null-geodesic trajectory.

For pure radial motion of photons there can be no cross-overs,
i.e. no particle can leave the region defined by the two
concentric boundary surfaces of the shell.

The equation of motion for a null-geodesic in the interior of the
holostar is given by:

\begin{equation} \label{eq:motion:photon}
\frac{dr}{dt} = \frac{r_0}{r} \sqrt{1-\frac{r_i^3}{r^3}}
\end{equation}

For $r \gg r_i$ the square-root factor is nearly one. Therefore
whenever the photon has reached a radial position $r \approx 10
r_i$, a negligible error is made by setting $r_i=0$, which
corresponds to pure radial motion.

Equation (\ref{eq:motion:photon}) with $r_i = 0$ has the solution:

\begin{equation} \label{eq:r(t)photon}
r(t) = \sqrt{2 r_0 t - r^2(0)}
\end{equation}

The radial distance between two photons, one travelling on the
inner boundary of the shell, starting out at $r(0) = r_i$, one
travelling on its outer boundary, starting out from $r(0) = r_i +
\delta r_i$, is given by:

\begin{equation} \label{eq:dr(t)photon}
\delta r(t) = \sqrt{2 r_0 t - (r_i + \delta r_i)^2} - \sqrt{2 r_0 t - r_i^2}
\end{equation}

If $r(t) \gg r_i$, Taylor-expansion of the square-root yields:

\begin{equation} \label{eq:dr(t)photon:appr}
\delta r \approx \delta r_i \frac{r_i}{r}
\end{equation}

In terms of the proper thickness $\delta l = \delta r \sqrt{A}$ of
the shell we find the following relation:

\begin{equation} \label{eq:dr/dl:photon}
\frac{\delta l}{\delta l_i} = \sqrt{\frac{r_i}{r}}
\end{equation}

Therefore, if the shell moves radially outward with the local speed of
light, its proper thickness changes according to an inverse square
root law.

\subsubsection{\label{sec:motion:shell:rad}Geodesic expansion of photons against the holostar pressure}

Whenever the proper radial thickness changes during the movement
of the shell, work will be done against the negative radial
pressure. The rate of change in proper thickness $d (\delta l(r))$
per radial coordinate displacement $dr$ of the shell is given by:

\begin{equation} \label{eq:dl:photon}
d (\delta l) = - \delta l_i \sqrt{r_i} \frac{dr}{2 r^{3/2}}
\end{equation}

Due to the anisotropic pressure (the two tangential pressure
components are zero) any change in volume along the tangential
direction will have no effect on the total energy. The purely
radial pressure has only an effect on the energy of the shell, if
the shell expands or contracts in the radial direction. The work
done by the negative radial pressure therefore is given by:

\begin{equation} \label{eq:dE:photon}
dE = - P_r 4 \pi r^2 d (\delta l) = - \delta l_i \sqrt{r_i} \frac{dr}{4 r^{3/2}}
\end{equation}

Because the radial pressure is negative, the total energy of the
shell is reduced when the shell is compressed along the radial
direction.

The total pressure-induced local energy change of the shell, when
it is moved outward from radial coordinate position $r_i$ to
another position $r \gg r_i$ is given by the following integral:

\begin{equation} \label{eq:dE2:photon}
\Delta E = \int_{r_i}^{r}{dE} = \frac{\delta l_i}{2}
\left(\sqrt{\frac{r_i}{r}} - 1\right)
\end{equation}

But $\delta l_i/2 = \rho_i \delta V_i$ with $\rho_i = 1 / (8 \pi
r_i^2)$ and $\delta V_i = 4 \pi r_i^2 \delta l_l$ is nothing else
than the original total local energy of the shell, $\delta E_i$.
Therefore we find the following expression for the final energy of
the shell:

\begin{equation} \label{eq:DeltaE:photon}
\delta E = \delta E_i + \Delta E = \delta E_i \sqrt{\frac{r_i}{r}}
\end{equation}

This result could also have been obtained by assuming the ideal
gas law $\delta E \propto P_r \delta V$ with $\delta V = 4 \pi r^2
\delta l$.

The total energy density in the shell therefore changes according
to the following expression:

\begin{equation} \label{eq:rho(r):photon}
\rho(r) = \frac{\delta E}{\delta V} =
 \frac{\delta E_i}{4 \pi r^2 \delta l_i} = \frac{\delta E_i}{\delta V_i} \frac{r_i^2}{r^2}
 = \rho(r_i) \frac{r_i^2}{r^2}
\end{equation}

We have recovered the inverse square law for the mass-density. The
holographic solution has a remarkable self-consistency. Any
spherical shell carrying a fraction of the total local energy,
moving inward or outward with the local velocity of light, changes
its energy due to the negative radial pressure in such a way, that
the local energy density of the shell at any radial position $r$
always remains exactly proportional to the actual local energy
density $\rho(r)$ of the (static) holographic
solution.\footnote{Strictly speaking, it was only shown, that the
energy-density follows an inverse square law, if the total
mass-energy within the shell moves outward (or inward) on a
null-geodesic trajectory. However, the argument can be applied to
any fraction of the total energy, if one postulates, that not the
total pressure, but only the partial pressure corresponding to the
moving fraction is used in equation (\ref{eq:dE:photon}), i.e.
$\rho = c / (8 \pi r^2) \rightarrow P_r = - c / (8 \pi r^2)$.}

This feature is essential, if the holographic solution is to be a
self consistent (quasi-) static solution.\footnote{The term
quasi-static is used, because internal motion within the holostar
is possible, as long as there is no net mass-energy flux in a
particular direction. A radially directed outward flux of (a
fraction of the) interior matter is possible, if this flux is
compensated by an equivalent inward flux. Note that the outward
and inward flowing matter must not necessarily be of the same
kind. If the outward flowing matter consists of massive particles
with a finite life-time, the inward flowing matter is expected to
carry a higher fraction of the decay products, which will be
lighter, possibly zero rest-mass particles.} A holostar evidently
contains matter. A fraction of this matter will consist of
non-zero rest-mass particles. These particles will undergo random
thermal motion. Furthermore, due to Hawking evaporation or
accretion processes there might be a small outward or inward
directed net-flux of mass-energy between the interior central core
region and the boundary. Even if there is no net mass-energy flux,
the different particle species might have non-zero fluxes, as long
as all individual fluxes add up to zero. If the internal motion
(thermal or directed) takes place in such a way, that the local
mass-density of the holostar ($\rho \propto 1/r^2$) is
significantly changed from the inverse square law in a time scale
shorter than the Hawking evaporation time-scale, the holostar
cannot be considered (quasi-) static.

Movement of an interior particle at the speed of light from $r_i
\simeq r_0 $ to the membrane at $r = r_h $ takes an exterior time
$t \simeq r_h^2/(2r_0) \propto M^2$. The movement of a particle
from $r_i = r_h/2$ to the membrane is not much quicker: $t \simeq
3 r_h^2/(8 r_0)$. In any case the time to move through the
holostar's interior is much less than the Hawking evaporation
time-scale, which scales as $M^3$. Therefore a necessary condition
for the holostar to be a self-consistent, quasi-static solution
is, that the radial movement of a shell of zero rest mass
particles should not disrupt the local (static) mass-density.

More generally any local mass-energy fluxes should take place in
such a way, that the overall structure of the holostar is not
destroyed and that local mass-energy fluxes aren't magnified to an
unacceptable level at large scales.

\subsubsection{\label{sec:ultrarel:gas}Geodesic expansion of photons against their radiation pressure}

The reader might not be satisfied with the assumption, that the
photons of a radially out- or in-moving shell should be subject to
the holostar-pressure. The anisotropic holostar-pressure is very
different from the isotropic pressure of an ultra-relativistic
gas. Although anisotropic pressures might become important at high
densities and temperatures, we shouldn't necessarily expect this
to be the case at the low end of the energy scale. Furthermore, at
low energy-densities and temperatures the mass-energy density
within any spherical thin shell of the holostar will consist only
out of a very small fraction of massless or extremely light
ultra-relativistic particles, such as photons or neutrinos. A much
higher fraction of particles will reside in the massive particle
species, such as baryons. At low temperatures one would rather
think, that the partial pressure of the baryons is negligible, so
that the massless particles only expand/contract against their own
radiation pressure. Under these circumstances it is far from
clear, whether the geodesic movement of radiation, expanding (or
contracting) against its own radiation pressure, will preserve the
inverse-square law for the energy density.

The remarkable - non-trivial - answer to this question is yes, as
will be shown in the following lines. Let us assume, that only a
fraction of the total energy content of a radially outmoving shell
consists out of massless particles, i.e.

$$\rho_\gamma = c \rho = \frac{c}{8 \pi r^2}$$

The partial pressure of the massless particles in the shell is
then given by the equation of state for an ultra-relativistic gas:

$$P_\gamma = \frac{\rho}{3} = \frac{c}{24 \pi r^2}$$

We have already seen, that a shell of massless particles contracts
in the radial direction by an inverse square-root
law.\footnote{This feature is independent of the fact, whether the
total mass-energy within a given shell or only a small fraction
(or just a single particle) moves inward/outward. The equations of
motion within the holostar solution are defined by the metric,
which is sensitive only to the total energy density, according to
the Einstein equations. Furthermore, the equivalence principle
tells us, that the motion of any one massless particle in a static
space-time (i.e. a space-time with a static metric!) doesn't
depend on the number of energy-density of the particles moving
subject to the metric (as long as the movement of the particles
doesn't change the metric).} Yet there is an expansion in the two
tangential directions proportional to $r^2$, so that the volume of
a shell of radially moving massless particles changes as:

$$V_\gamma = V_i \left( \frac{r}{r_i}\right)^{\frac{3}{2}}$$

From this we can calculate the energy change in the shell due to
the expansion of the ultra-relativistic particles against their
own - isotropic - partial pressure:

$$dE = - P_\gamma dV_\gamma = \frac{V_i}{r_i} \frac{c}{12 \pi r^2} \left( \frac{r}{r_i}\right)^{\frac{1}{2}} $$

The total energy at radial position $r$ is then given by:

$$E_\gamma = E_i + \int_{r_i}^r{dE} = E_i \left( \frac{r_i}{r}\right)^{\frac{1}{2}} $$

where $E_i$ is the original mass-energy in the shell at radial
position $r_i$, i.e. $E_i = \rho_\gamma(r_i) V_i$. We find, that
the energy of the shell changes with an inverse square-root law,
exactly as derived in the section before using the anisotropic
holostar-pressure.

The mass-energy density is obtained, by dividing $E_\gamma$ by
$V_\gamma$, which gives us:

$$\rho_\gamma = \rho_i \left( \frac{r_i}{r}\right)^{2} $$

Again we have discovered the inverse square law for the energy
density.

In the geometric optics approximation the photon number in the
shell remains constant, so that the individual energy of the
photons must change with $r$ in the same way, as the total energy
of the shell changes. Therefore the energy of a single photon, which is
proportional to its frequency, changes with an inverse square-root
law.

It is quite remarkable that the evolution of the energy-density of
a gas of radially outmoving ultra-relativistic particles is not
affected, whether we use the anisotropic holostar pressure to
calculate the pressure-induced energy-change or whether we use the
isotropic pressure derived from the equation of state for an
ultra-relativistic gas. In both cases the energy and frequency of
an ultra-relativistic particles changes with an inverse
square-root law.

For massive particles the analysis of the motion is more
complicated, as the pressure is expected to have a noticeable
effect not only on the local energy of the particles, but also on
the global space-time trajectory of the massive particles.
Whenever exterior forces - such as the pressure - are present, the
space-time trajectory of a massive particle doesn't follow a
geodesic. The extent of the deviation from geodesic motion
generally depends on the particle's energy. This is contrary to
massless particles, for which the constancy of the speed of light
guarantees, that their space-time trajectories are independent of
energy (at least for low densities).

\subsection{\label{sec:holographic:principle}Number of interior particles and holographic principle}

The self-consistency argument given in the above two sections
leads to some interesting predictions. Let us assume that the
outmoving shell consists of essentially non-interacting zero
rest-mass particles, for example photons or neutrinos (ideal
relativistic gas assumption). For a large holostar the
mass-density in its outer regions will become arbitrarily low.
Therefore the ideal gas approximation should be a reasonable
assumption for large holostars. Under this assumption the number
of particles in the shell should remain constant.\footnote{In
general relativity the geometric optics approximation implies the
conservation of photon number.} This allows us to determine the
number density of zero rest-mass particles within the holostar up
to a constant factor:

Let $N_i$ be the number of particles in the shell at the position
$r_i$. Then the number density $n(r)$ of particles in the shell,
as it moves inward or outward, is given by the respective change
of the shell's volume:

\begin{equation} \label{eq:numberdensity}
n(r) = \frac{N_i}{\delta V} =  \frac{N_i}{4 \pi r^2 \delta l} =
\frac{N_i}{4 \pi \sqrt{r_i} \delta l_i} \frac{1}{r^{\frac{3}{2}}}
= n(r_i) \left(\frac{r_i}{r}\right)^{\frac{3}{2}}
\end{equation}

Under the assumption that the interior matter-distribution of the
holostar is (quasi-) static, and that the local composition of the
matter at any particular $r$-position doesn't change with time,
self-consistency requires that the number density of zero
rest-mass particles per proper volume should be proportional to
the number density predicted by equation (\ref{eq:numberdensity}).

The total number of zero rest-mass particles in a holostar will
then by given by the proper volume integral of the number density
(\ref{eq:numberdensity}) over the whole interior volume:

\begin{equation} \label{eq:NumberParticles}
N \propto \int_0^{r_h}{\frac{dV}{r^{3/2}}} = \int_0^{r_h}{\frac{4
\pi r^2 \sqrt{\frac{r}{r_0}}dr}{r^{3/2}}} \propto r_h^2 \propto
A_h
\end{equation}

We arrive at the remarkable result, that the total number of zero
rest-mass particles within a holostar should be proportional to
the area of its boundary surface, $A_h$. The same result holds for
any concentric sphere within the holostar's interior. This is an -
albeit still very tentative - indication, that the holographic
principle is valid in classical general relativity, at least for
large compact self gravitating objects.

Under the assumption that the region $r < r_0/2 \approx
\sqrt{\hbar}$ can contain at most one particle (see the discussion
in \cite{petri/charge}), we find:

\begin{equation} \label{eq:N:photon}
N = \left(\frac{r_h}{r_0/2}\right)^2 \approx \frac{r_h^2}{\hbar} = \frac{A_h}{4 \pi \hbar} = \frac{S_{BH}}{\pi}
\end{equation}

$S_{BH}$ is the Bekenstein-Hawking entropy. The entropy per
particle $s$ is roughly equal to $\pi$. This quite close to the
entropy per particle of an ultra-relativistic gas, which - in the
case of zero chemical potential - amounts to roughly 4.2 for
fermions and 3.6 for bosons.

The holostar solution is not only compatible with the holographic
principle, it is the most radical fulfilment of this principle:
The number of degrees of freedom of the holostar scales with its
boundary area. But the fundamental degrees of freedom are not
localized exclusively at the boundary: The degrees of freedom of
the system can be identified with the holostar's well defined
interior matter-state.

Even more so, every relativistic particle in the holostar's
interior has a well defined entropy comparable to the entropy per
particle according to an ideal gas law. The sum of the entropies
of the individual particles is equal to the Hawking entropy of a
same sized black hole (up to a factor of order unity) under the
following assumptions:

\begin{itemize}
\item The holostar's entropy is dominated by zero rest-mass
particles \item The region $r <\approx r_0/2 \approx r_{Pl}$
contains one particle \item The entropy per particle is comparable
to the entropy per particles of a relativistic ideal gas
\end{itemize}

We will see later, that ultra-relativistic (zero rest mass)
particles are the dominant particle species (by numbers) in the
holostar's interior space-time. However, the assumption that the
holostar's entropy is dominated by the ultra-relativistic
particles is only correct in the radiation dominated central
region of the interior space-time. For large $r$ the interior
becomes matter-dominated. In the matter dominated era the entropy
is dominated by the massive species.

From equation (\ref{eq:N:photon}) one can estimate the
number-density $n_\gamma$ of the zero rest mass particles at high
temperatures. We find:

\begin{equation}
n_\gamma \approx \frac{\sqrt{r_0}}{2 \pi \hbar} \cdot \frac{1}{r^\frac{3}{2}}
\end{equation}

\subsection{\label{sec:Tlocal}Local radiation temperature and the Hawking temperature}

Under the assumption that the mass-energy density of the holostar
is dominated by ultra-relativistic particles, the mean energy per
ultra-relativistic particle can be determined from the energy
density $\rho \propto 1/r^2$ and the number-density given in
equation (\ref{eq:N:photon}).

\begin{equation} \label{eq:Tmean}
\overline{E}(r) = \frac{\rho(r)}{n(r)} \approx \frac{\hbar}{4} \frac{1}{\sqrt{r_0 r}} \propto \frac{1}{\sqrt{r}}
\end{equation}

The relation $E \propto 1/\sqrt{r}$ could have been obtained
directly from the pressure-induced energy-change of a geodesically
moving shell of zero-rest mass particles: As the number of
particles in the shell remains constant, but the shell's total
energy changes according to $\sqrt{r_i/r}$, the mean energy per
particle must change in the same way as the total energy of the
shell varies.

In a gas of relativistic particles in thermal equilibrium the mean
energy per relativistic particle is proportional to the local
temperature in appropriate units. This hints at a local radiation
temperature within the holostar proportional to $1 / \sqrt{r}$.
This argument in itself is not yet too convincing. It - so far -
only applies to the low-density regime in the outer regions of a
holostar, where the motion is nearly geodesic and thus interaction
free. It is questionable, if a temperature in the thermodynamic
sense can be defined under such circumstances.

However, there is another argument for a well defined local
radiation temperature with $T \propto 1 / \sqrt{r}$: At the high
pressures and densities within the central region of the holostar
all of the known particles of the Standard Model of particle
physics will be ultra-relativistic and their mutual interactions
are strong enough to maintain a thermal spectrum. The
energy-density of radiation in thermal equilibrium is proportional
to $T^4$. The energy density $\rho$ of the holostar is known to be
proportional to $1/r^2$. Radiation will be the dominant
contribution to the mass-energy at high temperature, so this
argument also hints at a local temperature within the holostar's
central region proportional to $1/\sqrt{r}$.

Therefore it is reasonable to assume that the holostar has a well
defined internal local temperature of its zero-rest mass
constituent particles {\em everywhere}, i.e. not only in the hot
central region, and that this temperature follows an inverse
square-root law in $r$.

This temperature is locally isotropic. This statement should be
self-evident for the high temperature central region of the
holostar, where the radiation has a very short path-length and the
interaction time-scale is short. But one also finds a locally
isotropic temperature in the outer regions of a holostar, where
the radiation moves essentially unscattered. Because of spherical
symmetry, only radiation arriving with a radial component of the
motion at the detector need be considered. Imagine a photon
emitted from the hot inner region of the holostar with an energy
equal to the local temperature at the place of emission, $r_e$.
Due to the square-root dependence of the temperature, the local
temperature at the place of emission $r_e$ will be higher than the
local temperature at the place of the detector $r_a$ by the
square-root of the ratio $r_a / r_e$. However, on its way to the
detector the photon will be red-shifted due to the pressure-effect
by exactly the same (or rather inverse) square-root factor, so
that its energy, when it finally arrives at the detector, turns
out to be equal to the local temperature at the detector. The same
argument applies to a photon emitted from the low-temperature
outer region of the holostar. Due to the pressure effect this
photon will be blue-shifted when it travels towards the detector.
Generally one finds, that the pressure induced red-shift (or
blue-shift) exactly compensates the difference of the local
temperatures between the place of emission $r_e$ and the place of
absorption $r_a$ of a zero rest mass particle.

Disregarding pressure effects, one could naively assume that an
individual photon emitted from an interior position $r_i$ would
undergo gravitational {\em blue} shift, as it moves "down" in the
monotonically decreasing effective potential towards larger values
of $r$. If this were true, a photon moving from a "hot" inner
position to a "cold" outer position would become even hotter. This
result is paradoxical. In fact, the apparent energy change due to
the naive application of the gravitational redshift-formula is
exactly opposite to the pressure-induced effect:

$$\frac{\nu}{\nu_i} = \sqrt{\frac{g_{00}(r_i)}{g_{00}(r)}} = \sqrt{\frac{r}{r_i}}$$

The naive application of the gravitational Doppler-shift formula
to the interior space-time of the holostar leads to grave
inconsistencies. In the holostar's interior the gravitational
Doppler-shift formula is not applicable. This has to do with the
fact, that its derivation not only requires a stationary
space-time, but also relies on the geodesic equations of motion,
which are only the "true" equations of motion in the absence of
exterior forces, i.e. in vacuum! Although particles move on
geodesics in the (rather unrealistic) case of a pressure-free
"dust-universe", this is not true when significant pressures are
present.\footnote{This fact can be experienced by anyone living on
the surface of the earth. None of us, except astronauts in orbit,
move geodesically. Geodesic motion means free fall towards the
earth's center. The pressure forces of the earth's surface prevent
us from moving on such a trajectory. From the viewpoint of general
relativity the earth's surface exerts a constant force
accelerating any object lying on its surface against the direction
of the "gravitational pull" of the earth.} Therefore one cannot
expect the gravitational Doppler-shift law to be applicable in
space-time regions where the pressure is significant. The pressure
not only affects the massive particles, but also the photons, as
the presence of matter affects the group- and phase-velocities of
the photons, which will deviate from the speed of light in vacuum
(although only marginally for the low densities encountered in the
present universe). This fact is well known in cosmology, where the
cosmological red-shift of the CMBR-photons is derived from the
energy loss due to the geodesic expansion of the CMBR-photons
against their own radiation pressure.

Note finally, that the frequency shift due to the interior
pressure applies to all zero rest-mass particles. Furthermore, the
pressure-induced frequency shift is insensitive to the route
travelled by the zero rest-mass particles. It solely depends on
how the volume available to an individual particle has changed,
i.e. only depends on the number-density of the particles which is
a pure function of radial position. If $r_a$ is the radial
position where a photon emitted from $r_e$ with
frequency $\nu_e$ is finally absorbed, one finds.

\begin{equation} \label{eq:Redshift1}
\frac{\nu_e}{\nu_a} = \sqrt{\frac{r_a}{r_e}} = 1+z
\end{equation}

\subsubsection{Geodesically moving radiation preserves the Planck distribution}

It is now an easy exercise to show, that any geodesically moving
shell of zero rest mass particles preserves the Planck-distribution:

The Planck-distribution is defined as:

$$ n(\nu, T) \propto \frac{\nu^2 d \nu}{e^{\frac{\nu}{T}}-1}$$

$n$ is the density of the photons, $\nu$ their individual
frequency and $T$ the temperature. The left side of the equation,
i.e. the number density of the photons with a given frequency,
scales as $1/r^{3/2}$ according to equation
(\ref{eq:numberdensity}). The right side of the equation has the
same dependence. The frequency $\nu$ of any individual photon
scales with $1/r^{1/2}$ according to equation
(\ref{eq:Redshift1}). The same shift applies to the frequency
interval $d \nu$. Therefore the factor $\nu^2 d\nu$ on the right
side of the equation also scales as $1/r^{3/2}$. The argument of
the exponential function, $\nu / T$ is constant, because both the
frequency of an individual photon $\nu(r)$, as well as the overall
temperature $T(r)$ have the same $r$-dependence.

We find that the Planck-distribution is preserved by the geodesic
motion of a non-interacting gas of radiation within the holostar.
This astonishing result directly follows from the holographic
metric and the effects of the negative radial pressure.

\subsubsection{A quick derivation of the Hawking temperature law}

The local inverse square-root temperature law for the holostar's
interior space-time allows us to derive the Hawking temperature
law: The zero rest-mass particles at the surface of the holostar
will have a local radiation temperature proportional to
$1/\sqrt{r_h}$. With the reasonable assumption that this is the
true surface temperature of the holostar, one should be able to
relate this temperature to the Hawking temperature of a black hole
of the same (asymptotic gravitating) mass. The Hawking temperature
is measured in the exterior space-time at {\em spatial infinity}.
As the exterior space-time of the holostar is pressure free, any
particle moving out from the holostar's surface to infinity will
undergo "normal" gravitational red-shift. The red-shift factor is
given by the square-root of the time-coefficient of the metric at
the position of the membrane, i.e. $\sqrt{r_0/r_h}$. Multiplying
the local surface temperature with this factor gives the
holostar's temperature at infinity. We find $T \propto 1/r_h = 1
/(r_+ + r_0)$. Disregarding the rather small value of $r_0$ the
Hawking temperature of a black hole has exactly the same
dependence on the gravitational radius $r_+ = r_h - r_0$ as the
holostar's local surface temperature, measured at infinity. We
have just derived the Hawking temperature up to a constant factor.
A simple dimensional analysis shows that the factor is of order
unity (see also the discussion in section
\ref{sec:holographic:principle} ). In \cite{petri/thermo} a more
definite relationship will be derived.

Turning the argument around one can use the Hawking temperature to
fix the local temperature at the holostar's boundary membrane. The
local temperature of the membrane will blue-shifted with respect
to the Hawking temperature $T_H$. Setting the blue-shifted Hawking
temperature equal to the local temperature of the membrane
$T(r_h)$ we find:

\begin{equation}
T(r_h) = T_{H} * \sqrt{\frac{B(\infty)}{B(r_h)}} = \frac{\hbar}{4
\pi \sqrt{r_h r_0}}
\end{equation}

Knowing the local temperature within the holostar at one point
allows one to determine the temperature at an arbitrary internal
position:

\begin{equation} \label{eq:Tlocal}
T = \frac{\hbar}{4 \pi \sqrt{r r_0}}
\end{equation}

\subsubsection{\label{sec:r0}A determination of the scale parameter $r_0$}

With the above equation for the local temperature, the unknown
length parameter $r_0$ can be estimated. Raising equation
(\ref{eq:Tlocal}) to the fourth power gives:

\begin{equation} \label{eq:T4/rho}
T^4 = \frac{\hbar^4}{2^5 \pi^3 r_0^2} \frac{1}{8 \pi r^2} =
\frac{\hbar^4}{2^5 \pi^3 r_0^2} \rho
\end{equation}

which implies:

\begin{equation} \label{eq:r0^2}
\frac{r_0^2}{\hbar} = \frac{\hbar^3}{2^5 \pi^3} \frac{\rho}{T^4} =
 \frac{\rho}{4 T} \left( \frac{\hbar} {2 \pi T} \right)^3
\end{equation}

Under the assumption, that we live in a large holostar of cosmic
proportions we can plug in the temperature of the cosmic microwave
background radiation (CMBR) and the mean matter-density of the
universe into the above equation. If the recent results from WMAP
\cite{WMAP/cosmologicalParameters} are used, i.e. $T_{CMBR} =
2.725 \, K$ and $\rho \approx 0.26 \rho_{crit}$, where
$\rho_{crit} = 3 H^2 / (8 \pi)$ is determined from the
Hubble-constant which is estimated to be approximately $H \approx
71 (km/s)/Mpc$, we find:

\begin{equation} \label{eq:r0^2:est}
r_0^2 \approx 3.52 \hbar
\end{equation}

which corresponds to $r_0 \approx 1.88 \sqrt{\hbar}$. Therefore
$r_0$ is roughly twice the Planck-length, which is quite in
agreement to the theoretical prediction $r_0 \approx 1.87 \, r_{Pl}$
at low energies, obtained in \cite{petri/charge}.

\subsection{\label{sec:massive:acc}Necessary conditions for nearly geodesic motion of massive particles}

In this and the following sections the radial motion of non zero
rest-mass (massive) particles will be analyzed in somewhat greater
detail. The main purpose of this analysis is to show, that as in
the case of photons, geodesic motion of massive particles is
self-consistent within the holostar solution, i.e. preserves the
energy-density $\rho \propto 1/r^2$. It cannot be stated clearly
enough, though, that for massive particles geodesic motion is at
best an approximation to the true motion of the particles within
the pressurized fluid consisting of massive particles and photons
alike. The radial pressure of the space-time will always exert an
acceleration on a massive particle, so that massive particles can
never move truly geodesically.\footnote{Strictly speaking, the
same argument applies to photons, as the group- and
phase-velocities are expected to differ from the local speed of
light in vacuum, due to the presence of matter in the interior
space-time.} However, we will see later, that whenever the
geodesic acceleration slightly dominates over the pressure-induced
acceleration, a massive particle will start to move outward and
its motion will become geodesical for all practical purposes,
whenever the particle has reached a radial position $r$ on the
order of a few multiples of its starting position $r_i$.

\subsubsection{Geodesic acceleration and pressure}

The motion of a massive particle in the holostar's interior is
subject to two effects: Geodesic proper acceleration and
acceleration due to the pressure.

The geodesic (proper) acceleration $g$ for a massive particle at
its turning point of the motion, i.e. at the position where it is
momentarily at rest with respect to the $(t, r, \theta, \varphi)$
coordinate system, is given by the following expression:

\begin{equation} \label{eq:g}
g = \frac{d \beta_r}{d \tau} = \frac{1}{2} \sqrt{\frac{r_0}{r}}
\frac{1}{r}
\end{equation}

In the holostar's interior the geodesic acceleration is
always radially outward directed, whereas it is inward-directed in
the exterior space-time. We find that the geodesic acceleration is
always directed towards the membrane. In a certain sense the
membrane can be considered as the true source of the gravitational
attraction.\footnote{Note, that the sum $\rho + P_r + 2 P_\theta$,
i.e. the "active gravitational mass-density", sometimes denoted as
the "true source of the gravitational field", is zero everywhere
except at the membrane, where it takes on a positive value.}

Due to the negative radial pressure, an interior (massive)
particle will also be subject to a radially inward directed proper
acceleration resulting from the "pressure force".

Under the rather bold assumption, that the negative radial
pressure in the holostar is produced in the conventional sense,
i.e. by some yet to be found "pressure-particles" moving radially
inward, which once in a while collide with the massive particles
moving outward, the momentum transfer in the collision
process\footnote{Alternatively one can argue solely in terms of
the pressure, without having to know its origin: When $r$
decreases, the density becomes higher due to the inverse square
law. The higher the density gets, the {\em lower} the radial
pressure will become, as the pressure is negative: $\rho = -P_r$.
The pressure-gradient is inward directed. Although the absolute
magnitude of the pressure gets higher towards the center, the
gradient - and therefore the force associated with this gradient -
points inward.} will result in an inward directed "net-force"
acting on the massive particles. The acceleration by the pressure
force $a_P$ can be estimated as follows for a particle in its
momentary {\em rest frame}:\footnote{If the particle has a
substantial tangential velocity at its turning point of the motion
$r_i$, the proper acceleration in the particle's rest frame must
be multiplied by the squared tangential $\gamma$ factor at the
turning point, i.e. $a_P \rightarrow a_P \gamma_i^2$. This
slightly complicates the derivation. But the same procedure
applies to the geodesic acceleration $g$. All tangential
$\gamma_i$ factors cancel, so the following derivation, which
assumes pure radial motion, still remains correct.}

\begin{equation} \label{eq:aP}
\frac{dp}{d\tau} = m \, a_P = P_r \sigma = - \frac{\sigma}{8 \pi r^2}
\end{equation}

$m$ is the rest mass of the particle and $\sigma$ it's rest frame
cross-sectional area with respect to the "force" mediated by the
"pressure-particles". The cross-sectional area will depend on the
characteristics of the field (or particles) that generates the
radial pressure. In particular $\sigma$ might depend on the
typical interaction energy at a particular $r$-value. If the
pressure is gravitational in origin, one would expect the
cross-sectional area to be roughly equal to the Planck-area.

For the following discussion I will assume, that $\sigma$ and $m$
(or rather the ratio $\sigma / m$) remain constant in the
geodesically moving frame. Note, that this must not necessarily be
so. Quantum field theory predicts, that the cross-sectional areas
of the strong, weak and electro-magnetic force vary with energy
and that the particle mass varies with energy. As the local
temperature in the interior holostar space-time depends on $r$,
one cannot rule out a priori that $\sigma / m = const$. In section
\ref{sec:nucleosynthesis} some evidence is presented, that $\sigma
/ m$ might depend on $r$.

The geodesic acceleration $g$ has a $1/r^{3/2}$-dependence,
whereas the pressure-induced acceleration $a_P$ follows an inverse
square law ($a_P \propto 1/r^2$). For large $r$ the geodesic
acceleration dominates over the pressure-induced acceleration, so
that we can expect nearly geodesic motion far away from the
center. The radial position from which geodesic motion will start
out can be derived from the condition $|g| >\approx |a_P|$. A
rough estimate is obtained by combining equations (\ref{eq:aP},
\ref{eq:g}):

\begin{equation} \label{eq:rmax:geodesic}
r \geq \left( \frac{\sigma}{4 \pi m} \right)^2 \frac{1}{r_0}
\end{equation}

However, there are some subtleties involved. Equation
(\ref{eq:rmax:geodesic}) refers to a stationary, non-relativistic
situation. If the particles move relativistically - individually
or collectively - the derivation brakes down, as $\dot{p} = m \,
a_P$ is only valid in the momentary {\em rest frame} of a
particle, whereas $g$ has been calculated in the stationary {\em
coordinate frame}. Both frames are not necessarily equivalent.

Two sub-cases of relativistic motion have to be considered.
Unordered individual motion of particles, corresponding to a
stationary situation at high temperatures, and highly relativistic
collective motion of particles, corresponding to highly
relativistic particle fluxes with respect to a stationary observer
at rest in the coordinate system.

We are primarily interested in the geodesic motion of particles,
which corresponds to the second sub-case.\footnote{Geodesic motion
is by definition collective: All particles with the same initial
data move on the same trajectory. For slightly different initial
conditions tidal forces become important. But these will be quite
low for large $r$.} The calculations will be done in co-moving
frame, so equation (\ref{eq:aP}) for the pressure-induced
acceleration remains unchanged. This result is not trivial. The
radial pressure $P_r = -1 / (8 \pi r^2)$ so far has only been
calculated in the {\em coordinate} frame. However, the
stress-energy tensor of the holostar space-time is radially
boost-invariant, and geodesic motion is nearly radial far away
from the turning point of the motion, as $\beta_\perp =
(r_i/r)^{3/2}$, whereas $\beta_r \simeq 1$. Therefore a radial
boost has no effect on the individual components of the
stress-energy tensor. In particular, a geodesically moving
particle will measure exactly the same principal pressures and the
same energy-density as a stationary observer.

In order to evaluate the condition $|g| \geq |a_P|$ for geodesic
motion in the co-moving frame, we have to transform the proper
acceleration $g$, which was evaluated in the frame of the
stationary observer, to the co-moving frame. The essential
parameter for the transformation is $\gamma$. Assuming nearly
geodesic motion one can use equation (\ref{eq:beta:r2:int_approx})
to calculate $\gamma$:

\begin{equation} \label{eq:gamma:ri}
\gamma^2 \simeq \frac{1}{1 - \beta_r(r)^2} = \frac{r}{r_i} \cdot
\gamma_i^2
\end{equation}

$r_i$ is turning point of the (geodesic) motion and $\gamma_i$ is
the (purely tangential) $\gamma$-factor at $r_i$. For the
following discussion $\gamma_i^2$ will be set to unity, which
corresponds to pure radial motion. Keep in mind, that for a more
sophisticated analysis one has to replace $r_i \rightarrow r_i /
\gamma_i^2$. This doesn't have any effect on the relations that
will be derived in this and the following sections.

$\gamma$ grows with the square root of $r$, so geodesic motion
becomes highly relativistic whenever one is far away from the
turning point of the motion $r_i$. For large $r$ the motion is
nearly radial, i.e. in the same direction as the geodesic
acceleration $g$. Let us denote the proper geodesic acceleration
in the co-moving frame by an overline. It is a well known result
from special relativity, that the proper acceleration transforms
as $\overline{g} = \gamma^3 g$, when the motion is parallel to the
acceleration, and as $\overline{g} = \gamma^2 g$, when the motion
is perpendicular to the acceleration. The condition for nearly
geodesic motion evaluated in the co-moving frame becomes

\begin{equation} \label{eq:g:mod}
\overline{g} = \gamma^3 g \geq a_P = \frac{\sigma}{8 \pi m r^2}
\end{equation}

The factor $\gamma^3$ has the effect to boost the geodesic
acceleration from a $1/r^{3/2}$ law to a constant
value\footnote{Keep in mind, that a geodesically moving observer
does not sense this "constant" geodesic acceleration. A
geodesically moving observer is by definition {\em unaccelerated}
in his local frame. The only force, that a (nearly) geodesically
moving observer feels, is the pressure induced acceleration, which
falls off with $1/r^2$ in the co-moving frame (and the tidal
acceleration due to the space-time curvature, which cannot be
transformed away). The geodesic acceleration is only "real" in the
frame of a stationary observer. Such an observer has to provide
some means by his own, such as rocket-power, to remain stationary.
Yet an observer firing his rockets does not feel the {\em
geodesic} acceleration, but rather the acceleration provided by
the engines of his rocket. Alternatively, it can be the
"non-gravitational" forces of the space-time itself, that keep an
observer stationary. In the holostar space-time this corresponds
to the situation, when the pressure-induced acceleration is equal
to the geodesic acceleration. (Note that this situation is very
familiar: It applies to any observer pressing his feet against the
solid ground of the earth) However, equivalently to the rocket
case it is {\em not} the geodesic acceleration, that the
stationary observer feels, but rather the "real" acceleration
provided by the pressure force! The "geodesic" acceleration is a
mere bookkeeping device which enables us to determine the
acceleration provided by the "real" forces (which are opposite to
the geodesic "force") in a stationary situation.}, so that the
ratio $|a_P| / |g|$ falls with an inverse square law. Geodesic
motion in a sense "bootstraps" itself. Consider a particle at the
position of (unstable) equilibrium with $|g| = |a_P|$. Let us
denote this position by $r_i$. Small fluctuations will displace
the particle from its equilibrium position. Whenever $r > \approx
r_i$ the outward directed geodesic acceleration will dominate over
the inward directed pressure-induced acceleration, as $g \propto
1/r^{3/2}$ and $a_P \propto -1/r^2$ (for low velocities). As the
motion becomes more and more geodesical the ratio $a_P / g$
quickly approaches an inverse square law, which means that the
pressure-induced acceleration becomes negligible with respect to
the geodesic acceleration whenever $r >\approx (2-3) r_i$.

Even if - for some unknown reason - the motion of a particle at $r
>\approx r_i$ were to remain non-relativistic for an extended
period of time ($\gamma \simeq 1$), the ratio of pressure-induced
acceleration to geodesic acceleration would yet follow an inverse
square root law, $|a_P| / |g| \approx \sqrt{r_i/r}$, so that
geodesic dominance is inevitable whenever an interior particle has
reached a radial position where $|g| >\approx |a_P|$. In any case
- geodesic motion, slow non-geodesic motion (or no motion at all)
- the pressure-induced acceleration falls faster with increasing
distance from the holostar's center than the geodesic
acceleration. We can be certain that the motion of a massive
particle will become geodesic to a very high degree of precision
at sufficiently large values of $r$.

The region of "almost" geodesic motion is characterized by $|g|
\gg |a_P|$. Combining equations (\ref{eq:g:mod}, \ref{eq:g},
\ref{eq:gamma:ri}) the condition $|g| \geq |a_P|$ for the start of
(nearly) geodesic motion can be expressed as:

\begin{equation} \label{eq:rgeodesic}
r^2 \geq \frac{\sigma r_i}{4 \pi m} \sqrt{\frac{r_i}{r_0}}
\end{equation}

In order to analyze the necessary conditions for geodesic motion
it is convenient to cast equation (\ref{eq:rgeodesic}) into a form
which involves the local radiation temperature $T_i$ at the
turning point of the motion $r_i$. Using equation
(\ref{eq:Tlocal}) we find:

\begin{equation} \label{eq:rgeodesic3}
\left( \frac{r}{r_i} \right)^2 \geq \frac{\sigma}{\hbar}
\frac{T_i}{m}
\end{equation}

The effective potential for geodesic motion in the interior
space-time always decreases with growing $r$, so that $r \geq r_i$
on any geodesic path. If we set $r = r_i$, equation
(\ref{eq:rgeodesic3}) gives rise to the following inequality,
which defines the necessary conditions for the onset of geodesic
motion:

\begin{equation} \label{eq:acc:equal}
\frac{m}{T_i} \geq \frac{\sigma}{\hbar}
\end{equation}

$r_i$ in equation (\ref{eq:rgeodesic3}) is the turning point of a
particle for true geodesic motion. Self consistency requires that
geodesic motion starts out roughly from this point. Therefore we
can set the left hand side of equation (\ref{eq:rgeodesic}) equal
to $r_i$, from which the turning point of nearly geodesic motion
can be calculated unambiguously:

\begin{equation} \label{eq:rgeodesic2}
r_i \simeq \left( \frac{\sigma}{4 \pi m} \right)^2 \frac{1}{r_0}
\end{equation}

This is - not quite unexpectedly - the same result as in equation
(\ref{eq:rmax:geodesic}).

If cross-sectional areas typical for the strong force ($\sigma_{S}
\approx 36 \pi \hbar^2/m_p^2 \approx 40 mb$) and $r_0 \approx 2 \,
r_{Pl}$ - as derived in section \ref{sec:r0} - are used, one finds

\begin{equation}
r_i \approx \left( \frac{9 \hbar^2}{m_p^3} \right)^2 \frac{1}{r_0}
\approx 10^{116} \, r_{Pl} \approx 10^{83} \, cm
\end{equation}

This is roughly a factor of $10^{55}$ larger than the radius
of the observable universe.

It is questionable whether the cross-sectional areas of the strong
or electro-magnetic forces should be used in equations
(\ref{eq:rgeodesic}, \ref{eq:rgeodesic2}). The pressure in the
holostar's interior space-time is gravitational in origin, the
interior equation of state is that of a radial collection of low
energy strings. The tangential extension of each string is equal
to the Planck area \cite{petri/string}. Therefore one would rather
expect that the cross-sectional area - with respect to the
"pressure force"-  should be comparable to the Planck area.

Furthermore, the cross-sectional area and particle mass in
equation (\ref{eq:rgeodesic3}) must be evaluated {\em at the
turning point of the motion} $r_i$. We will see later, that
geodesic motion in the holostar starts out from $r_i \approx r_0$.
Therefore we are talking about cross-sectional areas and particle
masses at roughly the Planck-temperature. It is reasonable to
assume that both $\sigma$ and $m$ are comparable to the
Planck-area and the Planck-mass at $r \approx r_0$.

If one sets $\sigma \approx \hbar$ one finds from equation
(\ref{eq:acc:equal}), that a massive particle starts to move out
geodesically from $r = r_i$ whenever the local radiation
temperature $T_i$ at the "turning/starting point" of the motion
$r_i$ falls below the particles rest mass (in units $c=k=1$).

For a particle with the mass of a nucleon ($m \approx 10^{-20}
\sqrt{\hbar}$) one finds $r_i \approx 10^{36} r_{Pl} \approx 17 \,
m$, roughly $0.5$ \% of the gravitational radius of the sun. The
local temperature of the holostar at this point is roughly $T_i
\approx 10^{13} K$, corresponding to a thermal energy of roughly
$1 \, GeV$, i.e. the rest mass of the nucleon. For an electron the
radial position of equal geodesic and pressure induced
acceleration will be larger by the squared ratio of the
proton-mass to the electron mass: $r_i \approx 5.8 \cdot 10^4 km$.
At this point the local temperature is roughly $T_i \approx 5
\cdot 10^9 K$, corresponding to an energy of $500 \, keV$, i.e.
the rest mass of the electron.

One important aspect has been glossed over, so far. The position
$r_i$, where a particle starts to move out geodesically, is a
position of unstable equilibrium. A particle that is displaced to
$r < r_i$ consequentially must move {\em inward}, as the inward
directed pressure-induced acceleration begins to dominate. All
particles in the region $r \leq r_i$ will be pushed towards the
center. It appears as if the holostar would collapse to a
singularity. This, however, is in conflict with the static nature
of the holostar metric and fields. There are two answers to this
apparent paradox.

The first answer is to assume that $r_i \approx r_0 \approx
r_{Pl}$. (We will see in section \ref{sec:Unruh} that $r_i = r_0$
can be proven rigorously.) Although we don't yet have the
necessary machinery for a convincing derivation of this
assumption, it is possible to give plausibility arguments: For a
particle at $r \approx r_0$ "inward" motion is not an option. The
uncertainty principle will not allow a definite localization of
any particle at $r \ll r_0$. See section \ref{sec:uncertainty} for
a thorough derivation. Furthermore, according to the results of
loop quantum gravity the region $r < r_0 \approx r_{Pl}$ is not
even well defined. No particle (or geometric entity) can have a
surface area less than the minimum area quantum of loop quantum
gravity, which is comparable to the Planck area. Therefore, from
the loop perspective, the region $r < r_0$ will contain very few
particles, no matter how many particles we try to "force" into
this region. The same result follows from string theory. The
equation of state of the holostar's interior matter is that of a
collection of radial strings. The transverse dimension of the
strings is equal to the Planck area \cite{petri/string}. It it not
possible to "compress" a string along its longitudinal direction
to a length that is much less than its transverse extension.
Therefore the region $r < r_0$ can contain only very few strings,
no matter how hard we try to compress the strings in their
longitudinal or transverse directions. We arrive at the not too
unexpected result, that it will be impossible to force any quantum
of matter (particle or string) into a region smaller than $r_0$.
This allows us to set up the following plausibility argument: If
equation (\ref{eq:rgeodesic2}) would give $r_i \gg r_0$, any
particles within the region $r \leq r_i$ would be pushed towards
the center. But there is a natural limit to the compression at $r
\approx r_0$ and the holostar is a static solution, so the
effective values of $m$ and $\sigma$ must adjust in such a way,
that $r_i \approx r_0$.

The second answer has to do with the fact, that (for $\sigma
\approx \hbar$) the temperature at the turning point of the motion
is comparable to the particle's rest mass. Therefore the
(unordered) motion of the particles at this point will become
relativistic. At relativistic energies we have to replace $m$ in
equation (\ref{eq:acc:equal}) with the total relativistic energy
of the particle $m \rightarrow E$, which leads to $E >\approx T
\cdot \sigma / \hbar$. The equal sign in the above equation refers
to the situation, when geodesic and pressure-induced acceleration
are equal, i.e. when the situation is stationary. At ultrahigh
temperature we find, that stationarity implies $E = \sigma / \hbar
\cdot T$, meaning that the total energy of an ultra-relativistic
particle must be proportional to its temperature $T$, with the
right proportionality factor $\sigma / \hbar$. But the relation $E
\propto T$ is well known from the thermodynamic relations for an
ultra-relativistic gas. The (thermodynamic) factor of
proportionality depends on the relative mixture of
ultra-relativistic fermions and bosons ($E \simeq 2.7 \, T$ for a
gas consisting purely out of photons, $E \simeq \, 3.15 \, T$ for a
pure fermion gas; a mixture has a factor somewhere between the two
extremes.). Whenever $\sigma / \hbar$ lies in the range $\approx
2.7-3.2$ the thermodynamic equations for an ultra-relativistic gas
guarantee a (nearly) stationary situation. If $\sigma / \hbar$ has
the right value, we can have a stationary situation over the full
high temperature range $r \leq r_i$. A somewhat more detailed
discussion of this feature of the holostar solution can be found
in the following section.

\subsubsection{A case for a gravitational origin of the pressure with a Planck-sized cross-sectional area}

The condition for the onset of geodesic motion is given by
equation (\ref{eq:rgeodesic3}), which reads

\begin{equation}
\left(\frac{r}{r_i}\right)^2 \geq \frac{\sigma}{\hbar} \cdot
\frac{T_i}{m}
\end{equation}

The above equation carries a very important message: The geodesic
and the pressure induced acceleration are equal whenever $r =
r_i$. This will be the case, when $m / T$ is equal to the
cross-sectional area $\sigma$ of the particle divided by the
Planck area. But $m / T$ is nothing else than the entropy per
particle $\sigma_S$ of a massive particle of an ideal gas in
thermodynamic equilibrium:

\begin{equation}
\sigma_S = \frac{m}{T}
\end{equation}

so that the condition for zero net-acceleration can be expressed as:

\begin{equation} \label{eq:sigma=s}
g + a_P = 0 \rightarrow \frac{\sigma}{\hbar} = \sigma_S
\end{equation}

Note that there is an ambiguity in the notation. $\sigma$ denotes
the cross-sectional area of a particle, but also its entropy.
Whenever it is necessary to distinguish between entropy and
cross-sectional area, the cross-sectional area will be shown
without subscript, whereas the entropy receives a subscript.

It is not altogether unreasonable to assume that a particle's
cross-sectional area might become comparable to its entropy. This
situation is well known for black holes. The cross-sectional area
for the capture of photons by a spherically symmetric black hole
is equal to $\sigma / \hbar = (27/16) \, A / \hbar = (27/4) \,
\sigma_S$.

For elementary particles, such as the electron or neutrino, one
expects that the cross-sectional area of the particle becomes
comparable to its entropy roughly at the Planck energy. At
ultra-high energies the ideal gas-approximation is nearly ideally
fulfilled. The entropy of an ultra-relativistic particle in the
ideal gas approximation is of order unity. For a gas with zero
chemical potential $\sigma_B = 3.6$ for a boson and $\sigma_F =
4.2$ for a fermion. $\sigma / \hbar = \sigma_S$ therefore requires
very small cross-sectional areas of order Planck area. However,
the cross-sectional areas of the strong, electro-magnetic and weak
force decrease with energy. Slightly below the Planck-energy the
cross-sectional areas of all forces, including gravity, become
nearly equal to the Planck area. This suggests, that $T_i \approx
T_{Pl}$, which requires $r_i \approx r_0$.

At the Planck-energy, i.e. in the central region $r \approx r_0$
of the holostar, we can be quite confident that $\sigma / \hbar
\approx \sigma_S$ holds, so that the pressure (as well as the
uncertainty principle) will prevent any further collapse to
regions smaller than $r_{Pl}$.

In the following discussion I will make the assumption, that the
cross-sectional area $\sigma$ - with respect to the "pressure
force" - is comparable to the Planck-area, i.e. $\sigma \approx
A_{Pl} = \hbar$, independent from the local temperature.

This appears to be in conflict with the well known result for the
strong, weak and electro-magnetic forces, for which the
cross-sectional area $\sigma$ depends on the collision energy in
the center-of-mass frame. These cross-sections are definitely
temperature dependent, as the average energy in the center-of-mass
frame increases with temperature. However, if the holostar
pressure has a gravitational origin with a nearly Planck size
cross-sectional area, it is more natural to assume that the
gravitational (=geometric!) cross-sectional area is unaffected by
the relative motion of the colliding particles.

The author is well aware, that the assumption of a (more or less
universal) non-zero, nearly Planck-size {\em geometric}
cross-sectional area for an elementary particle is in conflict
with the general belief (which has its foundations in conventional
QFT) that particles should be point-like. But the existence of
truly point like particles is challenged by the two major
competitors for a unified theory of quantum gravity. The results
of string-theory and of loop quantum gravity are quite clearly
incompatible with point-like structures.

In this respect it is quite remarkable, that the holostar solution
itself suggests that elementary particles might have a non-zero,
roughly Planck-sized boundary area, pointing to the possibility
that the cross-section associated with the negative pressure might
have a geometric origin. The smallest conceivable holostar is
given by $r_h = r_0$. The radius of the membrane cannot be
smaller, otherwise the gravitating mass is negative.\footnote{To
be more precise, the condition $r_h \geq r_0$ is only true for the
uncharged case, discussed in this paper. For the charged case it
can be shown \cite{petri/charge} that $r_h \geq r_0/2$. The equal
sign holds for an extremely charged holostar of minimal mass.} The
fact that $m=0$ for the "elementary" holostar solution with $r_h =
r_0$ suggests a particle interpretation. Elementary particles are
characterized by extremely small masses in natural units. For
instance, the mass of the proton is roughly $10^{-20}$ in Planck
units. The gravitating mass of a holostar with $r_h = r_0 \approx
2 r_{Pl}$ is zero, yet it has a non-zero boundary area $A = 4 \pi
r_0^2$, which is roughly equal to the Planck area. Its - purely
geometric - cross-sectional area is $\sigma \approx \pi r_0^2 =
A/4$.\footnote{One might rightfully question, whether the
classical formula $\sigma = A/4$ for the cross-sectional area of a
sphere remains valid at the Planck-scale. On the other hand, as $m
= 0$ for an "elementary holostar", the geometry is flat right down
to $r = r_h$, so in the context of classical general relativity
there appears to be no serious objection against this assumption.}
As the mass, charge and angular momentum of an "elementary"
holostar with $r_h = r_0$ are zero, it can only interact with
other particles by direct collision. Its cross-sectional area
$\sigma$ will thus be equal to its geometric cross-sectional area.
This area is independent of the energy of the colliding
particles.\footnote{One might call this the "billiard ball"
assumption, according to which the gravitational cross-sectional
area of a particle is equal to its {\em geometric} cross-sectional
area, defined by its (non-zero!) boundary area. This geometric
cross-sectional area is independent of the collision energy,
because the (geometric) cross-sectional area is perpendicular to
the motion of the particles, viewed in the center of mass frame.}
Remarkably, $\sigma / \hbar$ is identical to the "Hawking entropy"
of such an elementary holostar $\sigma_S = A / (4 \hbar)$, so that
equation (\ref{eq:sigma=s}) is trivially fulfilled.

With $\sigma \approx 3 \hbar$ we find from equation
(\ref{eq:acc:equal}), that whenever the local temperature of one
of the constituent particles of the holostar falls below roughly a
third of its rest-mass, the outward directed geodesic acceleration
becomes larger than the inward directed pressure-induced
acceleration, allowing the particle to move outward on a
trajectory which becomes more and more geodesical.

We can now answer the question, what happens in the region of the
holostar, where the temperature is higher than the rest-mass of
the particle. For the stationary case (no collective motion of
particles in a particular direction) the mass $m$ in equation
(\ref{eq:acc:equal}) should be replaced by the total mass-energy
of the particle, $m \rightarrow E = \sqrt{m^2 + p^2}$. Equation
(\ref{eq:acc:equal}), which is the condition of zero net
acceleration, then reflects the condition $E = (\sigma / \hbar) \,
T$. For an ideal ultra-relativistic gas we have $E = e \, T$ with
$e=const$ in the range $2.7 < e < 3.15$ for an ultra-relativistic
gas with zero chemical potential. If $\sigma / \hbar$ lies in the
same range, it might be possible to fulfil the condition
(\ref{eq:acc:equal}) over the whole temperature range $T > m$.

For example, for photons $E \simeq 2.7 \, T$. If $\sigma > 2.7 \,
\hbar$, the pressure-induced acceleration dominates. If $\sigma =
2.7 \, \hbar$ the net-acceleration in the frame of the observer at
rest in the ($t, r, \theta, \varphi$) coordinate system is nearly
zero. A gas of ultra-relativistic fermions with zero chemical
potential has $E \simeq 3.15 \, T$. Putting it all together, we
arrive at the conclusion, that for $\sigma \approx 3 \, \hbar$ the
holostar is nearly static in the high temperature regime.

In section \ref{sec:Gibbs=0} it will be shown that the free energy
$F$ in the holostar must be zero, if the holostar solution is to
reproduce the Hawking temperature and entropy. From $F=0$ one can
immediately deduce that $E = \overline{\sigma_S} \, T$, where
$\overline{\sigma_S}$ is the {\em mean} entropy per
particle.\footnote{Keep in mind, that $\overline{\sigma_S}$
denotes the average of the entropies of all ultra-relativistic
particles. In general bosons and fermions have different
entropies. Therefore the relation $E = \overline{\sigma_S} \, T$
doesn't apply to bosons or fermions individually, but rather to
the average of both species.} The condition for zero
net-acceleration in the high temperature central region of the
holostar will then be fulfilled exactly, if

\begin{equation} \label{eq:acc:equal:sigma}
\frac{E}{T} = \frac{\sigma}{\hbar} = \overline{\sigma_S}
\end{equation}

The temperature $T_i$, where a particle with rest mass $m$ starts
to move outward, is given by:

\begin{equation} \label{eq:T:out}
T_i = \frac{m}{\overline{\sigma_S}} \approx \frac{m}{3}
\end{equation}

It is not altogether clear, whether zero net-acceleration is
exactly achievable over an {\em extended} region of the holostar's
interior. For $\sigma / \hbar <\overline{ \sigma_S}$ the geodesic
outward directed acceleration of a particular particle is larger
than the pressure-induced deceleration. If the holostar solution
is combined with the results of loop quantum gravity and the
Immirzi parameter is fixed at $\gamma = \overline{\sigma_S}/(\pi
\sqrt{3})$ - see the discussion in \cite{petri/charge} - then
$\sigma/\hbar = \overline{\sigma_S}$, at least at the Planck
energy. At lower energies the {\em geometric} cross-sectional area
of the spin-1/2 fermions is expected to lie in the range $\pi
\sqrt{3/4} < \sigma / \hbar < \overline{\sigma_S}$ (see
\cite{petri/charge}). Therefore we expect the effective geometric
cross-sectional area $\sigma / \hbar$ for fermionic matter to be
always less than the mean entropy per particle
$\overline{\sigma_S}$, except at the central region $r \approx
r_0$. If this is essentially the right picture, the geodesic
acceleration for spin 1/2 fermions dominates over the
pressure-induced acceleration, up to Planckian temperatures $T
<\approx T_{Pl}$.

\subsection{A possible origin of the negative radial pressure}

Note that this section is highly speculative and trods into
uncharted territory without the appropriate guide. On the other
hand, the issues addressed here must be solved in one way or the
other, if the holostar is to be a truly self-consistent model, not
only for black holes comparable to the mass of the sun, but for
holostars of arbitrary size, up to and exceeding the observable
radius of the universe. It might well be that the solution to the
problems addressed in this section, if there is any, will turn out
to be completely different from what is presented here.

It is an astonishing coincidence, that the local temperature of
the holostar at the radial position $r_i$, i.e. where the net
acceleration (geodesic and pressure-induced) becomes nearly zero
for a particular particle, is roughly equal to the rest mass of
the particle. At least this is the case, if the result $\sigma /
\hbar \approx \sigma_S \approx \pi $, which was obtained
independently in \cite{petri/charge}, is used. This can be
regarded as an indication, that the negative radial pressure
within the holostar is not just a curious mathematical feature of
the solution, but could be a real, measurable physical effect, and
that the assumption that the pressure is gravitational in origin,
with a {\em geometric} cross-sectional area $\sigma$ comparable to
the Planck-area, is not too far off the track.

In fact, if the pressure were produced by a continuous flow of
particles moving radially inward, and that interact only very
weakly with the outflowing "ordinary matter", this could explain
the purely radial nature of the pressure. If the inflowing
"pressure-particles" carry a mass-energy equivalent to that of the
outflowing ordinary matter, this could at the same time explain
the mystery, that the holostar is a static solution which requires
that any outflow of mass-energy must be accompanied by an
equivalent inflow.\footnote{\label{fn:ord:matter}There is no
mystery, as long as the outmoving particles reverse their outward
directed motion in the exterior space-time and thus "swing" back
and forth between exterior and interior space-time (without
friction), as suggested by the holostar equations of motion. The
motion of the massive particles would be time-symmetric and any
outmoving particle would be confronted by a highly relativistic
flux of particles moving inward. The mystery arises, when we
identify the holostar with the observable universe. It will be
shown later, that any outmoving massive particle experiences an
isotropic radially outward directed Hubble-flow in its co-moving
frame. This is exactly what we ourselves observe. However, our
situation in the expanding universe appears time-asymmetric: We
haven't yet noticed a flow of particles hitting us head-on in our
"outward" directed motion (although we might already have noticed
the effects of the inflowing matter in the form of "dark energy",
i.e. a negative pressure). This must not altogether be a
contradiction. First, if the outflowing and inflowing matter moves
highly relativistically, as is expected from the holostar
equations of motion for a single massive particle, the
cross-sectional area for a collision between an outmoving "matter
particle" and an in-moving "pressure particle" can become almost
arbitrarily small, as $\sigma \propto \hbar^2/E^2$ for ordinary
matter. Secondly, the outflowing and the inflowing particles must
not necessarily be the same. The inflowing particles could be
decay products of the outmoving particles or altogether different
species. Any weakly interacting "dark matter" particle would
probably qualify. If the inflowing particles interact only weakly
with the outflowing "ordinary" matter, they could deliver an
inward-directed energy flow comparable to the outward directed
flow and a time-symmetric situation with respect to the net energy
flow could be restored.}

\subsubsection{Is the pressure produced by ordinary matter?}

The most conservative assumption is, that the (inflowing) pressure
particles are nothing else than the outflowing matter particles,
that have reversed their motion in the exterior space-time and are
now coming back. We wouldn't be able to notice the inflowing
matter, if the motion of out- and inflowing matter is highly
relativistic. This is actually the case in the holostar's
interior. Any massive particle moving nearly geodesically has an
extremely high $\gamma$-factor, growing with the square-root of
the radial coordinate value. For instance, a particle starting out
from the center $r_i \approx r_0 \approx r_{Pl}$ will have $\gamma
\approx 10^{30}$, when it has reached a radial position
corresponding to the current radius and density of our universe.
The cross-sectional area for the collision of an inflowing with an
outflowing electron in the center of mass-frame then is roughly
given by

\begin{equation}
\sigma \approx \frac{\hbar^2}{m_e^2 \gamma^2 } \approx 10^{-20} A_{Pl} \approx 10^{-60} b
\end{equation}

The cross-sectional area for a nucleon will be even lower. These
are utterly negligible cross-sections, so that for the low
densities encountered currently in our universe the inflowing
"pressure particles" will just pass through the outflowing matter
without any reaction at all: If one assumes, that the in- and
outflowing matter is distributed uniformly, i.e. most of the
matter does not reside in black holes, the probability for one
reaction to occur for all of the $N \approx 10^{80}$ particles in
the observable universe is roughly $10^{-30}$, if one is willing
to wait for a time comparable to the age of the universe.

It is quite likely, that the standard form of the high energy
cross-sectional area ($\sigma \propto 1/E^2$) has to be modified
for collision energies above the Planck energy.\footnote{For
larger collision energies a black hole will be formed, so that it
not clear, whether the formula $\sigma \propto \hbar^2 / E^2$ can
be extended to cross-sectional areas of sub-Planck scale, or
whether we have to modify the formula, so that the cross-sectional
area corresponds to the cross-sectional area of the intermediate
(virtual) black hole state, which is proportional to $E^2$. For
the second option one expects a cross-sectional area of $\sigma
\propto \hbar^2/E^2 + E^2$.} For $\sigma \approx \hbar^2 / E^2 +
E^2$, we still have a cross-sectional area of $\sigma \approx
10^{20} \, A_{Pl}$ at $\gamma \simeq 10^{30}$, meaning that the
collision rate is roughly $10^{-12} / s$, i.e. roughly 1 collision
every million years, for a volume comparable to the current
extension of the observable universe.

\subsubsection{Or do we need some new form of matter?}

As long as one does not know the exact nature of the pressure in
the holostar space-time, the assumption $\sigma \approx \hbar$
might turn out to be vastly incorrect. Without having solved the
equations of motion including pressure (with the correct value of
$\sigma$) one cannot exclude the possibility, that the motion of
the particles in the holostar turns out to be extremely
non-geodesic. In the most extreme case the true motion of the
particles might not even be highly relativistic, so that the
argument with respect to the low cross-sectional areas for
geodesically moving particles does not apply.\footnote{It is
doubtable, whether the holostar would remain a good model for a
uniformly expanding universe in the case of highly non-geodesic,
non-relativistic motion. As will be shown in section
\ref{sec:Hubble}, a geodesically moving observer experiences an
isotropic Hubble flow in his frame of reference. However, the
perceived isotropy of the local Hubble flow in the co-moving frame
is tied to the assumption, that the motion is geodesic: Only
(nearly) geodesic motion gives the right $\gamma$-factor, and thus
the "right" amount of Lorentz-contraction in radial direction of
the co-moving frame, so that the expansion-rates in the radial and
the tangential directions are equal.} If the holostar is to remain
a model for the universe under such - somewhat unlikely -
conditions, the inflowing "pressure particles" cannot be ordinary
matter, as a non-relativistic inflow of ordinary matter would not
have gone un-noticed on our presumably outward directed track.
Whenever "ordinary matter" and "pressure-particles" must be
regarded as distinct, we arrive at two conditions that should be
met by the "pressure particles":

\begin{itemize}
\item The pressure particles should interact very weakly with
ordinary matter (at least at energies below roughly 1 MeV) \item
the mass-energy density of the pressure particles should be
equivalent to the mass-energy density of ordinary matter
\end{itemize}

The first condition can - in principle - be fulfilled by several
particles. The graviton, the supposed messenger particle for the
gravitational force, looks like a suitable candidate. The second
condition suggests that supersymmetric matter might be the
preferred candidate for the pressure-particles, if they exist: It
seems quite improbable that massless gravitons, with only two
degrees of freedom, can deliver an energy flow exactly equal to
the energy flow of ordinary matter at any arbitrary radial
position within the holostar. But exact supersymmetry predicts
equal numbers\footnote{With equal numbers I don't mean equal
number densities per proper volume, but rather equal numbers of
degrees of freedom, i.e. equal number of particle species. The
number densities of bosons and fermions in thermodynamic
equilibrium in the holostar will be different, even if the number
of fermionic and bosonic species are equal (for a detailed
discussion see \cite{petri/charge}).} and masses of the
supersymmetric particles. If the "pressure particles" were the
supersymmetric partners of ordinary matter, it might be possible
to fulfil the second condition quite trivially.\footnote{If
supersymmetry is broken, the masses of the superparticles cannot
be much higher than the $W$- or $Z$-mass. In this case the
negative radial pressure could still be generated by an
appreciable, but not too high number-density of the lightest
superparticle (LSP).} Furthermore, for interaction energies below
the electro-weak unification scale supersymmetric matter and
ordinary matter are effectively decoupled. Supersymmetric matter
therefore fits both conditions well. Note also, that the holostar
solution assumes that the cosmological constant $\Lambda$ is
exactly zero. Exact supersymmetry is quite well suited to explain
a zero value for $\Lambda$. Therefore - from a purely theoretical
point of view - we would get a much more consistent description of
the phenomena, if supersymmetry were realized exactly, not only at
high energies and densities, but also in the low density / low
energy regions of a large holostar.

If the principal agents of the negative radial pressure consist of
supersymmetric matter, this requires an efficient mechanism which
converts the outflowing ordinary matter into the supersymmetric
"pressure-particles". This process is expected to take place in
the membrane. The membrane is in many respects similar to a
two-dimensional domain wall. Therefore the membrane could induce
interactions similar to the conversion processes that are believed
to take place when ordinary matter crosses a two-dimensional
domain wall. On the other hand, if the lightest supersymmetric
particle were light enough, ordinary matter could just decay into
the lightest superparticle anywhere on its way
outward.\footnote{The time of decay would have to be roughly equal
to the radius of the holostar. As the proton's lifetime is several
orders of magnitude larger than the current age of the universe,
this would require a rather large holostar/universe.}

Supersymmetry might also provide the solution to the problem, why
the "pressure particles" (bosons?) preferably move inward, whereas
ordinary matter (fermions?) moves outward.\footnote{In fact, such
a behavior might not necessarily require supersymmetry. If we set
the cross-sectional area of a particle equal to its entropy (times
the Planck area), as suggested by the Hawking-formula, we find
that the bosons with $\sigma_B \approx 3.6$ have a higher
cross-sectional area than the fermions, for which we find that
$\sigma_F \leq 3.2$ (in the supersymmetric case $\sigma_F \approx
2.9$, see the tables in \cite{petri/thermo}). The pressure induced
acceleration is proportional to the cross-sectional area of the
particles, so that bosons will be subject to a higher
pressure-induced acceleration than fermions. In the holostar's
interior the pressure-induced acceleration is always
inward-directed, whereas the geodesic acceleration is outward
directed. For ultra-relativistic particles the net-acceleration is
zero when $\sigma = \overline{\sigma_S} \hbar$, where
$\overline{\sigma_S}$ is the mean entropy per particle (averaged
over all particles, i.e. bosons, fermions and anti-fermions;
$\overline{\sigma_S} \simeq 3.23$ in a supersymmetric context). In
the holostar space-time it can be shown that $\sigma_F <
\overline{\sigma_S}$ and $\sigma_B > \overline{\sigma_S}$, always
\cite{petri/thermo}. Therefore the motion of the bosons will be
dominated by the inward directed pressure-induced acceleration,
whereas the motion of the fermions is dominated by the geodesic
outward directed acceleration, which naturally leads to the
conjecture, that fermions preferentially move outward and bosons
inward. Note that by this argument the anti-fermions would
preferentially move inward as well, because $\overline{\sigma_F} >
5.4$} Such behavior would be easier to understand if one could
think up a mechanism that explains this time-asymmetric situation:
If geodesic movement in the holostar were fully $T$-symmetric, the
time reversed process would be equally likely, in which case the
"pressure-particles" should be ordinary matter (or anti-matter).
In the following argument I make use of the fact, that the
membrane appears as the primary source of gravitational attraction
in the low density region of the holostar, at least for ordinary
matter. Could the membrane expel the "pressure particles", after
the interactions in the membrane have converted ordinary matter
into supersymmetric "pressure particles"? The gravitational force
is always attractive, unless we were able to find some form of
matter with $M^2 < 0$.

This is the point where supersymmetry might come up with an
answer: The Higgs-field, which is expected to give mass to the
particles of the Standard Model, is characterized by a quantity
$M^2$, which is usually identified with the mass squared of the
field. Whenever $M^2 > 0$, all particles of the Standard model are
massless. Whenever $M^2$ falls below zero, the Higgs-mechanism
kicks in. In the supersymmetric extension of the Standard Model
the dependence of $M^2$ on the energy/distance-scale can be
calculated. It turns out that $M^2$ is positive at the
Planck-scale and becomes negative close to the energy scale of the
Standard-Model. Therefore the condition $M^2 < 0$ will be
fulfilled in the problematic low-density region of the holostar,
i.e. whenever $T < M_{Higgs}$. The peculiar property $M^2 < 0$ of
the Higgs-field at low energies therefore might provide the
mechanism, by which supersymmetric matter is rather expelled from
the membrane, whereas ordinary matter is attracted. If
supersymmetry can actually provide such a mechanism, we could
understand why the holostar is a static solution, not only for
small holostars\footnote{For small holostars, where $M_H^2 > 0$ in
the whole interior, all of its constituent particles should be
massless. For massless particles the condition $T \approx E$ is
trivially fulfilled. If $\sigma / \hbar = \sigma_S = E / T$ the
pressure-induced inward directed acceleration and the geodesic
outward directed acceleration are equal throughout the whole
interior. In such a case the holostar is truly static,
time-symmetric and in thermal equilibrium. Note also, that an
extended static central region provides an excellent of
"protection" against continued gravitational contraction of the
whole space-time to a point-singularity.}, but also for an
arbitrarily large holostar, where a large fraction of the interior
matter is situated in the regime where the Higgs-field(s) have
$M^2 < 0$ and "ordinary matter" moves outward.

\subsection{\label{sec:motion:massive}Nearly geodesic motion of massive particles}

In the following sections we are interested in the low density
regions of the holostar, where the motion of massive particles
should become more and more geodesic, i.e. for $T \ll m$. The
geodesic equation of pure radial motion for a massive particle,
starting out at rest from $r=r_i$, is given by equation
(\ref{eq:beta:r2:m:radial}):

\begin{equation} \label{eq:motion:m}
\beta_r = \frac{dr}{dt} \frac{r}{r_0} = \sqrt{1- \frac{r_i}{r}}
\end{equation}

The above equation is only correct for pure radial motion.
However, we are interested in the characteristic features of the
motion {\em far away} from its turning point $r_i$. As has already
been remarked in section \ref{sec:eq:motion}, the general case of
geodesic motion is very well approximated by the above equation
(for $r \gg r_i$), if one replaces $r_i \rightarrow r_i /
\gamma_i^2$, where $\gamma_i$ is the tangential $\gamma$-factor at
the turning point of the motion. For pure radial motion $\gamma_i
= 1$.

Integration of the above equation gives:

\begin{equation} \label{eq:motion:m:t}
2 r_0 t = \sqrt{1-\frac{r_i}{r}}(r^2 + \frac{3 r_i}{2} r) +
\frac{3}{4} r_i^2 \ln{\left(\frac{2
r}{r_i}(\sqrt{1-\frac{r_i}{r}}+1)-1\right)}
\end{equation}

For $r \gg r_i$ the logarithm can be neglected and the
square-root can be Taylor-expanded to first order:

\begin{equation} \label{eq:motion:m:t2}
2 r_0 t \approx r^2 + r_i r
\end{equation}

or

\begin{equation} \label{eq:motion:m:r}
r \approx -\frac{r_i}{2} + \frac{1}{2} \sqrt{r_i^2 + 8 r_0 t}
\end{equation}

\subsubsection{\label{sec:motion:shell:mass}Geodesic motion of a thin spherical shell of massive particles}

Two massive particles separated initially at $r_i$ by a
radial coordinate separation $\delta r_i$ and moving geodesically
outward, will have a radial coordinate separation $\delta r(r)$
that tends to the following constant value when $r \gg r_i$

\begin{equation} \label{eq:delta_r:m}
\delta r(r) \rightarrow -\frac{\delta r_i}{2}
\end{equation}

The minus sign in the above equation is due to a "cross-over" of
the massive particles, which takes place at a very early stage of
the motion, i.e. where $r \approx r_i$. After the cross-over the
radial coordinate separation of the massive particles quickly
approaches the value $\delta r_i/2$ and remains effectively constant
henceforth.

Constant coordinate separation means that the proper radial
separation viewed by an observer at rest in the ($t, r, \theta,
\varphi$)-coordinate system, $\delta l$, develops according to:

\begin{equation} \label{eq:delta_l:m}
\delta l(r) = \delta l_i \sqrt{\frac{r}{r_i}}
\end{equation}

Therefore, quite contrary to the movement of the zero-rest mass
particles, an outmoving shell of massive particles expands along
the radial direction.

Due to the negative radial pressure the energy in the shell
increases as the shell expands radially. A similar calculation as
in the zero rest mass case gives the following total energy change
in the shell:

\begin{equation} \label{eq:E:massive}
\delta E(r) = \delta E_i \sqrt{\frac{r}{r_i}}
\end{equation}

The proper volume of the shell, as measured from an observer at
rest in the chosen coordinate system, changes according to:

\begin{equation} \label{eq:dB:massive}
\delta V(r) = \delta V_i \left(\frac{r}{r_i}\right)^{\frac{5}{2}}
\end{equation}

A factor proportional to $r^2$ comes from the proper surface area
of the shell, a factor of $r^{1/2}$ from the proper expansion of
the shell's radial dimension.

Note that exactly as in the case of zero rest-mass particles the
mass-energy density of the expanding shell, viewed from an observer at
rest, follows an inverse square law.

\begin{equation} \label{eq:rho:massive:shell}
\rho(r) = \frac{\delta E(r)}{\delta V} = \frac{\delta E_i
\sqrt{\frac{r}{r_i}}}{\delta V_i (\frac{r}{r_i})^{\frac{5}{2}}} =
\rho_i \left(\frac{r_i}{r}\right)^2
\end{equation}

Therefore the geodesic motion of a shell of massive particles also
is self-consistent with the static holographic mass-energy density. In
fact, this self-consistency is independent of the path of the
motion and the nature of the particle. It only depends on the
radial pressure which guarantees, that for any conceivable motion,
geodesic or not, the energy of the particles will change such that
the energy-density law $\rho \propto 1/r^2$ within the holostar is
preserved.

For instances, if the motion takes place in a way, that the proper
expansion of the shell in the radial direction is zero, no work is
done against the radial pressure. The energy within the shell will
remain constant. The proper volume of such a thin shell of
constant proper radial extension will be proportional to the
shell's spherical boundary area, i.e. $V \propto r^2$, so that the
energy-density within the shell evolves as $\rho \propto 1/r^2$.

Under the assumption that no particles are created or destroyed in
the shell, the number-density of massive particles develops
according to the inverse proper volume of the shell. For a
geodesically moving thin shell with $\delta r_i = const$ we find:

\begin{equation} \label{eq:n:massive}
n_m(r) = n_i (\frac{r_i}{r})^{\frac{5}{2}}
\end{equation}

\subsubsection{Does the motion of massive particles conserve energy?}

According to equation (\ref{eq:E:massive}) the energy of a
geodesically moving shell of massive particles increases with
$\sqrt{r/r_i}$. If we assume no particle creation in the
shell\footnote{We will see later, that this assumption is not
correct: Viewed from the frame of the co-moving observer a
geodesically moving shell expands isotropically against the
negative radial pressure. Due to the radial boost invariance of
the space-time the pressure in the co-moving frame is identical to
the pressure in the stationary coordinate frame. Expansion against
a negative pressure creates energy. If the mass of the massive
particles does not change, additional particles must be created.},
each massive particle must acquire an increasingly larger energy
$E = m \sqrt{r/r_i}$. Where does this energy come from?

The energy is nothing else than the energy of the motion of a
massive particle, as viewed by an asymptotic observer at rest with
respect to the $(t, r, \theta, \varphi)$ coordinate system.
According to equation (\ref{eq:gamma:ri}) the $\gamma$-factor of a
geodesically moving particle with respect to the stationary
coordinate frame is given by:

\begin{equation} \label{eq:gamma}
\gamma^2(r) = \frac{r}{r_i}
\end{equation}

The total energy $E(r)$ of a massive particle with rest mass $m$,
viewed by an observer at rest in the coordinate system, will then
be given by:

\begin{equation} \label{eq:E(r):massive}
E(r) = \gamma \,  m = m \sqrt{\frac{r}{r_i}}
\end{equation}

This is just the energy-increase per particle, which has been
derived from the pressure-induced increase due to the radial
expansion of the shell against the negative pressure.\footnote{In
the case of a non-zero tangential component of the motion one has
to replace $r_i \rightarrow r_i / \gamma_i^2$. This leads to $E =
m \gamma_i \sqrt{\frac{r}{r_i}} = E_i \sqrt{\frac{r}{r_i}}$. Now
$E_i = m \gamma_i$ is nothing else than the {\em total} energy of
the particle at the turning point of the motion, taking its
kinetic energy properly into account via its $\gamma_i$-factor.
Therefore energy is not only conserved for pure radial motion, but
for any type of geodesic motion of a massive particle.}

Therefore, from the perspective of an exterior observer, energy is
conserved not only locally, but globally in the holostar
space-time. This is remarkable, because global energy conservation
is not mandatory in general relativity. In fact, there exist only
a limited class of space-times (such as asymptotically flat
space-times) in which a global concept of energy can be rigorously
defined. Except for a a small class of symmetric space-times it is
generally impossible to define a meaningful concept of
gravitational energy.\footnote{So far no realistic cosmological
space-time has been found, in which global energy-conservation
holds. In the standard cosmological models global
energy-conservation is heavily violated in the radiation dominated
era. If the holostar turns out to be a realistic alternative to
the standard cosmological models, and energy is conserved globally
in the holostar space-time, one could use global energy
conservation as a selection principle to chose among various
possible solutions. This would lead to the requirement, that a
realistic space-time is quite likely one that possesses a global
Killing-vector field, from which an energy-conservation law can be
derived.}

\subsubsection{On the relative number-densities of charged particles and photons}

The assumption of the previous section, that the particle-number
remains constant in a geodesically expanding shell of massive
particles, might not be entirely correct. In section
\ref{sec:frames} it will be shown that geodesic expansion of the
massive particles against the negative pressure has the effect to
produce new particles in the co-moving frame. At least this will
be the case, if the rest mass of the particles doesn't change
during the expansion.

Yet genuine particle-production in the shell must obey the
relevant conservation laws. Some conservation laws, such as
conservation of lepton- and baryon-number, are empiric. Such
"empiric" conservation laws can be violated without severe
consequences for our established physical theories. Some
conservation laws, however, are linked to first principles, such
as local gauge-symmetries. One of these conservation laws is
charge-conservation. Therefore we expect, that particle creation
in the shell at least observes charge conservation. As charge is
quantized in units of the electron charge the difference between
positively and negatively charged elementary particles in the
shell must remain constant, so that the net number-density of
charged particles $n_+ - n_-$ will always scale with $1/r^{5/2}$
in the holostar space-time, even if the total number density of
neutral particle-antiparticle pairs $n_+ + n_-$ evolves
differently.

Note, that even if particle production in the frame co-moving with
the massive particles is taken into account, the number-density of
the massive particles declines much faster than the number-density
of the zero-rest mass particles. This is a consequence of the
fact, that a geodesically moving shell of photons is compressed in
the radial direction, whereas a geodesically moving shell of
massive particles expands. Although the expansion against the
negative pressure necessarily increases the local energy in the
shell, the energy generated by the expansion is not high enough to
produce massive particles in excess of the number of photons
entering the shell.\footnote{At least this is so, as long as the
rest-mass of the massive particles remains constant. In fact, the
rest mass of the massive particles would have to fall quicker with
$r$ than the red-shift of the photons, i.e. $m \propto 1/r^\alpha$
with $\alpha > 1/2$.} In section \ref{sec:frames} it will be
shown, that the number-density of geodesically moving massive
particles in the co-moving frame evolves as $\overline{n_m}
\propto 1/r^2$, if the rest-mass of the massive particles remains
constant.

In any case, the ratio of zero-rest mass particles to massive
particles will depend on $r$. The ratio must be independent of the
local Lorentz frame of the observer. Therefore, whenever the photons are
chemically and kinematically decoupled from the massive particles,
their ratio can be used as a "clock" by an observer co-moving with
the massive particles. However, any net momentum transfer between
massive and zero rest-mass particles will distort this relation.
Therefore this particular "clock" cannot be considered as highly
accurate.

\subsection{The Hawking entropy-area law}

It has already been shown, that the total number of zero rest-mass
particles in the holostar is proportional to the surface area of
its membrane. For an ideal gas of massless particles the entropy
per particle is constant. This suggests, that the total entropy of
the holostar is equal to the Hawking entropy, which also scales
with area.

Note, that due to the fact, that the holostar's temperature at
infinity is proportional to $1/r_h$, an {\em exterior} observer
will already be able to deduce an entropy $S \propto r_h^2$. The
observer does not need to know anything about the interior
space-time. He only needs to know the fundamental equation of
thermodynamics $\partial S / \partial E = 1 / T$ and the fact,
that the total energy of the holostar measured at infinity is
proportional to its "radius", i.e. $E = M = r_h/2$. With these
simple assumptions $S \propto r_h^2$ follows directly from
differential calculus. Here, however, we are not interested in the
perspective of an exterior observer, who is ignorant of the true
cause of the entropy of a compact self-gravitating body. Rather
the objective of this section is to determine the entropy of such
a body from its interior matter state, which is well defined in
the holostar space-time. The unique perspective of the interior
observer will enable us to shed some light on the true origin of
the Hawking-entropy.

\subsubsection{A holostar consisting predominantly out of ultra-relativistic particles}

For a large holostar the total number of non-relativistic
(massive) particles ($N_m \propto r_h^{3/2}$) can be neglected
with respect to the number of ultra-relativistic (zero rest-mass)
particles ($N \propto r_h^2$). The main contribution to the mass
of a large holostar comes from its outer low temperature regions.
Whenever the temperature becomes lower than the rest mass of a
particular particle, the number density of the non-relativistic
massive particles is thinned out with respect to the number
density of the yet relativistic particles. For a large holostar
the total number of particles is dominated by the zero rest-mass
particles. The same is true for a small holostar. For a small
holostar the internal local temperature is so high, that the
majority of massive particles will become ultra-relativistic, in
fact massless whenever the Higgs-mechanism fails to function,
because $M_H^2 > 0$. Therefore the dominant particle species - in
terms of numbers - of a large or small holostar will be
ultra-relativistic or zero rest mass particles.

Under the not too unreasonable assumption that the entropy of the
holostar is proportional to the number of its particles, one
recovers the Hawking entropy-area law for black holes up to a
constant factor. This factor is very close to unity, as can be
shown by the following simplified argument:

The number-density of the zero rest-mass particles scales with
$1/r^{3/2}$. The number-density has dimension $1/r^3$, so one
needs a dimensional quantity in order to get a meaningful
expression for the number-density. The only relevant dimensional
quantity in the holostar space-time is the scale parameter $r_0$,
which introduces a natural length scale. $r_0$ has been
experimentally determined in section \ref{sec:r0} as roughly twice
the Planck length. A reasonable ansatz for the number-density of
the mass-less particles, which should be correct within an order
of magnitude, is given by:

\begin{equation}\label{eq:number:density}
n = \frac{1}{(r r_0)^{\frac{3}{2}} }
\end{equation}

Let us denote the entropy per ultra-relativistic particle by
$\sigma$. For an ideal ultra-relativistic gas with zero chemical
potential $\sigma$ is equal to $s \simeq 3.6$ for bosons and
$\sigma \simeq 4.2$ for fermions. In any case, for an order of
magnitude estimate $\sigma \approx 3$ will be a correct enough
assumption.

The total entropy is then given by the proper integral of the
number-density multiplied by $\sigma$. We find:

\begin{equation}
S = \int_0^{r_h}{\sigma \, n  \, dV} \approx \int_0^{r_h}{
\frac{\sigma}{(r r_0)^{\frac{3}{2}} } \, 4 \pi r^2
\sqrt{\frac{r}{r_0}} dr} = \frac{\pi r_h^2}{\hbar} \cdot \frac{2
\sigma \hbar}{r_0^2}
\end{equation}

But $\pi r_h^2 / \hbar$ is nothing else than the Hawking-entropy
$S_{BH} = A / (4 \hbar)$ for a spherically symmetric black hole
with gravitational radius $r_h$. With $r_0^2 \approx 3.5 \, \hbar$
and with $\sigma \approx 3.6$ we get $S \approx 2 \, S_{BH}$ as an
order of magnitude estimate. A more definite relationship will be
derived in \cite{petri/thermo}. The fact that a rough order-of
magnitude estimate gives such a good agreement suggests quite
strongly, that holostar-entropy should in fact be equal to the
Hawking entropy.

\subsubsection{A holostar consisting predominantly out of massive particles}

Although the dominant particle species in the holostar - with
respect to particle-{\em numbers} - will be massless particles,
these particles must not necessarily dominate with respect to
their contribution to the local energy- or entropy-density. A large holostar,
who's local radiation temperature is well below the rest mass of
the nucleon, will be matter-dominated, with just a very low
contribution of ultra-relativistic particles such as neutrinos or
photons to its total mass-energy density.

It is instructive to calculate the entropy in the case, where the
holostar consists exclusively out of massive particles. The only
input that is needed is the assumption, that the holostar is in
thermal equilibrium, not only with respect to the radiation (which
is Planck-distributed, even when all interactions have ceased),
but also with respect to its massive particles. The assumption of
thermal equilibrium is reasonable, because the life-time of a
holostar is several orders of magnitude larger than its
evaporation time, except for Planck-size holostars.

To simplify the calculations, let us assume that the massive
particles consist of just one species with mass $m$. The argument
works equally well for a combination of massive particles with
different masses. The rest mass $m$ of the particles need not be
constant. The calculation for the total entropy is not changed,
when the rest mass is an (arbitrary) function of the radial
coordinate value $r$, i.e. $m = m(r)$.

If the particles with mass $m(r)$ are the {\em only} contributor
to the holostar's internal energy-density, their number-density
$n_m$ is given by:

\begin{equation} \label{eq:nm}
n_m = \frac{\rho}{m} = \frac{1}{8 \pi m(r) r^2}
\end{equation}

For a constant rest mass $m$ the number-density of the massive
particles in the matter-dominated era will thus scale with $n_m
\propto 1/r^2$.

It is a - apparently not very well known - result from microscopic
statistical thermodynamics, that the entropy of a massive particle
in thermodynamic equilibrium (with zero chemical potential) is
given by:

\begin{equation} \label{eq:sm}
\sigma_S = \frac{m}{T} = \frac{4 \pi m \sqrt{r r_0}}{\hbar}
\end{equation}

where equation (\ref{eq:Tlocal}) for the local radiation
temperature was used to obtain the the right hand side of the
above equation. For $m(r)=const$ the local entropy-density of
massive particles in the holostar space-time grows with the
square-root of the radial coordinate value $r$, due to the inverse
square-root dependence of the temperature. Note that $T \propto 1
/ \sqrt{r}$ follows solely from the geometric properties of the
holostar solution. The exact numerical factor, as written down in
equation (\ref{eq:Tlocal}), however, was obtained by setting the
holostar's temperature at infinity {\em equal} to the Hawking
temperature. Therefore the following calculation cannot claim that
the Hawking entropy area law is predicted with the right numerical
factor. This factor has already been incorporated into the
calculation by fixing the local temperature $T$ to the Hawking
temperature. Yet the {\em proportionality} of entropy and area $S
\propto A / \hbar$ is a genuine prediction derived solely from the
holostar's geometric properties.

With equations (\ref{eq:nm}, \ref{eq:sm}) the entropy of a
matter-dominated holostar can be calculated by a simple integral:

\begin{equation}
S = \int_0^{r_h}{\sigma_S \, n \, dV} = \frac{\pi r_h^2}{\hbar} =
\frac{A}{4 \hbar} = S_{BH}
\end{equation}

The integrand does not contain the local value of the mass $m(r)$.
Therefore the holostar's entropy is equal/proportional to the
Hawking entropy, even if the mass is an arbitrary radial function
of $r$. This observation will prove to be important in section
\ref{sec:nucleosynthesis}.

\subsection{\label{sec:s=g}On the identity of local entropy density and geodesic acceleration}

With the assumption that the holostar-entropy is {\em exactly
equal} to the Hawking entropy of a black hole with the same
extrinsic properties (mass, charge, angular momentum), one can
derive a truly remarkable relation between the entropy-density $s
= S/V$ and the geodesic acceleration $g$: With $S = \pi r^2 /
\hbar$ the entropy $\delta S$ within any thin spherical shell of
thickness $\delta r$ in the holostar's interior is given by

$$ \delta S = \frac{2 \pi r \delta r}{\hbar}$$

If we divide $\delta S$ by the proper volume of the shell $\delta
V = 4 \pi r^2 \delta r \sqrt{A}$ we get the following value for
the entropy-density $s$ at radial coordinate-position $r$:

\begin{equation} \label{eq:s=g}
s = \frac{\delta S}{\delta V} = \frac{1}{2 r
\hbar}\sqrt{\frac{r_0}{r}} = \frac{g}{\hbar}
\end{equation}

The highly non-trivial result is, that the local entropy-density
at radial position $r$ is nothing else, than the proper geodesic
acceleration of a {\em stationary} observer at this position,
measured in Planck units. In a certain sense one can say, that the
thermodynamic properties of the holostar space-time are determined
{\em globally and locally} by its geometric properties. This is
not totally unexpected. Jacobson was the first to point out, that
the Einstein field equations can be derived from thermodynamics
\cite{Jacobson} under some reasonable additional assumptions.

It is possible to relate the entropy-density $s$ to other
important parameters in the holostar space-time. If the holostar's
entropy is equal to the Hawking entropy, the holostar's
temperature at infinity {\em must} be equal to the Hawking
temperature as well.\footnote{Temperature and entropy are
conjugate variables. If one value is fixed, the other follows from
the relation $\delta S / \delta E = 1/T$.} The local radiation
temperature given in equation (\ref{eq:Tlocal}) was derived
exactly under this assumption. If one multiplies equation
(\ref{eq:s=g}) with the local radiation temperature one finds

\begin{equation} \label{eq:s=e/T}
s T = \frac{1}{8 \pi r^2} = \rho
\end{equation}

Note that the above result even remains valid, if the Hawking
temperature is changed by a constant factor (or equivalently, if
the holostar's temperature at infinity is not equal, but just
proportional to the Hawking temperature). It is well known, that
any rescaling of the Hawking temperature by a constant factor
induces the inverse factor in the entropy. This must be so,
because otherwise the thermodynamic relation $T \partial S /
\partial E = 1$ would not remain valid in the exterior space-time:
Rescaling the temperature at infinity $T \rightarrow \overline{T}
= c \, T$ has no effect on the total energy $E$ of a black hole or
holostar. Therefore the entropy $S \rightarrow \overline{S} = x \,
S$ must acquire a constant factor $x = 1/c$, so that the rescaled
product $\overline{T} \overline{S}$, evaluated in the {\em
exterior} space-time, remains unchanged. The same applies to the
product of the relevant interior quantities $s T$: The holostar's
interior entropy-density $s$ is directly proportional to the total
entropy $S$, measured by an observer in the exterior space-time
(the total volume $V$ of the interior space-time is unaffected by
any rescaling or $T$ or $S$). The interior temperature $T$ is
directly proportional to the temperature at infinity $T_\infty$,
because the blue-shifted temperature at infinity must be equal to
the local temperature at the holostar's boundary membrane.

Therefore equation (\ref{eq:s=e/T}) is a very general requirement,
independent of the type of interior matter. For the derivation of
equation (\ref{eq:s=e/T}) the only necessary assumptions are the
geodesic equations of motion, from which $T \propto 1 / \sqrt{r}$
for the radiation temperature follows (see section
\ref{sec:Tlocal}).  $T \propto 1 / \sqrt{r}$ - in consequence -
implies $T_\infty \propto 1/r$ and $S \propto r^2$, according to
the holostar equations. In fact, the three conditions $T \propto 1
/ \sqrt{r}$, $T_\infty \propto 1/r$ and $S \propto ^2$ are
interchangeable. Any one of these three conditions necessarily
leads to equation (\ref{eq:s=e/T}).

If we divide equation (\ref{eq:s=e/T}) by the number-density $n$
(and if we neglect the kinetic energy of the particles $E_{kin} =
3/2 T \ll m$), we get

\begin{equation} \label{eq:sigma=m/T}
\sigma_S = \frac{s}{n} \simeq \frac{m}{T}
\end{equation}

We have derived the correct thermodynamic relation for the entropy
of a massive particle (with nearly zero chemical potential) with
no reference to microscopic statistical
thermodynamics.\footnote{The exact formula for the thermodynamic
entropy of a massive particle in the ideal gas approximation is
$\sigma_S = m/T + 3/2 + (1-u)$, where $u$ is the chemical
potential per temperature $u = \mu / T$. Equation (\ref{eq:s=e/T})
predicts $\sigma_S = \epsilon / T$, where $\epsilon = m + 3/2 T$
is the {\em total} energy of the particle. If the holostar's
entropy per particle derived from equation (\ref{eq:s=e/T}) is to
be equal to the thermodynamic entropy of the particle, one gets
the prediction $u=1$.} The only ingredient are the Einstein field
equations (=classical geometry) and the Hawking entropy-area
formula (=quantum theory, i.e. evolution of a quantum field in the
exterior space-time). This again shows, that the field equations
of general relativity already contain all necessary information
for a complete thermodynamic description of a self-gravitating
system.

One can express equation (\ref{eq:sigma=m/T}) in a slightly
different form. For a massive particle in geodesic motion the
$\gamma$-factor of the motion, viewed by a stationary observer, is
given by $\gamma = \sqrt{r/r_i}$, where $r_i$ is the turning point
of the motion. We will see later, that $r_i \approx r_0$. The
momentum of the particle, viewed from the stationary frame, is $p
= \beta \gamma m$. Putting this all together and using equation
(\ref{eq:Tlocal}) for the local temperature and $\beta \simeq 1$
for $\gamma \gg 1$, the entropy per massive particle is given by

\begin{equation} \label{eq:sigma=pr0}
4 \pi \sigma_S \simeq \frac{r_0 p}{\hbar}
\end{equation}

Therefore the entropy per massive particle in the holostar
space-time is directly related to the product of the scale factor
$r_0$ with its momentum $p$, viewed from an observer in the
stationary coordinate frame.

\subsubsection{On the interpretation of a compact self-gravitating body as a massive particle}

Equation (\ref{eq:sigma=m/T}), which is the correct thermodynamic
equation for the entropy of a massive {\em particle} in
thermodynamic {\em equilibrium}, is quite compatible with the
black hole temperature formula. Consider an isolated spherically
symmetric black hole (or black hole type object, such as a
holostar) in an exterior vacuum space-time. Such a black hole type
object has an entropy of $S = 4 \pi M^2 / \hbar$. It's lifetime is
finite, due to Hawking evaporation, yet enormous compared to the
largest conceivable time-scales, such as the age of the universe.
When composite objects with much smaller life-time, such as stars
in a galaxy or a globular cluster, can be treated as approximately
in thermal equilibrium, there appears to be no fundamental reason
why the thermodynamics of a compact black hole type object should
be different from the thermodynamics of a (large) particle.

However, an isolated black hole type object cannot be in
thermodynamic equilibrium. It will necessarily evaporate due to
Hawking-radiation. In order to be in thermal equilibrium with its
surroundings, one must place the object into an environment with a
temperature comparable to the object's temperature (at infinity),
which is given by $T = \hbar / (8 \pi M)$ for a spherically
symmetric black hole type object.

If one identifies such a "thermalized" black hole type object with
a massive particle in a thermal bath and sets the temperature of
the thermal bath equal to the object's Hawking temperature, its
"particle type" entropy $\sigma_S$ (which includes the entropy of
the thermal bath) is given by equation (\ref{eq:sigma=m/T}), i.e.

\begin{equation}
\sigma_S = \frac{M}{T} = \frac{8 \pi M^2}{\hbar} = 2 S_{BH}
\end{equation}

This is double the entropy of an isolated black hole, which isn't
too surprising. In order to place an isolated black hole with mass
$M$ into a {\em stable} thermal environment at temperature $T$ one
needs additional matter to "create" the right temperature $T$, so
that the black hole does not evaporate. One - very simple way - to
create such an environment is to place a second black hole with
exactly the same mass at an infinite distance from the first black
hole. As the situation is symmetric, each of the two black holes
provide the correct temperature at infinity for the other black
hole, which allows both black holes to remain stationary.

A similar argument is this: Consider an isolated black hole with
mass $M$ and temperature $T$. Such a black hole will evaporate
eventually. Now break up the original black hole into two black
holes with mass $m = M/2$. The original entropy of the system
$S_1$ (with one isolated black hole) is given by ($\hbar=1$):

$$S_1 = 4 \pi M^2$$

Naively applying the black hole entropy-formula to the two black
holes of mass $m = M/2$ we find

$$S_2 = 2 \cdot  4 \pi m^2 = 2 \pi M^2 = \frac{S_1}{2}$$

According to the common belief the entropy of the broken up system
is a factor two smaller than the entropy of the of the initial
system. It appears, that the second law forbids such a process.

But it might not be correct to use the formula for an isoloated
black hole for the two broken up black holes. Both black holes
aren't isolated any more. Rather they reside in the thermal
environment provided by the other black hole. Any particle
evaporated from one black hole will eventually find its way to the
other. The situation is (quasi) stationary. As the mass is halved,
the temperature at infinity will be doubled with respect to the
original configuration $T_2 = 2 \, T = \hbar / (4 \pi M)$.
Therefore the entropy of each of the two broken up black holes is
given by $m / T_2 = 2 \pi M^2$. But this means that the total
entropy is nothing else than the original entropy:

$$S_2 = 4 \pi M^2 = S_1$$

Entropy appears to be conserved!

Evidently this argument is far too simplistic. It does not work,
when we break up the black hole in three or more pieces.
Furthermore we have broken spherical symmetry. Yet there is a way
out: We can superimpose a spherically symmetric black hole with a
spherically symmetric white hole. This preserves spherical
symmetry and generates a time-symmetric solution (The field
equations are time symmetric, and therefore do not prefer a black
hole over a white hole). In a certain sense the static holostar
solution is quite similar to the superposition of a white and a
black hole: The holostar solution can be regarded as the
superimposition of an expanding sector (the geodesically outmoving
particles) and a contracting sector (the geodesically inmoving
particles). Both sectors are essentially decoupled for large $r$,
because of the highly relativistic motion of the particles, which
lead to extremely low cross-sections for the collision of a
particle in one sector with a particle in the other sector.
However, in contrast to a white hole superimposed over a black
hole both sectors are not completely disjoint. Geodesically moving
particles in the holostar space-time can switch from the expanding
to the contracting sector: Any particle starting out from the
holostar's center swings back and forth between the center and the
angular momentum barrier situated in the exterior space-time (at
$r = 3 r_h / 2$, roughly half a Schwarzschild radius outside the
membrane). The probability for a switch from the expanding to the
contracting sector - or vice versa - is unity at the holostar's
center and nearly 1 at the photon angular momentum barrier in the
exterior space-time. Furthermore, any collision between outflowing
and inflowing particles can induce such a switch. The highly
relativistic motion and - consequentially - the low cross-sections
make such a switch quite improbable for the interior region,
unless one is quite close to the center.

A third - presumably more acceptable - argument can be devised for
the process, where the original (isolated) black hole is not split
in half, but rather a small fraction (a single particle or a small
black hole) with mass $m \ll M$ is removed to infinity. Such
processes continuously occur due to Hawking evaporation. What will
the entropy of the broken up system be? If we disregard the
back-reaction of the small mass $m$ on the large black hole $M$,
we find

$$S_2 = 4 \pi (M-m)^2 + \frac{m}{T} = 4 \pi M^2 - 4 \pi m^2 = S_1 + o(m^2)$$

Except for a small correction of order $m^2$ entropy is conserved
as well. One can go even one step further. The small correction
$S_1 - S_2$ is nothing else than the "Hawking entropy" of the
removed particle (or small black hole) in {\em isolation}: $\Delta
S = 4 \pi m^2$. One can argue, that this entropy should have been
added to $S_2$ in the first place: The small mass $m$ at infinity
is not large enough to prevent the original large black hole from
evaporating (this is a different situation than having two large
black holes of identical mass at large separation, or a white and
black hole superimposed over each other). Let us assume, that $M$
is constructed exclusively out of stable particles of mass $m$,
where $m \ll m_{Pl}$. The final result of the evaporation will be,
that we have a collection of {\em isolated} particles distributed
sparsely in a nearly flat space-time. But after the evaporation of
the large black hole there will be no well-defined temperature of
the space-time.\footnote{Most likely, the hypothetical end-state
of the evaporation will not meet the requirement for an internally
self-consistent space-time. It is quite unlikely, that the
evaporated particles don't clump together somewhere else in the
space-time. Self-consistency rather points to a scale-invariant,
self-similar space-time with a general $1/r^2$-law for the
matter-distribution, which allows significant clumping within any
hierarchical sub-structure. The matter-distribution in the
sub-structures will also be subject to a $1/r^2$-law. The
infinitely extended holographic solution might be a very simple
model for the - unclumped - large scale structure of such a
self-consistent space-time. A realistic truly self-consistent
description would have to extend the smooth holostar model to a
hierarchical, quasi-fractal model with self-similar
sub-structure.} The calculation of the thermodynamic entropy $m/T$
for the individual particles in a thermal bath fails. Yet the
isolated particles will still have an intrinsic entropy. Naively
applying the Hawking formula for any one particle one gets
$\sigma_S = 4 \pi m^2$.

Curiously, entropy conservation leads to the prediction, that the
number of "constituent" particles $N$ of the original large black
hole is proportional to $M^2$, i.e. $S_1 = S_2 \rightarrow N = M^2
/ m^2$.\footnote{Quite clearly all the arguments devised in this
section, especially the last, are not 100 \% watertight. One can
always object, that the final configuration, the physical process
leading from the initial to the final state, or even the initial
configuration itself (here: an isolated black hole in vacuum) does
not correspond to a "physically realistic" situation. This
objection must be taken very seriously. It applies to all sorts of
thought experiments in general relativity: The great difficulty in
general relativity is, that one cannot be certain whether a
particular {\em initial} configuration, on which the thought
experiment is based, corresponds to physical reality. General
relativity is a non-linear theory. Contrary to linear theories
like electro-magnetism, where the superposition principle holds,
valid initial configurations cannot be constructed just by placing
various known sub-structures ("smaller" solutions of the field
equations, such as particles or small black holes) at arbitrary
positions / velocities in order to construct a more complicated
state. Just to give an example: There is a famous thought
experiment, which - apparently - "proves" the existence of black
holes: Imagine a very large black hole. Place a shell of
low-density, low-entropy matter in a spherical shell distanced a
few gravitational radii from its - imagined - event horizon. If
the - imagined - black hole is large enough, the matter-density of
the shell can be made arbitrary low. The local physics of such a
shell will be very well known. Now let the shell collapse under
its own gravity. There is nothing in the local physics that can
prevent the shell from collapsing through the event horizon.
Inevitably a black hole will form. As soon as the shell passes the
apparent horizon, an enormous amount of entropy will be created.
The argument appears water-tight. But there is one very serious
catch: It doesn't address the crucial question, whether it is
physically possible to construct such a shell around a large
spherical void in the first place. It is quite obvious, that such
construction is physically impossible, at least in the universe we
live in. Any attempt to clear out such a huge void would fill it
up with the energy that was used for the removal process. Any
argument that is based on physically impossible premises, looses
much of its bite.}

\subsubsection{An upper limit for the mass of large black holes in the centers of galaxies}

With the assumption that the entropy of a large holostar (or black
hole) in a thermal bath is given by equation (\ref{eq:sm}) one can
derive an upper limit for the mass of the massive black holes in
the center of galaxies, which is quite close to the observations.

According to \cite{Peebles/Principles}, equation (5.146) there are
roughly $3 \cdot 10^8$ galaxies within the Hubble-length. Using
equation (\ref{eq:s=g}) one can calculate the total entropy within
the Hubble radius $r_H$ of an observer at radial position $r$ in
the holostar solution. We will see later (see section
\ref{sec:Hubble}) that the local Hubble-radius $r_H$ of a
geodesically moving observer at radial position $r$ is roughly
equal to $r_H \approx r$, so that the total entropy within his
Hubble-volume is roughly given by:

\begin{equation}
S_H \approx \frac{4 \pi}{3} r^3 s = \frac{2 \pi r^2}{3 \hbar}
\sqrt{\frac{r_0}{r}}
\end{equation}

With $r \approx 10^{61} r_{Pl}$ and $r_0 \approx 2 r_{Pl}$ this
evaluates to

\begin{equation}
S_H \approx 9 \cdot 10^{91}
\end{equation}

The entropy of all black holes at the centers of galaxies cannot
exceed this value. Under the assumption that every galaxy harbors
a black hole a rough order of magnitude estimate of the number of
large black holes $N_{H}$ within the Hubble-volume can be derived

$$N_{H} \approx 3 \cdot 10^8$$

which sets a maximum entropy for any of these black holes, given
by:

$$ S_{max} = \frac{S_H}{N_{H}} \approx 3 \cdot 10^{83} $$

However, the above equation assumes that the total mass of a
galaxy (or the total mass of the universe) is predominately
situated in the central black holes. This quite certainly is not
the case. Assuming that a large central black hole has a mass
which is considerably less than 1 \% of the mass of its host
galaxy, one gets

\begin{equation} \label{eq:S:max}
S_{max} \approx< 3 \cdot 10^{81}
\end{equation}

With the current value of the microwave-background temperature $T
= 2.725 K$ the entropy of a black hole of the mass of the sun is
determined by equation (\ref{eq:sm}):

$$S_\odot = \frac{M_\odot}{T} \approx 4.75 \cdot 10^{69}$$

This is roughly seven orders of magnitude less than the entropy
calculated by the Hawking formula. The microwave-background
temperature is constant within the Hubble-volume. Therefore the
entropy of a large black hole according to equation (\ref{eq:sm})
scales linearly with mass. We can express the maximum entropy in
equation (\ref{eq:S:max}) in units of the sun's entropy. The
maximum entropy can be converted to a maximum mass, which is given
by:

\begin{equation} \label{eq:BH:max:mass}
M_{BH} < 6 \cdot 10^{11} \, M_\odot
\end{equation}

The largest black holes that have been observed in active galactic
nuclei have mass estimates up to $3 \cdot 10^9 M_\odot$ \cite[p.
453]{Peacock/book}. This is still considerably less than the bound
given in equation (\ref{eq:BH:max:mass}). This is not unexpected,
due to the rough approximations in its derivation. For instance,
if one assumes that the fraction of black hole mass to the total
matter-content of the universe is considerably less than 1 \%
(which is quite certainly the case), one gets a bound which is
appropriately lower. Yet the bound in equation
(\ref{eq:BH:max:mass}) is close enough to the experimental value,
so that a mere numerical coincidence appears unlikely. If one
evaluates the entropy according to the Hawking-formula, ignoring
the thermal radiation bath, the mass-estimate fails miserably: One
gets a bound which is far too low: $M_{BH} < 100-1000 M_\odot$.

\subsection{Motion of massive particles in their own proper time}

In this section I will examine the equations of motion from the
viewpoint of a moving massive particle, i.e. from the viewpoint of
the co-moving material observer who moves geodesically. The
geodesic equation of radial motion for a massive particle,
expressed in terms of its own proper time, is given by:

\begin{equation} \label{eq:dr/dtau}
\frac{dr}{d\tau} = \sqrt{\frac{r_0}{r_i}}\sqrt{1-\frac{r_i}{r}}
\end{equation}

The radial coordinate velocity $dr / d\tau$ is nearly constant for
$r \gg r_i$. Integration of the above equation gives:

\begin{equation} \label{eq:tau2}
\tau = \sqrt{\frac{r_i}{r_0}}\left( r \sqrt{1-\frac{r_i}{r}} +
\frac{r_i}{2}
\ln{\left(\frac{2r}{r_i}\big(\sqrt{1-\frac{r_i}{r}}+1\big)-1\right)}\right)
\end{equation}

For large $r \gg r_i$ this can be simplified:

\begin{equation} \label{eq:tau}
\tau \cong \sqrt{\frac{r_i}{r_0}} \left(r + \frac{r_i}{2}
\ln{\frac{4 r}{r_i}}\right) \cong \sqrt{\frac{r_i}{r_0}} r
\end{equation}

The proper time it takes a material observer to move along a
radial geodesic trajectory through the holostar space-time is
proportional to $r$. This is very much different from what the
external asymptotic observer sees. The time measured by an
exterior clock at infinity has been shown to be proportional to
$r^2$.

Under the assumption that we live in a large holostar formula
(\ref{eq:tau}) is quite consistent with the age of the universe,
unless $r_i \ll r_0$.\footnote{As derived in section
\ref{sec:uncertainty}, it is quite unlikely, that any particle can
be emitted from $r_i \ll r_0$.} As can be seen from equation
(\ref{eq:tau}), it takes a material observer roughly the current
age of the universe in order to travel geodesically from the
Planck-density region at the holostar's center (i.e. at $r_i
\approx r_0$) to the low density region at $r \approx 10^{61}
r_0$, where the density is comparable to the density of the
universe observed today: For $r_i = r_0$ we find $\tau = r \approx
1.6 \cdot 10^{10} y$, if $r\approx 10^{61} r_{Pl}$. The proper
time of travel could be much longer: If a massive particle is
emitted (with zero velocity) from $r_i > r_0$, the proper time of
travel to radial position $r$ is larger than the former value by
the square root ratio of $r_i$ to $r_0$. Therefore the holostar
solution is compatible with the age of the oldest objects in our
universe ($\approx 1.3 - 1.9 \,  \cdot 10^{10} y$), but would also
allow a much older age.

Note, that the holostar solution has no need for inflation. The
"scale factor" $r$ develops proportional to $\tau$. The
"expansion", defined by the local Hubble-radius, also develops
proportional to $\tau$. Therefore any causally connected region
remains causally connected during the "expansion". The causal
horizon and the particle-horizon remain always proportional to
each other. Furthermore the number-density law $n_m \propto
1/r^{5/2}$ for massive particles indicates, that very massive
particles that have decoupled kinematically from the radiation at
an early epoch, such as magnetic monopoles, become very much
thinned out with respect to the radiation or the lighter
particles, such as baryons, which decouple much later.

\subsection{A linear and a quadratic redshift-distance relation}

From equation (\ref{eq:Redshift1}) a linear redshift-distance
relation can be derived, which is in some sense similar to the
redshift-distance relations of the standard Robertson Walker
models of the universe.

Imagine a concentric shell of material observers moving radially
outward through the holostar. Place two observers in galaxies at
the inner and outer surfaces of the shell and another observer in
a galaxy midway between the two outer observers. When the observer
in the middle reaches radial coordinate position $r_e$, the two
other observers are instructed to emit a photon with frequency
$\nu_e$ in direction of the middle observer.\footnote{All
observers could (at least in principle) synchronize their clocks
via the microwave background radiation or the total matter
density.} At this moment the proper radial thickness of the shell,
i.e. the proper distance between the two outer galaxies, shall be
$\delta l_e$. Let the three galaxies travel geodesically outward.
Some million years later (depending on how large $\delta l_e$ has
been chosen), the photons from the edge-galaxies will finally
reach the observer in the middle galaxy. The observer in the
middle determines his radial coordinate position at the time of
absorption, $r_a$. According to equation (\ref{eq:Redshift1}) the
photons will have been red-shifted by the squareroot of the ratio
of $r_e / r_a$. In order to derive the redshift-distance relation
we only need to know, how the proper distance between the galaxies
has changed as a function of $r$. According to equation
(\ref{eq:delta_l:m}) the proper radial distance grows proportional
to the square-root of the radial coordinate value, whenever the
galaxies move geodesically:

\begin{equation} \label{eq:RedshiftDistance}
1+z = \frac{\nu_e}{\nu_a} = \sqrt{\frac{r_a}{r_e}} = \frac{\delta
l_a}{\delta l_e}
\end{equation}

The final result is, that the light emitted from the distant
galaxies is red-shifted by the ratio of the proper distances of
the galaxies at the time of emission to the time of absorption.

However, this is the result that an observer at rest in the ($t,
r, \theta, \varphi$)-coordinate system would see. The co-moving
observer will find a different relation (see also the discussion
in the following section and in section \ref{sec:frames}). For the
co-moving observer the proper radial distance has to be multiplied
with his special relativistic $\gamma$-factor. If we denote the
proper separation between the galaxies in the system of the
co-moving observer with an overline, we find:

\begin{equation} \label{eq:RedshiftDistance:co1}
\frac{\overline{\delta l_a}}{\overline{\delta l_e}} =
\frac{r_a}{r_e}
\end{equation}

If we insert this into equation (\ref{eq:RedshiftDistance}) the
result is:

\begin{equation} \label{eq:RedshiftDistance:co}
(1+z)^2 = \left(\frac{\nu_e}{\nu_a}\right)^2 =
\left(\frac{T_e}{T_a}\right)^2 = \frac{\overline{\delta
l_a}}{\overline{\delta l_e}} = \frac{r_a}{r_e}
\end{equation}

This result might seem paradoxical. However, this is exactly what
the co-moving observer must see: The stress-energy tensor of the
holostar space-time is radially boost invariant. Therefore any
radial boost should not affect the local physics. This means that
in the co-moving frame, as well as in the coordinate frame, the
frequency of the photons should be proportional to the local
radiation temperature, i.e. $\nu \propto T$. The local radiation
temperature, however, depends on the inverse square-root of the
radial coordinate value: $T \propto 1/\sqrt{r}$. Due to Lorentz
contraction (or rather Lorentz-elongation in the co-moving frame)
the proper distance $\overline{\delta l}$ in the radial direction
between two geodesically moving massive particles develops
proportional to $r$ (see also the next section). Putting all this
together gives: $\overline{\delta l} \propto r \propto 1/T^2
\propto 1 /\nu^2 \propto 1/(1+z)^2$.

\subsection{\label{sec:Hubble}An isotropic Hubble flow of massive particles}

With the equations of motion for massive particles one can show,
that an observer co-moving with the massive particles within the
outmoving shell will see an isotropic Hubble-type expansion of the
massive particles with respect to his point of view.

First let us calculate the Hubble-flow viewed by the co-moving
observer in the tangential direction, by analyzing how the proper
distance between the radially moving observer and a neighboring
radially moving particle develops. I.e. observer and particle
always have the same $r$-coordinate value. Because of spherical
symmetry the coordinate system can be chosen such, that both move
in the plane $\theta = \pi/2$. The observer moves along the radial
trajectory $\varphi = 0$ and the neighboring particle along
$\varphi = \varphi_0$. The proper distance between observer and
particle is given by:

$$l = r \varphi_0$$

After a time $d\tau$ the distance will have changed due to the
radial motion of both particles:

$$\frac{dl}{d\tau} = \frac{dr}{d\tau} \varphi_0$$

The "speed" by which the particle at $\varphi_0$ moves away from the
observer is given by:

$$v = \frac{dl}{d\tau} = \frac{dr}{d\tau} \varphi_0 = \frac{dr}{d\tau} \frac{l}{r}$$

Therefore the Hubble-parameter in the tangential direction is
given by:

$$H_{\perp} = \frac{v}{l} = \frac{dr}{d\tau} \cdot \frac{1}{r}$$

Note that special relativistic effects due to the highly
relativistic motion of the co-moving observer don't have to be
taken into account, because the distances and velocities measured
are perpendicular to the direction of the motion.

For the derivation of the Hubble parameter in the radial direction
the Lorentz-contraction due to the relativistic motion has to be
taken into account. The proper radial separation of two particles
in geodesic motion, as seen by the observer at rest in the ($t$,
$r$, $\theta$, $\varphi$) coordinate system, develops as:

$$ l = l_i \sqrt{\frac{r}{r_i}}$$

This formula has to be corrected. Due to the relative motion of
the co-moving observer, the observer at rest in the ($t$, $r$,
$\theta$, $\varphi$) coordinate system will see a proper length,
which has been Lorentz-contracted. Therefore the co-moving
observer must measure a proper length which is larger by the
special relativistic $\gamma$-factor. In the system of the
co-moving observer the formula for the proper length (denoted by
barred quantities) is then given by:

$$\overline{l} = \overline{l_i} \frac{r}{r_i}$$

This gives:

\begin{equation}
H_r = \frac{\frac{d\overline{l}}{d\tau}}{\overline{l}} =
\frac{dr}{d\tau} \cdot \frac{1}{r} = H_{\perp}
\end{equation}

The radial and the tangential local Hubble-values are equal. They
just depend on $r$ and the "proper radial coordinate velocity"
$dr/d\tau$. Its value is given by equation (\ref{eq:dr/dtau}). It
is nearly constant for $r \gg r_i$. Therefore the isotropic
Hubble-parameter can finally be expressed as:

\begin{equation} \label{eq:Hubble}
H(r) = \frac{1}{r} \sqrt{\frac{r_0}{r_i}-\frac{r_0}{r}} \simeq
\frac{1}{r} \sqrt{\frac{r_0}{r_i}} \simeq \frac{1}{\tau}
\end{equation}

When we are far away from the starting point of the motion, i.e.
$r \gg r_i$, the product of the Hubble constant times the proper
time of travel is equal to one to a very high accuracy

\begin{equation} \label{eq:H:t=1}
H \tau = 1
\end{equation}

If the holostar is to serve as an alternative model for the
universe we have to interpret the time of travel $\tau$ as the age
of the universe. It is a remarkable experimental fact, that the
cosmological concordance model based on the recent
WMAP--measurements has determined $H \tau$ in the range $[0.96,
1.02]$, which is equal to the holostar prediction within the
measurement errors.

Note that the result $H \tau = 1$ is consistent with the fact,
that the active gravitational mass-density, i.e. the sum of $\rho$
and the three principal pressures is exactly zero in the holostar
space-time. It is well known, that in a string-dominated universe
$H t = 1$ follows from the zero active gravitational mass-density
of the strings (see for example \cite[p. 313-314]{Peacock/book}.
In local Minkowski-coordinates the gradient of the gravitational
acceleration is proportional to the active gravitational
mass-density. This means, that in the absence of exterior forces
any space-time with zero active gravitational mass-density will
have a deceleration parameter of zero, as viewed from the local
Minkowski frame of a co-moving observer. Zero acceleration is
particularly interesting with respect to the recent cosmological
observations. The supernova data \cite{Perlmutter/Schmidt, Riess}
indicate, that the present state of the universe is best described
by an average deceleration parameter of zero (in the case of a
true cosmological constant the deceleration parameter has
undergone a zero transition in the recent past at $z \approx
0.6$). The current observations have shown $H t \simeq 1$, which
is the relation that relates the Hubble constant and the age of
the universe in the case of (permanently) zero acceleration, i.e.
undecelerated expansion $r \propto t$.

\subsection{\label{sec:dL:z}Distance measurements in the holostar space-time}

In order to experimentally determine the geometry of the cosmos,
the angular diameter distance $d_A$ and the luminosity distance
$d_L$ and their relations to the redshift $z$ and the (local)
Hubble-constant $H$ are very important quantities.

\subsubsection{The angular diameter distance}

Let us first determine the angular diameter distance $d_A$ as a
function of redshift $z$. Consider two galaxies with the same
$r$-coordinate value, travelling radially outward. At radial
position $r_e$ a photon is emitted from one galaxy in the
"direction" of the other galaxy. From the viewpoint of the two
galaxies, the photon moves in the tangential direction. The photon
arrives at the second galaxy at radial position $r_a$. What is the
distance between both galaxies at the time of absorption?

We can answer this question in the coordinate-frame. Let us chose the
coordinate system such, that the plane defined by the motion of
the two galaxies corresponds to $\theta = \pi/2$. The galaxy that
emitted the photon will move on the ray $\varphi = 0$, whereas the
galaxy that absorbs the photon moves on the ray characterized by
$\Delta \varphi$. As the holostar space-time is nearly flat for
large $r$, we expect the angular diameter distance to be nothing
else than $d_A = r_a \Delta \varphi$.

Let's assume, that the particles that make up the galaxy started
out from $r_i \ll r_e$ so that the radial position $r_e$ of the
emission of a photon is much larger than the starting position
$r_i$. In this case we can use equation (\ref{eq:phi:j}) for the
particle trajectory and express $\Delta \varphi$ in terms of $r_e$
and $r_a$.

\begin{equation} \label{eq:delta:phi0}
\Delta \varphi = \int_{r_e}^{r_a}{d \varphi} \simeq r_e \,
\beta_\perp(r_e) \, \sqrt{\frac{r_e}{r_0}}
\int_{r_e}^{r_a}{\frac{dr}{r^2} } = \beta_\perp(r_e) \,
\sqrt{\frac{r_e}{r_0}} \left(1-\frac{r_e}{r_a}\right)
\end{equation}

$\beta_\perp(r_e)$ is the tangential velocity component of the
photon emitted at $r_e$. The motion of the galaxies, as viewed by
an observer at rest in the coordinate system, is highly
relativistic, if the galaxies move geodesically. The radial
gamma-factor for the geodesic motion of a massive particle (or any
gravitationally bound system, such as a galaxy) is given by

$$\gamma_r(r) = \sqrt{\frac{r}{r_i}}$$.

so that we can express the ratio $r_e/r_0$ in terms of the radial
gamma-factor of the galaxy at the time of emission, $\gamma_r(r_e)
= \sqrt{r_e/r_i}$ and the ratio appearing in the Hubble-law,
$r_i/r_0$:

$$\sqrt{\frac{r_e}{r_0}} = \gamma_r(r_e) \sqrt{\frac{r_i}{r_0}}$$.

The angular separation $\Delta \varphi$ between the two galaxies
can be calculated, when the radial position at the time of
emission $r_e$ and the radial position at the time of absorption
at $r_a$ is known:

\begin{equation} \label{eq:delta:phi3}
\Delta \varphi = \beta_\perp(r_e) \, \gamma_r(r_e)
\sqrt{\frac{r_i}{r_0}} \left(1-\frac{r_e}{r_a}\right)
\end{equation}

The photon at $r_e$ is emitted in the direction of the other
galaxy, which is at the same radial position at the time of
emission, as well as at the time of absorption. Naively one would
assume that the photon is emitted into the purely tangential
direction, which corresponds to $\beta_\perp(r_e) = 1$. However,
this is what the observer in the co-moving frame thinks. We are
doing the calculation in the coordinate-frame. The observer at
rest in the coordinate system sees a galaxy in highly relativistic
motion. An isotropic radiation source moving with a high
$\gamma$-value appears to the observer at rest as a source
strongly pointed into the direction of motion. From the viewpoint
of the observer at rest the photons are emitted into a narrow cone
closely centered around the direction of motion. The main lobe of
the cone lies at angle $\theta \simeq 1 / (2
\gamma$).\footnote{For high $\gamma$-values, such as is the case
here, we have $\theta = 1(/2 \gamma)$ with negligible error.} But
the angle $\theta$ of the cone is nothing else than the mean
tangential velocity component $\beta_\perp(r_e)$ of the photons at
the time of emission, as viewed by the observer at rest (at least
for small $\beta_\perp$). This means, that in the frame where we
are doing the calculation $\beta_\perp(r_e) \, \gamma_r(r_e)
\simeq 1/2$, which very much simplifies equation
(\ref{eq:delta:phi3}). Using equation
(\ref{eq:RedshiftDistance:co}) relating $r_a/r_e$ to the red-shift
factor $z$ we find:

\begin{equation} \label{eq:delta:phi}
\Delta \varphi = \frac{1}{2} \sqrt{\frac{r_i}{r_0}} \left(1 -
\frac{1}{(z+1)^2}\right)  = \sqrt{\frac{r_i}{r_0}} \frac{z
(1+\frac{z}{2})}{(1+z)^2}
\end{equation}

Note that the relation $\beta_\perp(r_e) \, \gamma_r(r_e) \simeq
1/2$ is quite independent from the fact, how large the radial
gamma-factor $\gamma_r(r_e)$ actually is at the time of
emission.\footnote{There might be a small constant correction
factor for low $\gamma_r(r_e) \approx 1$.} It is also independent
from the fact, how this radial gamma-factor arises, i.e. whether
it is the $\gamma_r$-factor of a galaxy in purely geodesic motion,
or if the galaxy follows a more complicated - non-geodesic -
space-time trajectory. The essential physical input for the
determination of $\Delta \varphi$ is, that the motion of the two
galaxies should be nearly radial. This is guaranteed whenever we
are far away from the turning point of the motion.

The holostar space-time is nearly flat for large $r$, so that the
angular diameter distance $d_A$ will simply given by:

\begin{equation} \label{eq:dA}
d_A(z) = r_a \Delta \varphi = \sqrt{\frac{r_i}{r_0}} r_a \frac{z
(1+\frac{z}{2})}{(1+z)^2} = \tau_a \frac{z
(1+\frac{z}{2})}{(1+z)^2}
\end{equation}

For small $d_A \ll \tau_a$ we find a nearly linear correspondence
between redshift and angular diameter distance. $\tau_a$ is the
proper time of the co-moving observer at the time of absorption.
For $d_A \rightarrow \tau_a/2$, when the angular diameter distance
approaches half the current "age" of the universe, $z \rightarrow
\infty$. This means, that there is a maximum value for the angular
diameter-distance of any object, and - consequentially - a minimum
solid angle: With the reasonable assumption that
there is a minimum cross-sectional area to any real physical
object, we find that the number of objects that an observer in the
holostar space-time can see is limited, in principle. Let us
assume, as suggested by loop quantum gravity, that any physical
object must at least be of Planck size, so that the minimum
cross-sectional area of any physical object is equal to the Planck
area, $A_{min} = \hbar$. Such an object subtends a solid angle in
the sky, which is given by:

\begin{equation} \label{eq:dOmega}
d \Omega = \frac{A_{min}}{{d_A}^2}
\end{equation}

By setting $d_A$ to the maximum angular diameter distance, we get
the minimum solid angle:

\begin{equation} \label{eq:dOmega:min}
d \Omega_{min} = \frac{A_{min}}{{{d_A}^2_{max}}} = \frac{4
\hbar}{\tau_a^2}
\end{equation}

The full solid angle of the sphere is $4 \pi$, so that the maximum
number of objects that can be seen in the sky is

\begin{equation} \label{eq:N:max}
N_{max} = \frac{4 \pi}{\Omega_{min}} = \frac{\pi \tau_a^2}{\hbar}
\end{equation}

Therefore the number of visible objects in the holostar-universe
is proportional to its squared age. As $\tau \propto r$ in the
holostar space-time, the maximum observable number of microscopic
objects for an observer at radial position $r$ is proportional to
the area $A = 4 \pi r^2$ of the concentric sphere with "radius"
$r$. For $\tau = r$ we find the remarkable result $N = A / (4
\hbar) = S_{BH}$, where $S_{BH}$ is the Hawking entropy of a
spherically symmetric black hole with boundary area $A$. Again we
have discovered the holographic principle already discussed in
section \ref{sec:holographic:principle}, albeit in a local
version. The holostar space-time shows a remarkable
self-consistency.

The angular diameter distance $d_A$ must not be confused with the
proper distance or with other distance measures. Different
distance measures, even if they are designed to give identical
results in flat space, usually differ in curved space-times. In
fact, by looking back to $d_A = \tau_a/2$ we don't look back a
proper distance $l = \tau/2$, we rather look back to infinite
redshift, i.e. to the limit of what can be principally
observed.\footnote{It will be difficult to discern any structure
at high $z$, because the opacity of the CMBR at $z \approx 1000$
will "block" our vision into any era that lies before the time of
CMBR-decoupling.}

Still one can set up an interesting argument by - falsely -
identifying the angular diameter distance $d_A$ with the proper
distance. Consider a sphere with proper radius $l = \tau/2$. With
the - incorrect - identification $d_A = l$ this sphere - in a
sense - defines the boundary of the "visible" universe: We will
not be able to resolve any two events lying beyond this sphere on
a telescope, because no information that would enable us to
discern two such events can reach us, due to the infinite
red-shift. The interesting observation now is, that any region
within this boundary was causally connected to any other region in
the past: The largest (proper) distance between two events is
attained for events situated at opposite positions at the boundary
sphere. A photon travelling from one event to the event on the
opposite side must pass through the whole "visible" universe.
Again, by - wrongly - identifying proper distance with angular
diameter distance, the time of travel will be twice the radius of
the "visible" universe,  i.e. $\tau$. But $\tau$ is nothing else
than the age of the universe, so that any region within the
"visible" universe was in causal contact with any other such
region at some time in it's finite past.

\subsubsection{The luminosity distance}

More important for the determination of large distances is the
luminosity-distance $d_L$. Whereas the angular diameter-distance
is related to the geometric trajectories of the photons, the
luminosity distance refers to the energy-flux. In a homogeneously
expanding FRW-universe, which is a very good approximation for the
geodesic motion of massive particles in the holostar space-time,
the luminosity distance and the angular diameter distance are
related by $d_L = (1+z)^2 d_A$ (see for example \cite[p.
92]{Peacock/book}). This relation is independent of the cosmology.
In the holostar space-time one therefore expects a similar
dependence:

\begin{equation} \label{eq:dL}
d_L(z) = (1+z)^2 d_A(z) = z (1+\frac{z}{2}) \tau_a
\end{equation}

Using the expression for the Hubble-constant (\ref{eq:Hubble}) we
find:

\begin{equation} \label{eq:Hubble:z}
H_a = \frac{z}{d_L} (1 + \frac{z}{2}) = const
\end{equation}

Keep in mind, that $H_a$ is the Hubble-constant at the present
time, defined via the well-known relation $H = \dot{r}/r = 1 /
\tau$. If we "define" the Hubble-constant under the assumption of
a linear luminosity-redshift relation for $z \ll 1$, which
corresponds to the original definition of the Hubble-constant, we
find that this value must change with redshift:

\begin{equation} \label{eq:Hubble:z:dL}
\widetilde{H} := \frac{z}{d_L}= \frac{1}{\tau_a} \frac{1}
{1+\frac{z}{2}}
\end{equation}

For $z \approx 0$ both definitions are identical, $\widetilde{H_a}
= 1/ \tau_a = H_a$. However, $\widetilde{H}$ gets lower for larger
$z$. The specific dependence of the luminosity distance $d_L$ on
$z$, as predicted in equations (\ref{eq:dL}, \ref{eq:Hubble:z},
\ref{eq:Hubble:z:dL}) should be measurable. The best method to
check these relations experimentally is to plot the luminosity
distance vs. its theoretical redshift-dependence. The same
standard candles should be used over the whole $z$-range. This
protects us against possible calibration errors in the first rungs
of the cosmological distance ladder.

\begin{figure}[ht]
\begin{center}
\includegraphics[width=12cm, bb=21 231 571 608]{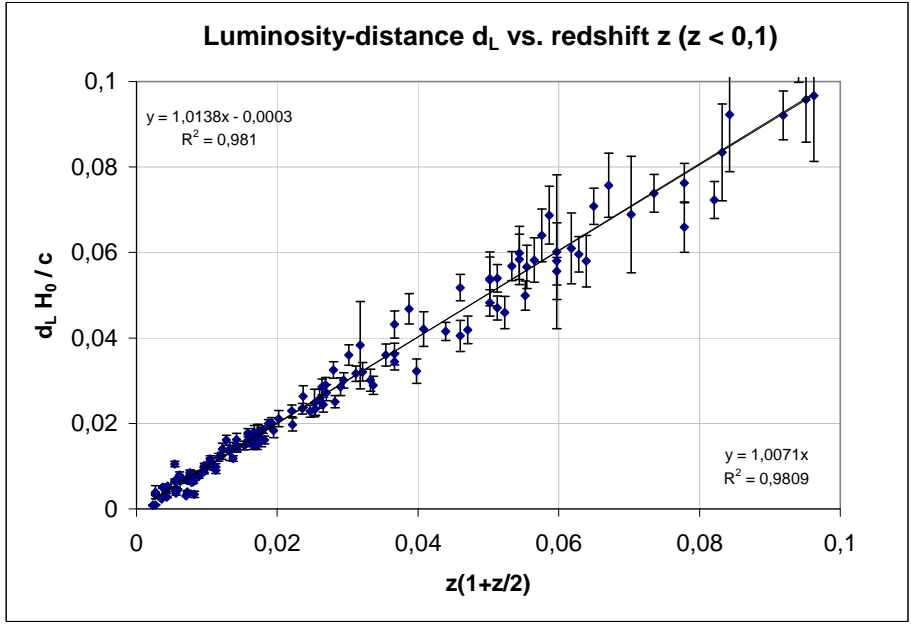}
\caption{\label{fig:supernova:small}  Luminosity distance $d_L$ as
a function of redshift $z$ for small $z$-values. Sample includes
130 supernovae with $z < 0.1$. Luminosity distances and error-bars
are taken from \cite{Tonry/Schmidt} and \cite{Barris/Tonry}. An
unweighted linear regression has been performed on the data, with
$y(0)$ fixed to zero (lower right) and $y(0)$ variable (upper
left). The respective coefficients of the regression line are
shown together with the regression coefficient $R^2$.}
\end{center}
\end{figure}

\begin{figure}[h]
\begin{center}
\includegraphics[width=12cm, bb=21 231 571 608]{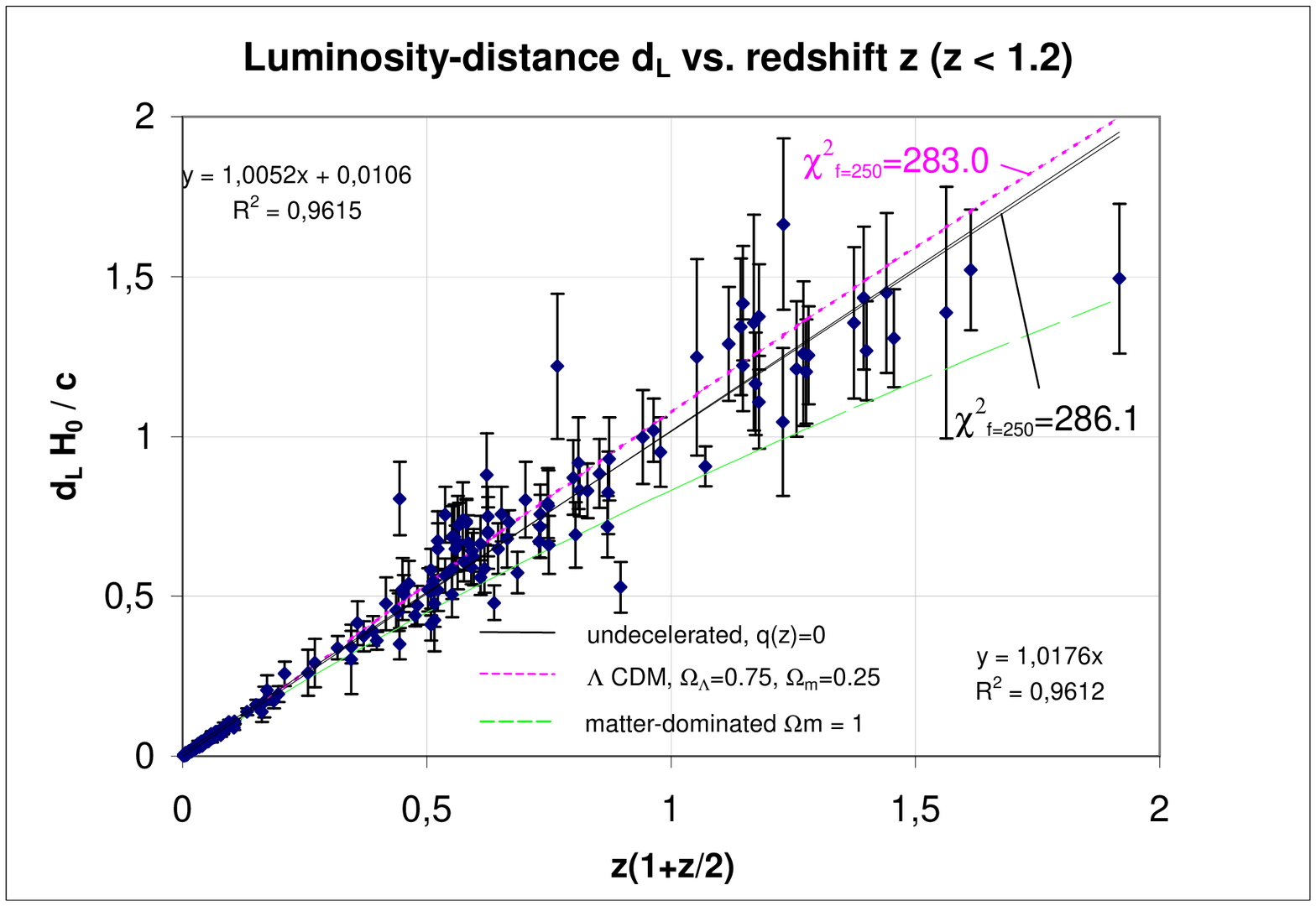}
\caption{\label{fig:supernova:large} Luminosity distance $d_L$ as
a function of redshift $z$; full range of $z$-values. Luminosity
distances and error-bars are taken from \cite{Tonry/Schmidt} and
\cite{Barris/Tonry}. The sample includes 253 supernovae with $z <
1.2$. The singular high $z$-supernova with $z = 1.755$ from
\cite{Tonry/Schmidt} is not shown. An unweighted linear regression
has been performed on the data, with $y(0)$ fixed to zero (lower
right) and $y(0)$ variable (upper left). The respective
coefficients of the fit-curves are shown together with the
regression coefficient $R^2$. The prediction from the standard
$\Lambda$-CDM model with $\Omega_\Lambda=0.75$ and $\Omega_m =
0.25$ is shown as dotted line, the prediction for a flat
matter-dominated model with $\Omega_m = 1$ is shown as dashed
line.}
\end{center}
\end{figure}

Such standard candles are provided by the SN-1a supernova
measurements. Using the recent results of \cite{Barris/Tonry} and
\cite{Tonry/Schmidt} we find a linear relation between $d_L(z)$
and $z(1+z/2)$ over the whole $z$-range, as can be seen from
Figures \ref{fig:supernova:small} and \ref{fig:supernova:large}.
Note, that the slope of the regression-line is identical for the
low-$z$ data-set (Figure \ref{fig:supernova:small}) and the full
data-set (Figure \ref{fig:supernova:large}) within the errors.
This would not be the case, if $d_L$ depended on a power of $1+z$
additionally to the $z(1+z/2)$-dependence.\footnote{On the other
hand, $d_L \propto z(1+z/2)/\sqrt{1+z}$ seems to fit the data in
Figure \ref{fig:supernova:large} (full $z$-range) somewhat better.
The catch is, that it is not possible to fit the relation $d_L
\propto z(1+z/2)/\sqrt{1+z}$ with the same coefficients to the low
and the high $z$ data-sets. The slope of the low $z$ data-set is a
factor of 1.3 larger than that of the full $z$ data-set. This
discrepancy lies far outside the scatter of the data-points, so
that a dependence $d_L \propto z(1+z/2) (1+z)^{\alpha}$ can be
ruled out with high statistical confidence level, except for very
low values of $|\alpha| < 0.1$.}

$d_L \widetilde{H_a} = z(1+z/2)$ is the dependence of the
luminosity-distance in a FRW-universe with zero deceleration
parameter. $\widetilde{H_a}$ is the value of the Hubble-constant
measured via the redshift-distance relation at low redshift. The
holostar model of the universe thus predicts a deceleration
parameter of zero, i.e. permanently undecelerated expansion. This
result is not unexpected. It could have been inferred from the
fact, that the active gravitational mass-density within the
holostar's interior space-time is exactly zero, and that the
proper geodesic acceleration approaches zero rapidly for large $r$
($g \propto 1/r^{3/2}$ in the coordinate frame). Note also, that -
as was already pointed out in section \ref{sec:Hubble} -
permanently unaccelerated expansion leads to the prediction $H
\tau = 1$, which is quite consistent with the WMAP-observations.

Figures \ref{fig:supernova:small} and \ref{fig:supernova:large}
show, that the supernova data are compatible with an expansion,
that was unaccelerated over the whole $z$-range $0 < z < 1.2$,
well within the measurement errors. Permanently unaccelerated
expansion corresponds to a strictly linear dependence in Figures
\ref{fig:supernova:small} and \ref{fig:supernova:large}. Due to
the large errors it is virtually impossible to decide between the
holostar-model (permanently unaccelerated expansion) and the
$\Lambda$-CDM models with $\Omega_\Lambda + \Omega_m \approx 1$
and $\Omega_m$ in the range $0.2 < \Omega_m < 0.4$. For instance,
the theoretical curve for the $\Lambda$-CDM model with $\Omega_m =
0.25$ and $\Omega_\Lambda = 0.75$ is shown in Figure
\ref{fig:supernova:large} with a dotted line. This line differs
from the holostar (straight line) prediction substantially only at
large $z > 2$, where there are not yet enough data available.

A statistical analysis shows, that it is practically impossible -
at least with the current available data and using standard
statistical techniques - to make a definite decision in favor of
the $\Lambda$-CDM or the holostar model, if one takes the
measurement errors into account.

If one uses all data-points of the combined sample of
supernova-data (\cite{Tonry/Schmidt} and \cite{Barris/Tonry},
discarding the three lowest $z$ supernovae, which show very large
deviations from the global Hubble-flow) and calculates the
$\chi^2$-values with respect to the {\em magnitudes}\footnote{The
luminosity distance is determined by the measurement of {\em
fluxes} (a quadratic measure of inverse distance), not magnitudes
(a logarithmic distance measure). Therefore determining $\chi^2$
with respect to the magnitude-deviations appears to be in conflict
with the process by which the data have been measured. However,
the essential requirement for a $\chi^2$-fit is, that the
measurement errors follow a {\em Gaussian} distribution. It is
quite clear that flux-measurements, especially for low $z$, are
{\em not} Gaussianly distributed. The magnitude-deviations appear
somewhat closer to a Gaussian distribution than the fluxes, so
magnitudes were used in the fit. In any way, if the measurement
errors are non-Gaussian, a $\chi^2$-fit can and will become a very
bad likelyhood estimator, in which case the question whether the
$\chi^2$-fit was performed with respect to the measured or derived
quantities is very much academic. The systematic
(not-controllable!) uncertainties due to the non-applicability of
the $\chi^2$-procedure itself usually will be much larger than any
(controllable!) uncertainty introduced by calculating the
$\chi^2$-value with respect to a derived quantity. What has to be
taken into account though, is that the normalization of the error
will be different for the derived quantities. A factor of 2-3 is
not uncommon. Therefore the absolute $\chi^2$-values, which depend
crucially on the correct normalization, have to be interpreted
with care. Despite the uncertainty in the absolute value it is
still possible is to compare the relative $\chi^2$-values for
different theoretical models. Except for the unlikely case that
none of the theoretical models provides a fairly good fit to the
data, one can use the $\chi^2$-value of the best-fitting model to
normalize the $\chi^2$-values. As long as the other models have
their normalized $\chi^2$-value within $(1-2) \, \sigma$ of the
best fit model, one cannot say with high statistical confidence
that any one model is ruled out.} predicted by both models, one
finds $\chi^2 = 286.1$ for the holostar $q(z)=0$ model and
$\chi^2=283.0$ for the $\Lambda$-CDM model with $\Omega_m = 0.25$
and $\Omega_\Lambda = 0.75$.

\begin{figure}[h]
\begin{center}
\includegraphics[width=12cm, bb=21 231 571 608]{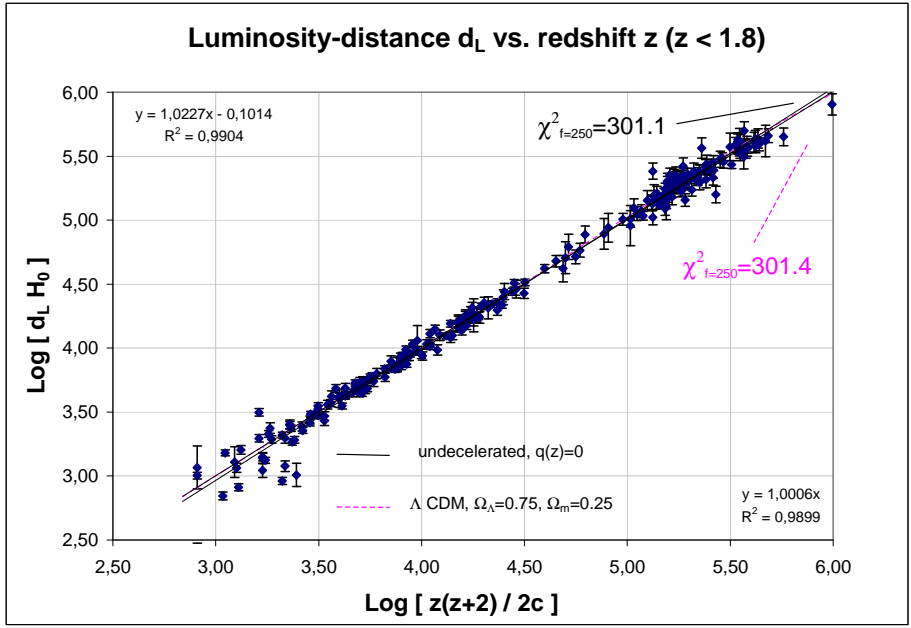}
\caption{\label{fig:supernova:log} Luminosity distance $d_L$ as a
function of redshift $z$ in a logarithmic plot; full range of
$z$-values. Luminosity distances and error-bars are taken from
\cite{Tonry/Schmidt} and \cite{Barris/Tonry}. The sample includes
all 253 supernovae with $z < 1.755$. An unweighted linear
regression has been performed on the logarithmic data, with $y(0)$
fixed to zero (lower right) and $y(0)$ variable (upper left). The
coefficients of the fit-curves are shown together with the
regression coefficient $R^2$. The prediction from the standard
$\Lambda$-CDM model with $\Omega_\Lambda=0.75$ and $\Omega_m =
0.25$ is shown as dotted line.}
\end{center}
\end{figure}

For large $N$ the $\chi^2$ distribution is quite well approximated
by a Gaussian distribution, meaning that the expected outcome of a
$\chi^2$-test is $f \pm \sqrt{2f}$, where $f$ are the number of
degrees of freedom of the distribution. Taking 250 supernova-data
one expects $\chi^2$ to lie in the range $250 \pm 23$. Both models
have their $\chi^2$ roughly $1.5 \, \sigma$ higher than expected.
However, one must keep in mind that the (unnormalized)
$\chi^2$-test only gives correct numbers, when the measurement
errors are Gaussian and if one knows the exact value of their
Gaussian variance. If the experimenter cannot determine the exact
$\sigma_1$ Gaussian error, the $\chi^2$-value comes out wrong by a
constant factor. If the measurement errors are non-Gaussianly
distributed (for example because of non-controllable systematics),
the Gaussian variance is not a well defined quantity. This is
quite evidently the case, as can be seen from the logarithmic plot
in Figure \ref{fig:supernova:log}.

Especially the low red-shift data ($N \leq 30$) are problematic. A
high percentage of the low-z magnitudes show relative deviations
more than a factor 10 higher than the measurement errors. This is
incompatible with a Gaussian distribution and points to
systematics, such as badly or not compensated peculiar velocities.
It is clear, that these data-points will blow up the
$\chi^2$-value. In a more sophisticated analysis it would be
prudent to omit the first 30 or so low $z$ data from the sample.
This brings down the $\chi^2$ values somewhat closer to the
expected range. For $N = 224$, i.e. leaving out the first 29
data-points, we find $\chi^2_{holo} = 230.7$ and $\chi^2_{\Lambda
CDM} = 228.2$.

Although it is - in principle - possible to normalize the
$\chi^2$-distribution by rescaling the experimentally determined
measurement error, I have not done so. Any such rescaling is
model-dependent and it is hard to escape unwanted selection
effects. The reader therefore has to keep in mind, that the
$\chi^2$-values quoted in this paper might be off by a constant
factor, which is estimated to lie in the range $0.7$ to $1.2$.
However, when one tests different theoretical models by a
$\chi^2$-test, without being sure that the measurement errors are
Gaussianly distributed and without being able to put a strong
prior on the expected theoretical model, it is not so much the
absolute $\chi^2$-values, but rather the difference of the
$\chi^2$-values that is important. Both the holostar- and the
$\Lambda$-CDM model fit the data fairly well and their respective
$\chi^2$-values, although somewhat higher than expected, are so
close, that it is not possible to exclude one or the other model
by today's available data.\footnote{The fact, that both models fit
the data nearly optimally is a robust indicator, that the absolute
$\chi^2$-values of both models mark the center of the "true"
$\chi^2$-distribution, normalized to the "true" Gaussian error.}
With 220-250 data-points the expected variance of the
$\chi^2$-test lies between 20-23. Both models differ by $2.5 - 3$,
i.e. roughly one eight of the $\chi^2$-variance. Such a small
difference cannot be regarded as statistically significant.

\begin{figure}[h]
\begin{center}
\includegraphics[width=12cm, bb=21 231 571 608]{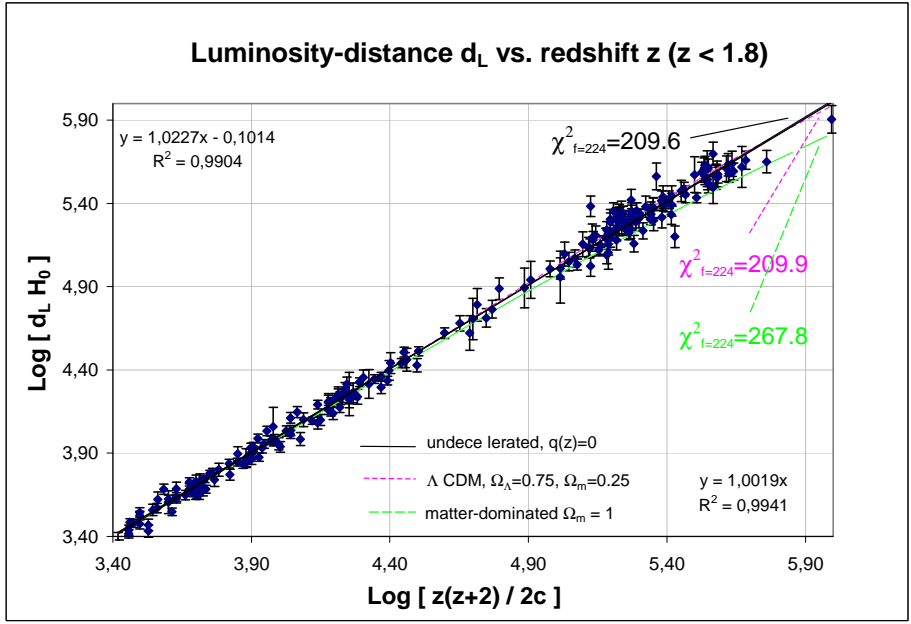}
\caption{\label{fig:supernova:log:high} Luminosity distance $d_L$
as a function of redshift $z$ plotted logarithmically; the 29
lowest $z$-values were omitted. Luminosity distances and
error-bars are taken from \cite{Tonry/Schmidt} and
\cite{Barris/Tonry}. The sample includes 224 supernovae with $z
\leq 1.755$. An unweighted linear regression has been performed on
the data, with $y(0)$ fixed to zero (lower right) and $y(0)$
variable (upper left). The respective coefficients of the
fit-curves are shown together with the regression coefficient
$R^2$. The prediction from the standard $\Lambda$-CDM model with
$\Omega_\Lambda=0.75$ and $\Omega_m = 0.25$ is shown as dotted
line.}
\end{center}
\end{figure}

In the case of a non-gaussian error distribution a somewhat more
robust estimator for the relative error is the absolute deviation.
In Figures \ref{fig:supernova:log} and
\ref{fig:supernova:log:high} the $\chi^2$-values were estimated,
by summing up the absolute deviations (observed - expected
magnitudes) divided by the quoted measurement error, rather than
their squares. For large $N$ one can expect this sum to be
approximately Gaussian distributed. For a Gaussian distribution
the standard deviation is roughly a factor of $1.12$ higher than
the absolute deviation. If one estimates a $\chi^2$-value by
summing up the absolute errors, one will generally underestimate
the true $\chi^2$-values by approximately this factor.

This can be seen in the $\chi^2$-values of Figure
\ref{fig:supernova:log:high}. For the 224 high $z$ data points the
holostar-model receives $\chi^2 = 209.6$, whereas the
$\Lambda$-CDM model has $\chi^2 = 209.9$. Both values are very
close to each other and lie 7 \% below the expected value $224$
for a genuine $\chi^2$-test.

\begin{figure}[h]
\begin{center}
\includegraphics[width=12cm, bb=21 231 571 608]{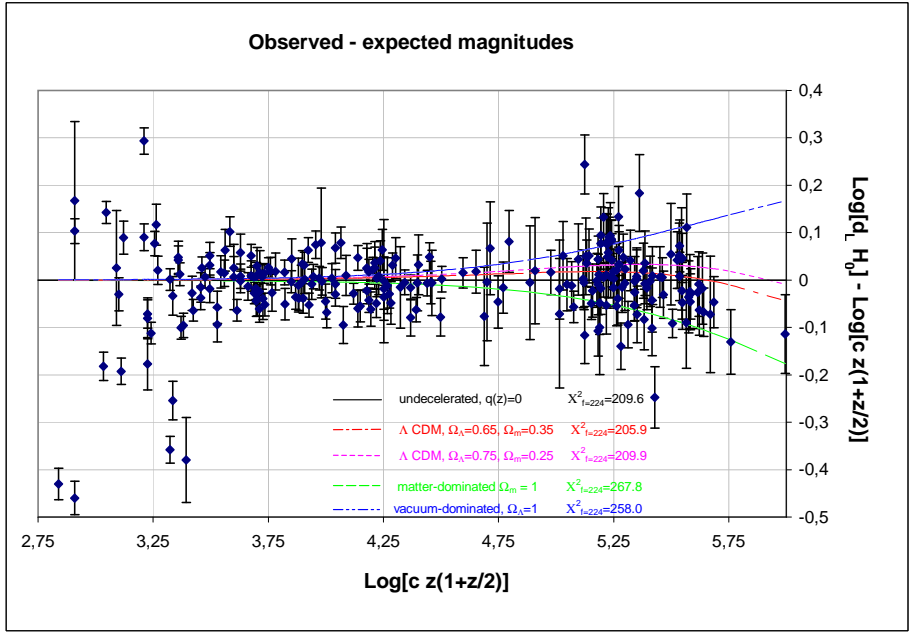}
\caption{\label{fig:supernova:log:err} Observed minus expected
magnitudes for the full range of $z$-values. The magnitudes and
error-bars are taken from \cite{Tonry/Schmidt} and
\cite{Barris/Tonry}. The sample includes all 253 supernovae with
$z \leq 1.755$. The quoted $\chi^2$-values refer to the 224
highest $z$-supernovae in the sample. The reference-model (0-line)
corresponds to permanently undecelerated expansion. The respective
curves for the flat Standard $\Lambda$-CDM model with
($\Omega_\Lambda=0.75, \Omega_m = 0.25$) and
($\Omega_\Lambda=0.65, \Omega_m = 0.35$), as well as the curves
for a flat matter-dominated universe (lower curve) and a flat
vacuum-dominated universe (upper curve) are shown.}
\end{center}
\end{figure}

The results of the analysis are summarized in Figure
\ref{fig:supernova:log:err}, where the observed magnitudes minus
the magnitudes expected for permanently undecelerated expansion (=
holostar-model) are plotted against red-shift on a logarithmic
scale. The holostar-model corresponds to the horizontal line $y =
0$. The predictions for the flat $\Lambda$-CDM models with
$\Omega_m = 0.25$ and $\Omega_m = 0.35$ are shown as dotted and
dashed-dotted curves. The curves expected from a flat
matter-dominated model ($\Omega_m = 1$) and a flat
vacuum-dominated model ($\Omega_\Lambda = 1$) are shown as well
(lower and upper curve).

The flat $\Lambda$-CDM model with $\Omega_m = 0.35$ gives the best
$\chi^2$-value. The matter-dominated and vacuum dominated models
are clearly ruled out by the data. Their respective
$\chi^2$-values lie three standard deviations higher than the best
fitting models.

Although the flat $\Lambda$-CDM model with $\Omega_m = 0.35$ seems
to be slightly preferred, the marginally different $\chi^2$-values
between the holostar- and the flat $\Lambda$-CDM models in the
range $0.2 < \Omega_m < 0.4$ can hardly be regarded as
statistically significant, even if one would be willing to accept
extremely low confidence levels. It is clear, that in order to
rule out one or the other model more high $z$ supernova data are
required. For $z > 2$ the luminosity red-shift relations of the
standard $\Lambda$-CDM models and the holostar-model of the
universe differ substantially.

One must also keep in mind, that the so called $K$-corrections
relating the observed supernova-intensity in a fixed frequency
band to the total bolometric intensity integrated over the whole
frequency range is somewhat model-dependent (see for example
\cite{Buchalter/2004}). Without analysis of the model-dependencies
one cannot exclude the possibility of a $z$-dependent systematic
error, if one naively interprets the supernova-data in the context
of the holostar space-time using the standard $K$-corrections
derived in a different context.

The luminosity-distance $d_L$ plotted in Figures
\ref{fig:supernova:small} and \ref{fig:supernova:large} is
normalized to the Hubble-length $c/\widetilde{H_0}$. The
Hubble-length cannot be determined by comparing the relative
luminosities of the standard candles to their red-shift. In order
to determine $\widetilde{H_0}$ the absolute magnitude of the
standard candles must be known. The absolute magnitude of the
SN-1a supernovae has been determined from the first rungs of the
cosmological distance ladder, and therefore is subject to the
errors within these rungs. On the other hand, a single accurate
distance measurement to a nearby supernova, such as SN 1987A, can
provide an independent absolute calibration. An analysis of the
extended gas envelope of SN 1987A in the large magellanic cloud
provides such a means. See \cite{Panagia/2003}.

\subsection{\label{sec:delta:rho}Structure formation in the holostar space-time}

The Hubble law given in equation (\ref{eq:Hubble}) contains an
unknown parameter, the (nearly constant) proper radial coordinate
velocity $dr/d\tau \approx \sqrt{r_0/r_i}$. If the holostar is an
appropriate model for the universe, $dr/d\tau$ can be determined
by comparing the measured Hubble constant with our current radial
coordinate position $r$, which can be estimated from the
mass-density. This will be done in section \ref{sec:measurement}
and gives $r_i \approx r_0$, a result that has been hinted already
in section \ref{sec:massive:acc}. However, it would be quite
helpful if the parameter $dr / d\tau$ could be determined by some
independent method. The evolution of the density-fluctuations from
the time of decoupling to the structures we find today might
provide such a means:

The fluctuations in the microwave background radiation have been
determined to be roughly equal to $\delta T / T= 10^{-5}$. These
small temperature fluctuations are interpreted as the relative
density fluctuations at the time of decoupling, i.e. at the time
when the microwave background radiation temperature was believed
to be roughly 1000 times higher than today. In the adiabatic
approach (instantaneous decoupling) the respective fluctuations in
the baryon-density at the time of decoupling are $\delta = \delta
\rho_B / \rho_B \approx 3 \, \delta T / T$.\footnote{This estimate
is based on the assumption, that decoupling happened very fast,
almost instantly. Today's refined estimates take into account,
that the decoupling took longer. In this case the ratio $\delta
\rho_B / \rho_B$ at decoupling is lower than in the simple
adiabatic model up to a factor of 10 (see for example
\cite{Silk}).} These fluctuations are believed to have been the
seeds of the structure formation process, from which the large
scale distribution of galaxies, as we see them today, have formed.
The density contrast in the distributions of galaxies on a large
scale is roughly of order unity today: $\delta_{today} = \delta
\rho / \rho \simeq 1$. In the standard cosmological models this
evolution of the density fluctuations is quite difficult to
explain. The problem is, that in the dust approximation (no
significant pressure, velocity of sound nearly zero), which for a
long time was assumed to be the correct description of the
universe after decoupling, the fluctuations grow with $\delta
\propto t^{2/3}$. In this picture the universe is matter-dominated
after decoupling. In a matter-dominated FRW-model we find $r
\propto t^{2/3}$. The temperature $T$ and the scale factor $r$ are
related by $T \propto 1/r$. Combining these dependencies we get:

\begin{equation}
\delta \propto t^{2/3} \propto r \propto \frac{1}{T}
\end{equation}

The above formula predicts $\delta_{today} \approx 10^{-2}$, which
is roughly two orders of magnitude less than the observed value.

In order to explain the rather large density fluctuations today,
the standard cosmological model has been extended to encompass
cold dark matter and - lately - so called "dark energy".

In the holostar universe the evolution of the density-fluctuations
can be explained quite easily. Only one parameter, the nearly
constant radial coordinate velocity $dr/d\tau$ which appears in
equation (\ref{eq:Hubble}) for the Hubble value need to be
adjusted.

\subsubsection{The evolution of density perturbations in the dust-case}

Let us first consider the dust case, i.e. the evolution of the
density fluctuations in the holostar by deliberately ignoring the
effects of the pressure. The reader should be aware, that this is
an unrealistic assumption in the holostar's interior space-time.
The negative radial pressure, equal in magnitude to the
matter-density, will always exert a substantial influence on the
large scale structure formation. The effect of the pressure will
be studied in the subsequent section. We will see that the results
for the dust case are almost identical, except for slightly
different numerical factors, to the realistic case which includes
the pressure.

Far from the starting point of the motion the expansion in the
holostar universe (viewed by an observer in the co-moving frame)
is very similar to the expansion in the standard Robertson Walker
models. The expansion is nearly isotropic. The energy density is
matter-dominated over an extended period of time. Therefore the
usual standard model formula for the evolution of the density
fluctuations should be applicable. In a matter-dominated Robertson
Walker universe with zero pressure, the density fluctuations
evolve according to the following differential equation (see for
example \cite{Peacock/book}):

\begin{equation} \label{eq:delta}
\ddot{\delta} + 2 \frac{\dot{r}}{r} \, \dot{\delta} - 4 \pi \rho
\, \delta = 0
\end{equation}

For the holostar we find:

\begin{equation}
r = \sqrt{\frac{r_0}{r_i}} \tau
\end{equation}

\begin{equation}
\frac{\dot{r}}{r} = \frac{1}{\tau}
\end{equation}

\begin{equation}
4 \pi \rho = \frac{1}{2r^2} = \frac{r_i}{2 r_0}\frac{1}{\tau^2}
\end{equation}

With these relations equation (\ref{eq:delta}) reduces to the
following differential equation:

\begin{equation} \label{eq:delta:2}
\ddot{\delta} + \frac{2}{\tau} \dot{\delta} - \frac{r_i}{2 r_0} \frac{1}{\tau^2} \delta = 0
\end{equation}

The equation can be solved by the following ansatz:

\begin{equation}
\delta = \tau^n
\end{equation}

This gives a quadratic equation for $n$

\begin{equation}
n^2 + n -\frac{r_i}{2 r_0} = 0
\end{equation}

which can be solved for $n$:

\begin{equation}
n = -\frac{1}{2} \pm \sqrt{\frac{1}{4} + \frac{r_i}{2 r_0}}
\end{equation}

In the holographic universe we have the following dependencies:

$$r \propto \tau \propto \frac{1}{T^2}$$

Therefore we can express $\delta$ as a function of temperature
$T$:

\begin{equation} \label{eq:delta:T}
\delta \propto \tau^n \propto \frac{1}{T^{2n}} \propto
\frac{1}{T^{\epsilon}}
\end{equation}

with

\begin{equation} \label{eq:exponent:delta}
\epsilon = -1 \pm \sqrt{1+\frac{2 r_i}{r_0}}
\end{equation}

The exponent $\epsilon$ in the above equation can be estimated
from the known ratio of the density contrast $\delta_{dec} /
\delta_{today}$ and the ratio of the decoupling temperature to the
CMBR-temperature.

In the holostar the temperature at decoupling will be larger than
in the standard cosmological model, because the number ratio of
baryons to photons doesn't remain constant. If radiation and
matter do not interact after decoupling (no re-ionization!), the
baryons move geodesically and the baryon number remains constant
in the co-moving volume, the baryon-density will scale as $1/r^{3}
\propto T^6$ in the frame of the co-moving material observer.
However, according to the discussion in section \ref{sec:frames}
geodesic expansion against the negative pressure leads to particle
creation in the co-moving frame, so that the number-density of the
massive particles scales with $1/r^2$, if their rest mass remains
constant.\footnote{This can be seen from the holostar-energy
density $\rho =1 / (8 \pi r^2)$, which is the same in the
co-moving frame as in the coordinate frame due to the radial boost
invariance of the space-time. If the co-moving mass-energy density
is matter-dominated, the number density of the massive particles
is nothing else than $n_m = \rho / m = 1 / (8 \pi m r^2)$, which
leads to an inverse square law whenever $m=const$.} The
baryon-density will scale as $1/r^2 \propto T^4$. Under these
circumstances the Saha-equation yields a temperature at decoupling
of roughly $T_{dec}\approx 4 900 K$, when no dark matter component
is assumed and the baryon-density today is set to the total
matter-density as determined by WMAP
\cite{WMAP/cosmologicalParameters}, i.e. $\rho_B = \rho_m \approx
2.5 \cdot 10^{-27} kg / m^3$.\footnote{This matter density
corresponds to $\Omega_m \approx 0.26$ or roughly $1.5$ nucleons
per cubic meter at a critical density $\rho_c$ determined from a
Hubble value of $H \simeq 71 km/s / Mpc$.} A decoupling
temperature of $4 900 K$ corresponds to a red-shift $z_{dec}
\simeq 1790$. In order to achieve $\delta_{today} \approx 1$ from
$\delta_{dec} \approx 3 \cdot 10^{-5}$ at $z_{dec} = 1790$, the
exponent $\epsilon$ in equation (\ref{eq:delta:T}) must be roughly
equal to $1.48$. For more realistic scenarios in which decoupling
doesn't happen instantaneously $\delta_{dec}$ is estimated to be
lower, up to a factor of 10 \cite{Silk}. For a very slow
decoupling scenario with $\delta_{dec} \approx 0.3 \cdot 10^{-5}$
we require $\epsilon \approx 1.7$.

For $r_i = 4 r_0$ we get $\epsilon = 2$, i.e. $\delta \propto
1/T^2$, which quite certainly is too high. For $r_i = 2 r_0$ we
find $\epsilon = \sqrt{5}-1 \simeq 1.24$, which is too low. A good
compromise is given by $r_i \approx 3 r_0$, which provides a
fairly good fit for moderately slow decoupling:

\begin{equation}
r_i = 3 r_0 \,  \rightarrow \,
\delta \propto \frac{1}{T^{\sqrt{7}-1}} \simeq \frac{1}{T^{1.65}}
\end{equation}

The value $r_i = 3 r_0$ is interesting, because for this value the
critical mass-energy density in the standard cosmological model,
which is given by $\rho_c = 3 H^2 / (8 \pi)$ is exactly equal to
the energy density of the holostar, which is given by $\rho = 1 /
(8 \pi r^2)$.

We find that the evolution of the density fluctuations requires
$r_i \approx 3 r_0$ and thus $H \approx 0.6 / r)$. However, we
cannot expect highly accurate results from the above treatment.
Although it might be appropriate to ignore the pressure after
matter and radiation have become completely decoupled (which might
be as low as $z \approx 100$), it is not possible to neglect the
pressure in the holostar universe completely. Quite certainly this
will not be the case before "recombination", i.e. for $z
> 1800$. If kinematic decoupling happens around $z \approx 100$,
the dominant contribution to the evolution of $\delta$ from $z
\approx 1800$ (recombination) to $z=1$ (today) falls into an era,
where the radiation pressure cannot be neglected.

\subsubsection{The evolution of density perturbations in the radiation dominated era}

As has been stated before, ignoring the pressure in the holostar
space-time is unrealistic. In order to incorporate pressure into
the treatment above, we need to know how the anisotropic pressure
will manifest itself in the co-moving frame. In
\cite{McManus/Coley} universes with anisotropic pressure were
studied. The authors found, that an anisotropic pressure will
appear as an isotropic pressure for the observer co-moving with
the cosmic fluid. The isotropic pressure in the co-moving frame,
$P$, is related to the anisotropic pressure components as $P =
(P_r + 2 P_\theta)/3$. If this relation is applied to the
holostar, we find $P = -1/(24 \pi r^2) = -\rho/3$. This is the
equation of state for an isotropic string gas. Except for the sign
the pressure is that of normal radiation. Nonetheless, the sign of
the pressure shouldn't influence the square of the velocity of
sound, which is the relevant parameter for the density evolution
in a Robertson Walker universe including pressure. Furthermore,
the analysis in section \ref{sec:ultrarel:gas} showed that it
doesn't matter, whether one uses the anisotropic holostar pressure
or the isotropic radiation pressure of an ultra-relativistic gas
in order to calculate the pressure-induced energy change
experienced by radiation in geodesic motion. Therefore - whenever
matter and radiation cannot be treated as kinematically decoupled
- let us make the ansatz, that the density fluctuations in the
holostar evolve just as the density fluctuations of a radiation
dominated universe. The evolution equation (\ref{eq:delta}) then
has to be replaced by:

\begin{equation} \label{eq:delta:p}
\ddot{\delta} + 2 \frac{\dot{r}}{r} \dot{\delta} - 4 \pi \rho \frac{8}{3}
\delta = 0
\end{equation}

If the above equation is expressed in terms of $r_i/r_0$, the only
difference to the dust case in equation (\ref{eq:delta:2}) is to
replace $r_i/r_0 \rightarrow (8 r_i)/(3 r_0)$, so that the
exponent in equation (\ref{eq:delta:T}) is given by:

\begin{equation} \label{eq:exponent:delta:p}
\epsilon = -1 \pm \sqrt{1+\frac{16 r_i}{3 r_0}}
\end{equation}

In order to achieve an exponent of roughly $1.65$ we need $1+16
r_i/(3 r_0) \approx 7 $, which requires

\begin{equation}
\frac{r_i}{r_0} \approx \frac{9}{8}
\end{equation}

Note, that the above value is very close to unity. For $r_i = r_0$
we find

\begin{equation}
\delta \propto \frac{1}{T^{\sqrt{19/3}-1}} \simeq \frac{1}{T^{1.52}}
\end{equation}

which is a good compromise between models assuming instantaneous
decoupling ($\epsilon \approx 1.4$) or very slow decoupling
($\epsilon \approx 1.7$).

Equation (\ref{eq:exponent:delta:p}) is quite sensitive to the
value of $r_i / r_0$. It seems quite clear that $r_i$ cannot
possibly lie outside the range $r_0/2 < r_i < 2 r_0$ in any era,
where radiation pressure is significant. If the temperature at
decoupling was $4 900 K$ and decoupling took place very fast, the
best fit would be $r_i/r_0 \simeq 0.9$. For non-geodesic motion
(after decoupling) one expects that the decoupling temperature
will be lower, somewhere in the range between $3 500$ and $4 900
K$. A lower decoupling temperature and non-geodesic motion both
require a higher exponent in equation (\ref{eq:delta:T}), so that
$r_i \simeq r_0$. For non-adiabatic (i.e. not instantaneous)
decoupling the baryonic density fluctuations at decoupling have
been estimated to be lower (see for example \cite{Silk}), placing
$\delta_{dec}$ in the range $0.3 \cdot 10^{-5} < \delta_{dec} < 3
\cdot 10^{-5}$, in which case $r_i/r_0$ should be very close to
unity.

Whatever the exact value of $r_i/r_0$ might be, we are drawn to
the conclusion that the massive particles making up the matter in
our universe must have moved nearly geodesically from $r_i \approx
r_0$. Although the internal logic of the holostar space-time
requires $r_i \approx r_0$, it is difficult to believe this
important result without further experimental
evidence.\footnote{\label{fn:ri}Note also that $r_i \approx r_0$
indicates, that the particles must have had extremely high radial
velocities at decoupling: With the assumption that the number of
particles remains constant during geodesic expansion the order of
magnitude of the relativistic $\gamma$-factor at decoupling can be
roughly estimated: $\gamma_{dec} \approx \sqrt{r_{dec}/r_i}
\approx 1.7 \cdot 10^{27}$. Such high velocities at decoupling are
only conceivable, if the massive particles that constitute the
matter of the universe today have truly originated from $r_i
\approx r_0$. This possibility is discussed in section
\ref{sec:phi}.}

An accurate estimate for the exponent in equation
(\ref{eq:delta:T}) and therefore an accurate prediction for the
Hubble-value can only be made if the true equations of motion of
massive particles in the holostar space-time, including pressure,
are used. However, this requires the inclusion of quantum effects:
According to section \ref{sec:massive:acc} the deviation from pure
geodesic motion due to the pressure-forces is completely
irrelevant except in the very first stage of the motion, i.e. $r_i
\leq r <\approx (2-3) \, r_i$. But with $r_i \approx r_0 \approx 2
\, r_{Pl}$ the motion starts out roughly a Planck distance from
the center, where the discreteness of the geometry and quantum
effects will come into play. A detailed analysis of the motion of
particles in the holostar, including the region close to its
center, therefore must be left to future research.\footnote{There
is also the possibility, that the massive particles move in a way
that keeps the proper radial expansion of an outmoving shell zero.
Such a motion would be compatible with the inverse square-law for
the matter-density, as the expansion only proceeds in the two
tangential directions, i.e. $V \propto r^2$. No work will be done
against the negative radial pressure in this case, so that the
number of massive particles in the shell remains constant.
Pressure-induced particle-production, which would be necessary
otherwise (see the discussion in section \ref{sec:frames}), is not
mandatory.}

\subsection{\label{sec:phi}On the angular correlation of the CMBR}

As has already hinted in footnote \ref{fn:ri} the experimentally
determined ratio $r_i/r_0 \approx 1$ only makes sense, if the
particles in the universe - as we see it today - have originated
from a few Planck-distances from the holostar's center. A rather
conservative scenario within this line of thought is, that a -
presumably massive - precursor particle was created close to the
center and then moved out on a nearly geodesic trajectory
henceforth. Nearly geodesic motion through the hot interior region
of the holostar space time, starting out from $r \approx r_0$,
seems only conceivable for a massive particle of roughly Planck
mass which only interacts weakly with the other particles.
Somewhere on its trajectory, not too far from the radial
coordinate value where the temperature has dropped below the
electro-weak unification scale, the particle (or its intermediate
decay products) must have decayed into ordinary matter, endowing
the stable final products (electrons, protons, neutrons and
neutrinos - or more generally the light quark/leptons) with the
high momentum gathered on its outward track. This scenario -
although well within the limits of standard physics - is quite an
extravagant claim, for which it would be helpful to have another -
independent - verification. The missing angular correlation of the
microwave background radiation at large angular separations might
provide such a means.

The low quadrupole and octupole moments of the CMBR as determined
from of the WMAP data \cite{WMAP/cosmologicalParameters}
demonstrate, that there is practically no correlation between the
fluctuations in the microwave background radiation at angular
separations larger than approximately $60^\circ$, which
corresponds to roughly $1$ radian. This feature can be quite
effortlessly explained by the motion of massless and massive
particles in the holostar space-time

\subsubsection{Limits on the angular motion of massive particles}

As can be seen from equation (\ref{eq:Nrev}) particles that were
emitted just a few Planck distances from the center have limited
angular spread. According to the scenario proposed beforehand let
us assume, that a massive particle is created with not too high a
tangential velocity $\beta_i$ close to the center of the holostar.
The particle must not necessarily have a long life-time. Any decay
products will follow the original trajectory. The maximum angular
spread of the particle (or of its decay products) can be
calculated:

\begin{equation} \label{eq:implicit:beta}
\varphi_{max} = \beta_i \sqrt{\frac{r_i}{r_0}}
\int_{0}^1{\frac{dx}{ \sqrt{1 - x \left(1 -
\beta_i^2(1-x^2)\right)}}}  =  \beta_i
\sqrt{\frac{r_i}{r_0}} \, \xi(\beta_i^2)
\end{equation}

$\xi = \xi(\beta_i^2)$ depends on the tangential velocity
$\beta_i$ at the turning point of the motion. As remarked in
section \ref{sec:eq:motion}, $\xi$ lies between $1.4 < \xi \leq 2$,
depending on $\beta_i$. Equation \ref{eq:implicit:beta} gives an
implicit relation for $\beta_i^2$ , which can be solved
iteratively for $\beta_i^2$, whenever the maximum angular spread
$\varphi_{max}$ and the ratio $r_i/r_0$ is known.

$\varphi_{max}$ and $r_i/r_0$ can be determined experimentally.
However, there is a subtlety involved in the experimental
determination of $r_i/r_0$ from the characteristics of the
expansion. The derivation of equation (\ref{eq:Hubble}) for the
Hubble-value relies on the - implicit - assumption, that the
motion started out from $r_i$ at rest. This is unrealistic. At
$r_i \approx r_0$ there is an extremely high temperature, so that
even a particle of nearly Planck mass will have an appreciable
velocity at its (true) turning point of the motion, $r_i$.
Therefore the ratio $r_i/r_0$ in equation (\ref{eq:Hubble})
doesn't refer to the true turning point of the motion $r_i$, but
rather to a fictitious "zero-velocity" turning point
$\widetilde{r_i}$, which describes the radial part of motion far
away from the turning point. Both values are related: In section
\ref{sec:eq:motion} it has been shown, that the radial part of the
motion of a particle with an appreciable tangential velocity at
its turning point is nearly identical to the motion of a particle
that started out from a somewhat smaller radial coordinate value,
whenever $r \gg r_i$. The relation between the true turning point
$r_i$ of the motion and the apparent (zero-velocity) turning point
$\widetilde{r_i}$ measured at large distances is given by:

\begin{equation}
r_i = \gamma_i^2 \widetilde{r_i}
\end{equation}

where $\gamma_i$ is the special relativistic $\gamma$-factor of
the particle at its true turning point of the motion.

The experimentally determined ratio $r_i/r_0$ in the Hubble
equation (\ref{eq:Hubble}) or in the equation for the density
evolution (\ref{eq:delta:p}) therefore rather refers to
$\widetilde{r_i}/r_0$, whereas $r_i$ in equation
(\ref{eq:implicit:beta}) refers to the true turning point of the
motion. In section \ref{sec:delta:rho} this ratio has been
estimated as $\widetilde{r_i}/r_0 \approx 1$, if the density
perturbations found in the microwave background radiation $\delta
\approx 10^{-5}$ evolve according to equation (\ref{eq:delta:T})
in combination with equation (\ref{eq:exponent:delta:p}) to the
value observed today. Let us denote the experimentally determined
ratio with $\kappa$:

\begin{equation}
\kappa = \frac{\widetilde{r_i}}{r_0} = \frac{1}{\gamma_i^2}
\frac{r_i}{r_0}
\end{equation}

The equation for the maximum angular spread then reads:

\begin{equation}
\varphi_{max}^2 = \beta_i^2 \xi^2(\beta_i^2) \frac{r_i}{r_0} =
\beta_i^2 \gamma_i^2 \xi^2(\beta_i^2) \kappa
\end{equation}

so that:

\begin{equation} \label{eq:determine:betai}
\beta_i^2 \gamma_i^2  = \frac{\beta_i^2}{1-\beta_i^2} \simeq
\frac{\varphi_{max}^2}{\xi^2(\beta_i^2) \kappa}
\end{equation}

Whenever we know $\kappa$ and $\varphi_{max}$, the tangential
velocity at the true turning point of the motion $\beta_i$ follows
from equation \ref{eq:determine:betai}) . Knowing $\beta_i$ the
true turning point of the motion can be determined

\begin{equation}
r_i = \gamma_i^2 \widetilde{r_i} = (1 + \beta_i^2 \gamma_i^2)
\kappa r_0
\end{equation}

where the relation $\gamma^2 = 1 + \beta^2 \gamma^2$ was used.

For $\varphi_{max} = 60^\circ = \pi /3 \approx 1$ the maximum
angular spread is nearly unity (in radians). For $\kappa = 1$ we
find $\xi \approx 1.77$. With these values

\begin{equation}
\beta_i^2 \gamma_i^2 \simeq 0.319
\end{equation}

so that

\begin{equation}
\beta_i^2 = \frac{\beta_i^2 \gamma_i^2}{1+ \beta_i^2 \gamma_i^2}
\simeq 0.242
\end{equation}

and

\begin{equation} \label{eq:ri:true}
\frac{r_i}{r_0} \simeq 1.319
\end{equation}

If we know $\beta$ and the momentum $p$ of a massive particle at
the radial position $r_i$, where the particle's motion has become
nearly geodesical, we can determine the mass of the particle. The
momentum of any massive particle is related to its mass $m_0$ by

\begin{equation}
p = \beta \gamma m_0
\end{equation}

To determine $m_0$ we need $p$. Any massive particle at the
holostar's center will be immersed in a very hot radiation bath of
ultra-relativistic or zero rest mass particles. It seems
reasonable to assume, that at position $r_i \approx r_0$, where
the particle starts to move out geodesically, the thermal
radiation will have imprinted its momentum on the massive
particle. In the holostar the mean momenta $\overline{p}_\gamma$
of the ultra-relativistic particles are proportional to the
interior radiation temperature, given by equation
(\ref{eq:Tlocal}).

\begin{equation}
\overline{p}_\gamma = \sigma_S T_\gamma = \frac{\sigma_S}{4 \pi}
\frac{\hbar}{\sqrt{r_i r_0}}
\end{equation}

In \cite{petri/thermo} it is shown, that the factor of
proportionality $\sigma_S$ is equal to the entropy per particle
(which is slightly larger than $\pi$ for a great variety of
circumstances). Putting all equations together and using the
relation $r_0^2 / \hbar \approx 4 \sigma_S / \pi$ at the
Planck-energy proposed in \cite{petri/charge} we find:

\begin{equation}
m_0 \simeq \frac{1}{8 } \sqrt{\frac{\sigma_S}{\kappa \pi}}
\frac{\sqrt{\hbar}}{\beta_i \gamma_i^2}
\end{equation}

With $\kappa = 1$ and $\varphi_{max} = 1$ the mass of the
precursor particle $m_0$ is given by:

\begin{equation}
m_0 \simeq 0.201 \, m_{Pl}
\end{equation}

We have used $\sigma_S = 3.38$, which is the mean entropy per
ultra-relativistic particle, when the ultra-relativistic particles
at the Planck-temperature are predominantly fermions. For a more
detailed discussion of how $\sigma_S$ can be derived from
microscopic statistical thermodynamics see \cite{petri/thermo}.

\subsubsection{The preon - a massive particle of roughly Planck mass?}

It appears, that in order to explain the missing angular
correlation in the CBMR at large angular separations we require a
heavy precursor particle of roughly a fifth of the Planck mass.
Lets call this particle the "preon".

Let us assume that the preon is created close to the holostar's
center. Due to $\overline{p}_\gamma = \sigma_S T_\gamma =
\sqrt{\sigma_S / \pi}/4 \approx 0.561 \, m_0 \approx 0.113 \,
m_{Pl}$ the preon-production will be somewhat inhibited at at $r_i
\approx 1.3 \, r_0$, as an energy of $E =
\sqrt{\overline{p_\gamma}^2 + m_0^2} \approx 0.231 \, m_{Pl}$ is
required. However, at $r = r_0/2 \approx r_i/2.6$ there is just
enough energy available to create preons in (in pairs) in
substantial numbers, either by the collision of any two massive or
massless particles of the surrounding radiation bath, or -
slightly more efficiently - by Unruh radiation: At a radial
coordinate position $r = r_0/2$ the local radiation temperature
$T_\gamma$ will be up by a factor of $\sqrt{2.6}$ with respect to
the temperature at $r_i$, so that the mean momentum of the
radiation will be given by $\overline{p}_\gamma \approx 0.183 \,
m_{Pl}$. The mean energy of the (massive) preon at $r_0/2$ will be
higher as well, but not with the same factor as the radiation,
because of its non-negligible rest-mass. We find: $E_0 \approx
\sqrt{\overline{p_\gamma}^2 + m_0^2} \approx 0.272 \, m_{Pl}$. The
mean momentum of the radiation quanta doesn't yet suffice to
produce the preon efficiently in pairs, at least not with the
right momentum. However, the mean energy of the two radiation
quanta is almost large enough to produce two preons with zero
momentum.

Presumably the most efficient way to produce preons is via Unruh
radiation between $r_0/2 < r < r_0$. The Unruh-temperature at $r =
r_0/2$ will be twice the radiation temperature (see section
\ref{sec:Unruh}). The energy of a particle produced by Unruh
radiation therefore is $E \approx 0.366 \, m_{Pl} \approx 1.35 \,
E_0$. The production of a preon by Unruh radiation poses no
problem, even if the high chemical potential of the
ultra-relativistic particles, $\mu > 1.34 \, T_\gamma$, in the
holostar is taken into account.\footnote{For the relevance of
chemical potentials and the thermodynamics of highly relativistic
matter states see \cite{petri/thermo}.}

So far the case $\kappa = \widetilde{r_i/r_0} = 1$ was discussed.
The mass of the preon $m_0$ depends on $\kappa$, albeit only
moderately. If $\kappa$ lies in the range $0.5 < \kappa < 4$ we
find the following mass-range:

\begin{equation} \label{eq:preon:mass:range}
0.164 < m_0 < 0.243
\end{equation}

in Planck units.

In Table \ref{tab:preon:mass} the mass $m_0$ and the tangential
velocity $\beta_i$ of the preon is given for several values of $\kappa$:

\begin{table}[h]
\begin{center}
\begin{tabular} {c||c|c|c}
$\kappa$ & $\beta_i$ & $\xi$ & $m_0$ \\ \hline
1/2 & 0.647 & 1.65 & 0.165 \\
3/4 & 0.553 & 1.72 & 0.188 \\
1   & 0.490 & 1.77 & 0.201 \\
3/2 & 0.405 & 1.83 & 0.218 \\
2   & 0.353 & 1.87 & 0.228 \\
5/2 & 0.316 & 1.89 & 0.234 \\
3   & 0.288 & 1.91 & 0.238 \\
4   & 0.250 & 1.93 & 0.243 \\
\end{tabular}
\linebreak
\\
\caption{\label{tab:preon:mass}preon mass $m_0$ as a function of
$\kappa$; $\varphi_{max} = 56.8^\circ$}
\end{center}
\end{table}

If equation (\ref{eq:delta:p}) for the density evolution (with
pressure) is correct, $\kappa$ should lie in the range $3/4 <
\kappa < 5/4$, so that the bound for the preon mass will be
roughly $0.19 < m < 0.21$ in Planck units. $\kappa \approx 1$ also
fits better with the current measurements of the Hubble-constant
(see section \ref{sec:measurement}).

We find, that if the current expansion rate of the universe and
the evolution of the density perturbations after decoupling can be
(approximately) described by the holostar solution, one requires a
new particle. Its properties can be quite accurately inferred from
the above discussion: Its mass should be less than one quarter of
the Planck mass. Its exact value shouldn't lie too far outside the
mass range given by equation (\ref{eq:preon:mass:range}). The
particle - or its decay products - should interact only weakly
with the other particles at energies above the electro-weak scale.

\subsubsection{\label{sec:preon:2}An independent estimate for the preon mass}

It is possible to estimate the mass of the preon by
another, independent argument. According to the discussion in
section \ref{sec:uncertainty} the proper volume occupied by a single
particle at the holostar's center is given by

\begin{equation}
V_1 = \int_0^{r_0/2}{dV} = \frac{8 \pi}{7 \sqrt{2}}
\left(\frac{r_0}{2}\right)^3 \approx \frac{8 \pi}{7 \sqrt{2}}
V_{Pl}
\end{equation}

On its way outward the volume available to the preon will become
larger, which enables it to decay into lighter particles, such as
neutrons, protons and electrons. The decay is expected to conserve
mass-energy locally. What will the energy-density of the stable
decay products be at large $r$-values? To simplify the
calculations it is convenient to consider a thought-experiment, in
which the proper expansion in the radial direction is zero. This
thought-experiment simplifies the calculations as we don't have to
take into account any changes of the internal energy in a radially
expanding volume due to the non-zero radial
pressure.\footnote{Keep in mind that the actual motion of the
preon might be different from the thought-experiment. However, the
thought-experiment allows us to neglect the effects of the
negative radial pressure without changing the physical results
obtained from a more realistic treatment. If the pressure-effects
are included in the total energy-balance, the actual (geodesic?)
motion of the preon yields the same results as the method of
"constant internal energy" in this thought-experiment.} Any motion
leaving the internal energy of a spherically outmoving shell of
particles unaffected, requires that the proper thickness of the shell
remains constant, i.e. the volume develops as the proper surface
area of the concentric spherical shell, i.e. $V \propto r^2$.

At the radial position $r = 9.18 \cdot 10^{60} \, r_{Pl}$, which
corresponds to the radius of the observable universe today, the
volume $V_1$ will have expanded to:

\begin{equation} \label{eq:V:today}
V_{today} = \left(\frac{r}{r_{Pl}}\right)^2 V_1 \simeq 1.0 \cdot 10^{18} m^3
\end{equation}

Assuming local mass-energy conservation and assuming that the
mass-energy of the preon ends up predominantly in nucleons (i.e.
no significant dark matter component; energy-contributions of
electrons, neutrinos and photons negligible with respect to the
baryons), the total number of nucleons within this volume will be
given by:

\begin{equation} \label{eq:N:proton:from:preon}
N_{n} = \frac{E_{preon}}{m_p} \approx 1.68 \cdot 10^{18}
\end{equation}

if the mass-energy of the preon is assumed to be $E_{preon} =
p_\gamma(r_0) \simeq \sqrt{\sigma_S /\pi} \, m_{Pl}/8 = 0.13 \,
m_{Pl}$. From equations (\ref{eq:V:today},
\ref{eq:N:proton:from:preon}) the number-density of nucleons in
the universe today can be estimated as:

\begin{equation}
n_{n} = 1.68 \frac{1}{m^3}
\end{equation}

amounting to roughly $1.7$ nucleons per cubic meter. This is very
close to the number-density of nucleons in the universe derived
from the total matter-density determined by WMAP assuming no
significant dark matter component, $n_n = 1.48 / m^3$ (see section
\ref{sec:measurement}). Therefore a preon-mass in the range
between $0.1$ to $0.2$ Planck masses is quite consistent with the
findings in the observable universe today.

\subsubsection{An estimate for the amplitude of the density perturbations}

The assumption that the preon eventually decays into nucleons at a
temperature slightly below the nucleon rest-mass, enables us to
give a - very crude - estimate for the absolute value of the
density contrast $\delta$ at the time of baryogenesis.
Astoundingly this crude estimate fits quite well with the
experimentally determined values of the density contrast today
$\delta_{today} \approx 1$ and at the time of decoupling
$\delta_{dec}\approx 10^{-5}$:

The red-shift $z_b$ where the mean energy of the radiation is
equal to the nucleon rest-mass is given as:

\begin{equation}
z_{b} \approx \frac{m_p/3.37}{T_{CMBR}} \approx 1.19 \cdot 10^{12}
\end{equation}

At this redshift the number-density of the nucleons $n_b$ in the
holostar space-time will be higher than today by a factor of
$z_b^4$, i.e.

\begin{equation}
n_b \approx \frac{1.5}{m^3} z_b^4 \approx \frac{3 \cdot
10^{48}}{m^3}
\end{equation}

which corresponds to a matter-density of $\rho_b \approx 5 \cdot
10^{21} kg/m^3$. This value is roughly a factor of thousand higher
than the typical neutron star density and four orders of magnitude
higher than the typical density of stable nuclei.

If the nucleons in our universe originate from the preon at
roughly this redshift, one would expect a density-contrast
$\delta_b$ on the order of the nucleon to preon mass at this time.
With a preon mass $m_{preon} \approx 0.15 \, m_{Pl}$ we find

\begin{equation}
\delta_b \approx \frac{m_p}{m_{preon}} \approx 5 \cdot 10^{-19}
\end{equation}

As has been shown in section \ref{sec:delta:rho} the density
contrast evolves as a power-law with respect to the redshift. For
$\kappa=1$ the exponent is given by $\epsilon = \sqrt{19/3}-1
\simeq 1.517$ in the radiation dominated era, so that:

\begin{equation}
\delta \propto \frac{1}{z^{1.517}}
\end{equation}

Therefore we can "predict" the density-contrast at any redshift $z
< z_b$ from the density-contrast at the time of baryogenesis. We
find:

\begin{equation}
\delta_{dec} = \delta_b \left(\frac{z_b}{z_{dec}}\right)^{1.517}
\approx 1.2 \cdot 10^{-5}
\end{equation}

with $z_{dec} \approx 1800$ and

\begin{equation}
\delta_{today} = \delta_b \left(\frac{z_b}{1}\right)^{1.517}
\approx 1.05
\end{equation}

However, keep in mind that as long as not much is known about the
particle spectrum at energies above the electro-weak scale this
"prediction" stands on more than shaky ground.

\subsubsection{A string explanation for the CMBR power spectrum}

The string-character of the holostar space-time allows an
alternative explanation, not only for the absolute value of the
density perturbations but also for the scale-invariant spectrum of
the CMBR-fluctuations found by WMAP. As this has been discussed
elsewhere, I will just give a very short summary of the
presentation in Peacock \cite[p. 316-321]{Peacock/book}:

It is easy to see, that the holostar's interior space-time can be
interpreted as a collection of strings, aligned in the radial
direction. For a string-dominated universe one finds the following
rough estimate for the horizon-scale amplitude $\delta_H$:

\begin{equation}
\delta_H = \left(\frac{\delta \rho}{\rho}\right)_H \approx \left(\frac{E_{GUT}}{E_{Pl}}\right)^2
\end{equation}

As $E_{GUT} \approx 10^{-3} E_{Pl}$ numbers close to the
characteristic value $\delta_H \approx 10^{-5}$ can arise quite
naturally.

Because of the way in which the string network scales, the density
perturbations generated by stringy matter have an approximately
scale invariant spectrum. However, string structure formation is
generally non-Gaussian in nature. Therefore it might be possible
to detect a non-Gaussian string signature in the small-scale CMBR
anisotropies

\subsubsection{Limits on the angular motion of photons - an estimate for the maximum angular correlation distance}

So far the maximum angular correlation distance of the microwave
background radiation ($\varphi_{max} \approx 60^\circ$) has been
put in by hand, as determined from the observations. It would be
nice, if this value could be derived by first principles. There
appears to be a way to do this. For this purpose let us consider
the angular spread of zero mass particles in the holostar. The
maximum angular spread for a photon emitted from $r_i$ is given
by:

\begin{equation} \label{eq:max:angle:photon}
\varphi_{max} = \frac{\sqrt{\pi}}{3}
\frac{\Gamma(\frac{1}{3})}{\Gamma(\frac{5}{6})}
\sqrt{\frac{r_i}{r_0}}\simeq 1.4022 \sqrt{\frac{r_i}{r_0}}
\end{equation}

Let us consider photon-pair production (or the production of any
other massless particle in pairs) by Unruh radiation. This process
will be most efficient if the Unruh-temperature is high. In
section \ref{sec:Unruh} the following relation between radiation
temperature and Unruh-temperature will be derived:

\begin{equation}
\frac{T_U}{T_\gamma} \simeq \frac{r_0}{r}
\end{equation}

For $r_i = r_0/2$ the Unruh temperature is twice the radiation
temperature, so that photon pair production by the Unruh-effect
should be a common occurrence in the central region of the
holostar. Unruh creation of photon pairs is most efficient at the
most central conceivable location available for a particle within
the holostar space-time. According to the discussion in section
\ref{sec:uncertainty} there are many good reasons to believe, that
$r_i = r_0/2$ is the "closest" conceivable position that any one
particle can occupy at the  holostar's center. If we insert the
ratio $r_i/r_0 = 1/2$ into equation (\ref{eq:max:angle:photon}) we
get:

\begin{equation} \label{eq:max:angle:photon:2}
\varphi_{max} = \frac{\Gamma(\frac{1}{3})}{\Gamma(\frac{5}{6})}
\sqrt{\frac{\pi}{18}} \simeq 0.9915 = 56.81^\circ
\end{equation}

Voil$\grave{a}$.

Finally it should be noted, that $\varphi_{max} \approx 1$ makes
the "expansion" of particles that move radially outward in the
holostar space-time nearly equal in the radial and the tangential
directions: $\delta l_\perp = r \varphi_{max}$, whereas $\delta
l_r \approx r$.

\subsubsection{\label{sec:uncertainty}Applying the uncertainty principle to the holostar's central region}

If one compares the theoretical prediction for the Hubble
constant, the theoretical prediction for the evolution of the
density perturbation amplitudes and maximum
correlation angle with the cosmological observations, all methods
favor $r_i \approx r_0$, meaning that the particles making up our
universe must have originated from a region $r_0/2 < r_i < 2 r_0$.
Is it conceivable, that the particles could have originated from
an interior position very much closer to the center?

It is easy to see that the answer is no. Any particle located
close to the holostar's center will have a mean momentum
comparable to the Planck-momentum. However, any particle must be
subject to the Heisenberg uncertainty relation. If we set the
momentum uncertainty $\Delta p$ of the particle equal to its mean
(thermal) momentum at $r_i$, and the uncertainty in position
$\Delta r_i$ equal to to $r_i$, the uncertainty relation reads:

\begin{equation}
\Delta p \, \Delta r_i = \frac{\sigma_S}{\pi} \frac{\hbar}{4}
\sqrt{\frac{r_i}{r_0}} \geq \frac{\hbar}{2}
\end{equation}

With $\sigma_S \approx \pi$ we find the following inequality for
$r_i/r_0$:

\begin{equation}
 \frac{r_i}{r_0} \geq 2
\end{equation}

Quite evidently this is not an exact result, as $\Delta p \neq p$
and $\Delta r_i \neq r_i$. Still one can expect this relation to
be correct within a factor of two or three. We find that any
particle in the holostar's center will not be able to occupy a
region (or originate from a region) much smaller than $r \approx
r_0$.

There are some other independent reasons to believe, that $r_0$ -
or rather $r_0/2 \approx \sqrt{\hbar}$ - provides a universal
cut-off for the region that can by occupied by any one particle in
the holostar space-time.

First, it does not seem possible that more than one particle will
be able to enter - or occupy - the region bounded by the smallest
possible area quantum of loop quantum gravity, which turns out to
be roughly equal to $4 \pi (r_0/2)^2$ (see \cite{petri/charge}).
Therefore the smallest "separation" between two particles should
be roughly equal to $r_0$, as far as semi-classical reasoning can
still be trusted at the Planck scale.

Second, in \cite{petri/thermo} the number of particles within a
spherical concentric region in the holostar space-time has been
determined as $N = (\pi / \sigma_S) \, (r^2/\hbar) \simeq r^2 /
\hbar$ (as $\sigma_S \approx \pi$). This relation was derived in
the context of microscopic statistical thermodynamics, which
generally requires macroscopic $N$. Extrapolating this relation to
the Planck-scale using the experimental estimate $r_0 \approx 2
\sqrt{\hbar}$ from equation (\ref{eq:r0^2:est}) we find that $N
\simeq 1$ for $r =r_{Pl} \approx r_0/2$.

A third argument is this: In \cite{petri/charge} it has been
shown, that $r_0/2$ is the radial position of the membrane of an
elementary extremely charged holostar with zero (or negligible)
mass. A holostar with $r_h=r_0/2$ is the smallest holostar
possible: Any holostar with $r_h < r_0/2$ necessarily has negative
mass.

Therefore any particle coming from the holostar's hot central
region will appear to have originated from $r_i \approx r_0$ for
an observer situated a large distance from the center. Whether the
particle was created at the hot central region, or whether an
inmoving particle merely reversed its motion at $r \approx r_0$
(or was reflected from the center by a small angle) is quite
irrelevant to the general picture.

\subsection{\label{sec:measurement}Estimating cosmological parameters from the radiation temperature}

In this section the total local mass-energy density $\rho$, the
local Hubble value $H$, the radial coordinate value $r$ and the
proper time $\tau$ in a "holostar universe" will be determined
from the temperature of the microwave background radiation.

The total energy density in the holostar universe can be
determined from the radiation temperature, whenever $r_0^2$ is
known. It is given by equation (\ref{eq:T4/rho}):

\begin{equation}
\rho = \frac{2^5 \pi^3 r_0^2}{\hbar^4} T^4
\end{equation}

There is some significant theoretical evidence for $r_0^2 \approx
4 \sqrt{3/4} \hbar$ at the low energy scale. However $r_0^2 = 4
\hbar$ or a value a few percent higher than $4 \hbar$ might also
be possible (for a somewhat more detailed discussion see
\cite{petri/charge}). With $r_0^2 = 4 \sqrt{3/4} \hbar$ and
$T_{CMBR} = 2.725 K$ we find:

\begin{equation}
\rho = 2.425 \cdot 10^{-27} \frac{kg}{m^3} = 1.450 \cdot
\frac{m_p}{m^3} =  4.702 \cdot 10^{-124} \rho_{Pl}
\end{equation}

This is almost equal to the total matter-density of the universe
determined by WMAP \cite{WMAP/cosmologicalParameters}:

\begin{equation}
\rho_{WMAP} = 2.462 \cdot 10^{-27} \frac{kg}{m^3}
\end{equation}

From the matter density the radial coordinate position $r$ within
the holostar can be determined.

\begin{equation}
r = \frac{1}{\sqrt{8 \pi \rho}} = 9.199 \cdot 10^{60} r_{Pl} =
1.575 \cdot 10^{10} ly
\end{equation}

This value is quite close the radius of the observable universe.

The local Hubble-constant can be determined from the matter
density via equation (\ref{eq:Hubble}):

\begin{equation}
H = \sqrt{8 \pi \rho \frac{r_0}{r_i}}
\end{equation}

With $r_i/r_0 = 1$ and the matter density determined beforehand we
find:

\begin{equation} \label{eq:H:r0qg}
H = 2.021 \cdot 10^{-18} \, [\frac{1}{s}] = 62.36 \,
[\frac{km/s}{Mpc}]
\end{equation}

The Hubble-constant comes out quite close to the value that is
used in the concordance model by the WMAP-group
\cite{WMAP/cosmologicalParameters} with $H = 71 \, km/s \, / Mpc$.
This is an encouraging result. Note, however, that the value of
the Hubble constant is model-dependent. It is possible to relate
the Hubble value $H_s$ of the various versions of the standard
cosmological models to the Hubble value $H_h$ of the holostar
solution via the mass-density. In the standard cosmological models
we find:

\begin{equation}
\rho_m = \Omega_m \rho_c = \Omega_m \frac{3 H_s^2}{8 \pi}
\end{equation}

For the holostar the mass-density is given by

\begin{equation}
\rho_m = \frac{r_i}{r_0} \frac{H_h^2}{8 \pi}
\end{equation}

Setting the mass densities equal, the local Hubble-values can be
related:

\begin{equation} \label{eq:Hubble:standard:holo}
H_h^2 = \Omega_m \frac{3 r_0}{r_i} H_s^2
\end{equation}

Both values are equal, if $\Omega_m = r_i/(3 r_0)$. This is almost
the case for $\Omega_m \approx 0.26 \approx 1/4$ according to WMAP
and $r_i/(3 r_0) \approx 1/3$ as determined in section
\ref{sec:delta:rho}. This result is quite robust. If we determine
$H_h$ from $H_s$ via equation (\ref{eq:Hubble:standard:holo}) for
various combinations of $\Omega_m$ and $H_s$ that have been used
in the past\footnote{Some years ago $\Omega_m \approx 0.3$ and
$H_s \approx 65 km/s / Mpc$ was a common estimate.} we get the
result of equation (\ref{eq:H:r0qg}) with an error on the order of
a few percent.

From the local Hubble-value determined in (\ref{eq:H:r0qg}) the
proper time $\tau$, which can be interpreted as the "age" of the
universe, can be derived:

\begin{equation}
\tau = \frac{1}{H} = 9.180 \cdot 10^{60} t_{Pl} = 1.57 \cdot
10^{10} y
\end{equation}

This is somewhat larger than the result recently announced by the
WMAP group, $t = 1.37 \cdot 10^{10} y$. Keep in mind, however,
that the WMAP-result for the age of the universe is strongly
correlated with the prior on the Hubble-constant via the relation
$H t \approx 1$. In fact, it is the product $H t$ which has been
determined to the remarkably high accuracy stated by the WMAP
group. In order to determine $H$ or $t$ individually to the same
accuracy, independent measurements for $H$ (or $t$) are required,
such as provided by the concordance model.

If we set $r_i/r_0 = 3/4$ we find an almost perfect agreement of
the holostar's predictions for the local Hubble value $H_h =
71.85$ (km/s)/Mpc and the current proper time $\tau = 13.6 \, Gy$
compared to the values determined by WMAP
\cite{WMAP/cosmologicalParameters}, $H = 71$ (km/s)/Mpc and $t =
13.7 \, Gy$.

The holostar solution is quite compatible with the recent findings
concerning the large scale structure and dynamics of the universe.
The recent WMAP results are reproduced best by setting $\kappa =
r_i/r_0 = 3/4$. From the evolution of the density perturbations
(see section \ref{sec:delta:rho}) one would rather expect $r_i
\simeq r_0$. It is quite clear that $r_i/r_0$ cannot lie very much
outside the range $3/4 < r_i/r_0 < 1$, which corresponds to an age
of the universe in the range between $13-16 \, Gy$.

Note that the above age comes close to the ages of the oldest
globular clusters. If $r_i/r_0$ is chosen larger than unity, the
proper time $\tau$ in the holostar universe will be larger. For
$r_i/r_0 \approx 2$ the holostar universe can easily accommodate
even the old estimates for the ages of the globular clusters,
which not too far ago have been thought to lie between $13$ to
$19$ billion years. The problem with such an assignment is, that
the Hubble value comes out far too low. For $r_i = 2 r_0$ we find
$H \approx 44$ (km/s)/Mpc, which appears incompatible with today's
experimental measurements, even if the large systematic errors in
calibrating the cosmological distance scale are taken into
account.

From a theoretical point of view $r_i = r_0$ appears as the most
preferable choice (see for instance the discussion in section
\ref{sec:Unruh}). However, this choice is based on an
extrapolation of the classical geodesic equations motion from
large distances back to roughly a Planck distance from the
holostar's center. It is questionable whether classical reasoning
at the Planck-energy scale will remain valid, so that one should
expect some moderate adjustment due the quantum nature of
space-time at this scale. Therefore it is too early to make a
definite numerical prediction for the ratio $r_i/r_0$ (measured
via the Hubble-constant at late times). If $r_i/r_0$ can be pinned
down theoretically we are in the very much desirable position to
make a precise prediction for the Hubble value, which should be an
order of magnitude better than today's measurement capabilities.
Therefore any significant advances in the understanding of the
motion of particles in the holostar universe could be of high
practical value for the development of a "high precision
cosmology" in the near future.

\subsection{\label{sec:entropy:density}A determination of the local entropy density}

Another important quantity that can be determined from the the
radiation temperature is the entropy density $s$. According to
equation (\ref{eq:s=g}) the entropy-density is given by:

\begin{equation}
s = \frac{g}{\hbar} = \frac{1}{2 r \hbar} \sqrt{\frac{r_0}{r}}
\end{equation}

Replacing $r$ with the radiation temperature via equation
(\ref{eq:Tlocal}) we find:

\begin{equation} \label{eq:s:r0}
s = \frac{4 r_0^2}{\hbar} \left(\frac{2 \pi T}{\hbar}\right)^3
\end{equation}

With $r_0^2 / \hbar = 4 \sqrt{3/4}$ and inserting the
CMBR-temperature we get the following prediction for the local
entropy-density:

\begin{equation}
s = \frac{5.817 \cdot 10^{12}}{m^3} = \frac{2.454 \cdot
10^{-92}}{r_{Pl}^3}
\end{equation}

It appears, that this entropy density is quite in conflict with
the entropy-density in our universe, which - according to the
common belief - is thought to be dominated by the entropy in the
CMBR-radiation. The entropy-density of the CMBR is roughly three
orders of magnitude lower, $s_\gamma \approx 1.5 \cdot 10^{9} /
m^3$. However, this point of view does not take the entropy of the
nucleons (and electrons) properly into account. It is a -
strangely - utterly ignored result from microscopic statistical
thermodynamics, that the entropy $\sigma_m$ per massive particle
in the non-relativistic limit is given by:

\begin{equation} \label{eq:s:m}
\sigma_m = \frac{m-\mu}{T} + \frac{5}{2} =
\frac{\epsilon-\mu}{T}+1
\end{equation}

in units ($c=k=1$). See \cite[p. 277]{Peacock/book} for an
approximate derivation when $\mu = 0$ and \cite{petri/asym} for a
full derivation. $\epsilon$ is the total energy per particle,
including the kinetic energy $\epsilon = m + (3/2) \, T$.

If one assumes $\mu \approx \epsilon \approx m$ for massive
particles, the entropy per massive particle remains
small.\footnote{There is not much experimental evidence for $\mu
\approx m$. At low energies, i.e. in chemical reactions between
different massive particles (=molecules) one does not measure the
absolute value of the chemical potential, but rather the
difference of the chemical potentials of the reacting species. At
high energies the chemical potentials of the particles are usually
taken to be zero.} In a self-consistent space-time, i.e. a
space-time that can be considered to be a closed thermodynamical
system to a good approximation, this assumption quite likely is
not correct. For instance, for a spherically symmetric black hole
we know that the entropy is a function of the energy alone $S =
S(E)$. The number of particles within a black hole is undefined.
But the chemical potential is defined as the energy change at {\em
constant} entropy, when a particle is added to the system. For a
spherically symmetric black hole this energy-change is zero:
Adding a particle {\em at constant entropy} does not change the
total energy of the system, because of the definite relation $E =
E(S)$. Therefore the chemical potential of the "interior
particles" must be zero.\footnote{Of course, for a black hole the
number of interior particles is undefined!} Something similar
applies to the holostar. A holostar has a definite number of
particles - at least on average and at high temperatures. But
again there is a definite relation between total mass $M$,
temperature at infinity $T$ and entropy, i.e. $S \propto M^2
\propto M / T$. As long as the total mass-energy of the holostar
does not change, the total entropy remains constant. Therefore any
{\em internal} fluctuations of particle-numbers neither change the
total energy, nor the total entropy of the whole system. The
chemical potential should be zero.

Somewhat more formally:

$$dE = \frac{\partial E}{\partial S} dS + \frac{\partial E}{\partial N} dN + \frac{\partial E}{\partial V} dV + X_i dx_i$$

$X_i dx_i$ refer to exterior parameters of the system, such as
angular momentum or charge. For a closed system the exterior
(conserved) quantities need not be considere. If $E$ only depends
on $S$, as is the case for the total energy and entropy of a black
hole or holostar, the partial derivatives of $S$ with respect to
$N$ and $V$ must be zero. But $\partial E /
\partial N$ at fixed $S, V$ is nothing else than the definition
for the chemical potential $\mu$. Therefore - seen from a {\em
global} perspective, the chemical potential of any interior
particle must be zero in thermodynamic equilibrium.

It therefore seems reasonable to assume, that that the chemical
potential of the massive particles is zero in the holostar's
interior space-time. In this case the constant factor $5/2$ in
equation (\ref{eq:s:m}) can be neglected, when the temperature is
well below the rest-mass of a particular particle, so that

$$\sigma_m \simeq \frac{m}{T}$$

Therefore in the matter-dominated era the entropy-density in the
holostar universe is given by:

\begin{equation} \label{eq:s:total}
s = \left(\frac{m_p n_p + m_n n_n + m_e n_e}{T} + n_\gamma
\sigma_\gamma + n_\nu \sigma_\nu\right)
\end{equation}

The contribution from the three known stable massive particle
species (protons, neutrons and electrons) and the two known
massless species (photons and neutrinos) to the entropy-density
was included. If the reader likes, he can double the
photon-contribution in order to include a "graviton" contribution.

The entropy-density of the CBMR-radiation $n_\gamma \sigma_\gamma$
is given by:

\begin{equation} \label{eq:s:rad}
n_\gamma \sigma_\gamma =  \left( \frac{2 \pi T}{\hbar}\right)^3
\frac{f_\gamma}{4 \pi \cdot 3^2 \cdot 5}
\end{equation}

$f_\gamma=2$ is the number of degrees of the CMBR-photons. Under
the assumption that the neutrinos have zero chemical potential, we
have to use a weighting factor of $7/8$ for every fermionic degree
of freedom.\footnote{Note, that the assumption of zero chemical
potential is false for ultra-relativistic fermions in the holostar
space-time. Other weighting-factors have to be used, which are
generally higher than 7/8. See \cite{petri/thermo} for a
derivation of the non-zero chemical potentials and a table for the
appropriate weighting factors.} Therefore the entropy-density of
the radiation, including three neutrinos with one helicity state,
is given by using equation (\ref{eq:s:rad}) and replacing
$f_\gamma \rightarrow f_{rad} = 2 + 3 \cdot 7/8$. In order to
calculate the entropy-density we need to know the number-densities
of protons, neutrons and electrons. If our universe consists out
of cold dark matter (CDM), as is suggested by the observations, we
furthermore need to know the masses and number-densities of the
dark matter. However, equation (\ref{eq:s:total}) tells us, that
for non-relativistic matter it suffices to know $\sum{m_i \cdot
n_i}$, which is nothing else than total energy-density of all
forms of non-relativistic matter $\rho_m = \rho_{CDM} + \rho_b +
\rho_e$. We find:

\begin{equation} \label{eq:s:total:rho}
s = \left(\frac{\rho_m}{T} +  \left( \frac{2 \pi
T}{\hbar}\right)^3 \frac{f_{rad}}{4 \pi \cdot 3^2 \cdot 5}\right)
\end{equation}

For radiation (with zero chemical potential) the entropy-density
and the energy-density is related by:

\begin{equation} \label{eq:s:e}
s = \frac{4}{3} \frac{\rho_{rad}}{T}
\end{equation}

Inserting this into equation (\ref{eq:s:total:rho}) gives:

\begin{equation} \label{eq:s:total:rho:2}
s = \frac{\rho_m + \frac{4}{3} \rho_{rad}}{T}
\end{equation}

The total matter-density has been very accurately determined by
WMAP. Using $\rho_m \approx 0.26 \rho_c = 2.462 \cdot 10^{-27} kg
/ m^3$, $\rho_{rad} = 1.073 \cdot 10^{-30} kg / m^3$ and the
measured value for the CMBR-temperature we find experimentally:

\begin{equation} \label{eq:s:total:experimental}
s = \frac{5.891 \cdot 10^{12}}{m^3} = \frac{2.485 \cdot
10^{-92}}{r_{Pl}^3}
\end{equation}

The experimental value is almost equal to the theoretical
calculation.

\subsection{\label{sec:Gibbs=0}On the relation between energy- and entropy-density and the free energy}

Equation (\ref{eq:s:total:rho}) contains a very interesting
message.  In the matter-dominated era of our universe we can
neglect the radiation-contribution to the entropy-density. In this
case the entropy-density is nothing else than the energy-density
divided by $T$:

\begin{equation} \label{eq:sT:rho}
s \simeq \frac{\rho}{T}
\end{equation}

Equation (\ref{eq:sT:rho}) allows us to determine the free
energy $F$ in the interior holostar space-time.

\begin{equation} \label{eq:F=0}
F = E - S T = V (\rho - s T) = 0
\end{equation}

We find the remarkable result, that the free energy is minimized
to zero, whenever $s = \rho / T$. Note that this is a local
relation: The free-energy {\em density} $f = \rho - s T$ is zero
everywhere in the interior space-time.

From equation (\ref{eq:s:total:rho:2}) one would assume, that the
relation $s = \rho / T$ is only valid approximately in the
matter-dominated era, when the radiation contribution to the total
energy-density can be neglected. This will be the case whenever
the average interior radiation temperature lies below the
nucleon-rest mass, which requires $M > 0.01 M_\odot$. For any
smaller - radiation dominated - holostar we expect, that the
entropy-density is higher by a factor of $4/3$. But this is in
gross conflict with the general result derived in section
\ref{sec:s=g}, equation (\ref{eq:s=e/T}), according to which $s =
\rho / T$, {\em independent} of the type of the interior matter.
Any deviation from this result would lead to an entropy (or
temperature at infinity) which is not even proportional to the
Hawking entropy or temperature.

Whereas for large matter-dominated holostars the proportionality
between the holostar's entropy and the Hawking entropy is obvious,
small radiation dominated holostars with $s T = 4/3 \rho$ appear
to be incompatible with the Hawking entropy-area formula. Although
one might expect some modification of the Hawking-formula at very
high energies, close to the Planck energy, a modification that
sets in for black holes with roughly 1 \% of the mass of the sun
(or less) seems awkward.

Even worse, it does not help to replace the factor $4$ in the
Hawking entropy-area formula with a different factor. This does
not work, because equation (\ref{eq:s:total:rho:2}) not only
contains the entropy, but also the temperature: $s T = \rho$. Any
constant rescaling of the Hawking entropy leaves the product $s T$
undisturbed.\footnote{This follows from the well-known
thermodynamic identity $\partial S / \partial E = 1/T$, which must
be valid in the exterior space-time.}

It doesn't help either, to change the total interior
matter-density by the unwanted factor $4/3$. The interior
matter-density of the holostar is completely fixed. The holostar's
remarkable properties arise from a delicate cancellation of terms
in the Einstein equations, which only takes place for the
"special" matter-density $\rho = 1 /(8 \pi r^2)$.

How then can the requirement $s T = \rho$ be achieved in the
radiation dominated era? A complete answer cannot be given in this
paper. The interested reader is referred to \cite{petri/thermo}
for a detailed discussion. There it will be shown, that the
ultra-relativistic fermions must acquire a non-zero chemical
potential in the holostar space-time, which changes the relative
contributions to the energy- and entropy-densities of the
fermionic and bosonic degrees of freedom in a way that $s T =
\rho$ is achieved for the total energy- and entropy density. The
required value for the chemical potential depends on the ratio of
fermionic to bosonic degrees. In general $\mu$ cannot be expressed
in a closed form, but must be found numerically by an implicit
function.

However, when the fermionic and bosonic degrees of freedom are
equal, the formula become very simple. If every bosonic degree of
freedom is paired with a fermionic degree of freedom, as suggested
by supersymetry, the non-zero chemical potential of the fermions
has the effect to enhance the entropy-density of a fermionic
degree of freedom exactly by a factor of 3/2 over a bosonic degree
of freedom. The energy-density is enhanced by a factor of
7/3.\footnote{Note, that these factors are quite different from
the case when the fermions have a zero chemical potential: here a
fermionic degree of freedom always has a lower entropy- and energy
density as a bosonic degree of freedom. The factor is 7/8.}
Therefore the entropy-density on the left side of equation
(\ref{eq:s:total:rho:2}), which only contains the bosonic
contribution (with zero chemical potential) has to be replaced by:

\begin{equation}
s_{rad} = \frac{4}{3} \frac{\rho_{rad}}{T} \rightarrow s_{SUSY} =
\frac{4}{3} \frac{\rho_{rad}}{T} \left(1 + \frac{3}{2}\right) =
\frac{10}{3} \frac{\rho_{rad}}{T}
\end{equation}

whereas the energy-density has to be replaced by

\begin{equation}
\rho_{rad} \rightarrow \rho_{SUSY} = \rho_{rad} \left(1 +
\frac{7}{3}\right) = \frac{10}{3} \rho_{rad}
\end{equation}

so that

\begin{equation}
s_{SUSY} = \frac{\rho_{SUSY}}{T}
\end{equation}

Supersymmetry - or rather a (model-dependent) non-zero value for
the chemical potential of the fermions - has the remarkable effect
to modify the fermionic contribution to the energy- and entropy
densities with respect to the bosonic contributions in such a way,
that the requirement $s = \rho / T$ is met even in the radiation
dominated era. This ensures, that the holostar has an entropy
exactly proportional to the Hawking entropy, not only for large
holostars with $M > M_\odot$, but for any conceivable size.

One can turn the argument around: Hawking's derivation for the
black hole entropy / temperature is extremely robust. The only
input is the evolution of a quantum field in the exterior
space-time. The approximations are expected to brake down for
Planck-sized black holes, but not for black holes with 1 \% of the
mass of the sun. Viewed from the exterior space-time a holostar
has exactly the same gravitational field as a black hole, except
for a Planck-sized region outside its gravitational radius.
Therefore, from the viewpoint of Hawking's analysis in the
exterior space-time, the holostar should have an entropy exactly
equal to the Hawking entropy, with the possible exclusion of
Planck-sized holostars. But this requires, that supersymmetry (or
at least non-zero chemical potentials of the fermions) must become
important in the {\em interior} space-time, whenever the
temperature is high enough, that the interior matter consists
predominantly out of radiation.

\subsection{On the conservation of energy and entropy in the interior and exterior space-time}

The holostar is a static solution with a constant gravitating mass
$M$. Its entropy has been shown to be proportional to the Hawking
entropy, i.e. $S \propto M^2$. For a large holostar the mass and
entropy are nearly constant, because the Hawking evaporation time
$t \propto M^3 / \hbar$ is immensely large. Therefore entropy and
entropy are conserved globally in the holostar space-time.

Is entropy and energy conserved locally as well? At first sight
entropy and energy-conservation appear to be violated from the
local viewpoint of a geodesically moving observer. For such an
observer the volume expands isotropically as $r^3$, the total
energy-density scales with $1/r^2$ and the entropy-density with
$1/r^{3/2}$ according to equation (\ref{eq:s=g}). As the local
observer sees more and more of the holostar space-time (his
Hubble-radius increases with $r =\tau$), he sees an ever
increasing energy $E \propto r$ and an increasing entropy $S
\propto r^{3/2}$ within the observable part of the universe,
defined by his local Hubble-volume. Both energy and entropy
increase by different power laws in $r$.

But $E$ and $S$ have exactly the dependencies on $r$ that are
required from local energy/entropy conservation. The important
thing to realize is, that the local observer is not sitting in a
closed system.\footnote{This is contrary to a homogeneously
expanding FRW-space-time. In such a space-time there can be no net
energy-, entropy or particle-inflow into or out of a co-moving
volume, so any large co-moving volume can be regarded as closed.}
Rather, as the observable part of the universe expands, work is
done against the negative pressure and entropy is produced by the
(pressure-induced) change in energy at a specific well defined
temperature. If the pressure-induced work is taken properly into
account we find, that energy is conserved locally: With $V \propto
r^3$ and $P \propto -1/r^2$ we find $dE = - P dV \propto +dr$, so
that $E \propto r$, which is exactly the energy dependence that a
local observer sees within his Hubble-volume. The same argument
applies to the entropy. If we take the entropy into account that
is produced by the increase in energy via the well known relation
$dS = dE / T$, we find for $E \propto r$ and $T \propto 1 /
\sqrt{r}$, that $dS \propto \sqrt{r} dr$, so that $S \propto
r^{3/2}$, exactly as required. For a detailed discussion, which
analyzes the phenomena from the frame of the co-moving and a
stationary observer, see section \ref{sec:frames}.

Therefore we get the remarkable result, that both energy and
entropy are conserved globally and locally in the holostar
space-time. Whereas energy-conservation should pose no conceptual
problem, entropy-conservation seems to be in conflict with the
second law of thermodynamics $\Delta S \geq 0$. However, this is
only an apparent contradiction. Quite clearly the second law
allows reversible processes with $\Delta S = 0$.\footnote{The
knowledgeable reader will be aware that the more strict version of
the second law $\Delta S > 0$ (or the Boltzmann H-theorem),
although widely accepted as a true law of physics, still has the
status of an empirical statement. All attempts to prove $\Delta S
> 0$ from first principles have failed. This is not too
surprising, because all microscopic processes are time-reversible.
The holostar solution provides a solution to this puzzle: The
holostar solution as a whole is time-reversible and its entropy is
constant (neglecting Hawking radiation, which is a very slow
process for large holostars). Yet the interior matter of the
holostar solution "splits" into two sectors, the outmoving and
inmoving particles. These sectors become "separated from each
other" at a very early stage of the motion: The highly
relativistic motion of the outmoving with respect to the inmoving
particles decreases the cross-sections in a way, that there is
virtually no interaction between the two sectors at late times.
The expanding sector corresponds to an increase in entropy, the
contracting sector to a decrease. Both sectors are superimposed
over each other at every space-time point. Although $\Delta S$ is
clearly positive in the expanding sector, the contracting sector
exactly compensates this change, so that $\Delta S = 0$ {\em
locally!} at every space-time point.} Yet the entropy increases
for any outward moving observer. Not knowing the reason for this
increase, any sensible observer must assume that there is a
general law of the form $\Delta S > 0$. But this is just an {\em
apparent} increase in entropy, tied to the observer's state of
motion. An inward moving observer, ignorant of the existence of
the outmoving observer, will assume $\Delta S < 0$. An observer
who knows the total picture will conlude that $\Delta S = 0$.

The assumption, that the sign of $\Delta S$ might be related to
the arrow of time or the expansion of the universe, is not new.
Remarkably this more or less intuitive association receives a
(partial) justification and a quite unexpected resolution: $\Delta
S$ is in fact related to the expansion, but the statement that
$\Delta S > 0$ when $\Delta \rho < 0$ is rather a
misconception\footnote{yet showing a remarkably good intuition by
the people who fathered this "misconception"!}, based on our
ignorance of the large scale structure of the universe. If the
entropy-production due to the pressure-induced energy increase is
taken properly into account, and if one realizes that the energy-
and entropy increase in the "expanding" sector of the interior
space-time is associated with an associated decrease of equal
magnitude in the "contracting" sector, $\Delta S = \Delta E = 0$
not just globally, but also locally!

\subsection{Local matter distribution and self-similarity}

For a very large holostar of the size of the universe, its outer
regions will consist of low-density matter, which should be quite
comparable in density and/or distribution to the matter we find in
our universe. Such matter can be expected form local hierarchical
sub-structures comparable to those found in our universe, as long
as the scale of the sub-structures remains compatible with the
overall mass-energy distribution $\rho \propto 1/r^2$ of the
holostar.

On not too large a scale some local regions might collapse,
leaving voids, others might expand, giving rise to filamental
structures. However, any local redistribution of mass-energy has
to conserve the total gravitational mass of the holostar, its
total angular momentum and - quite likely - its entropy, i.e. its
total number of particles. These exterior constraints require that
regions of high matter-density must be accompanied by voids.

Furthermore one can expect, that the local distribution of matter
within a holostar will exhibit some sort of self-similarity, i.e.
the partly collapsed regions should follow the global $1/r^2$-law
for the local mass-density. This expected behavior is quite in
agreement with the observations concerning the mass-distribution
in our universe: The flat rotation curves of galaxies, as well as
the velocity dispersion in galaxies and clusters of galaxies hint
strongly, that the matter distribution of the local matter in
galaxies and clusters follows an $1/r^2$-law. As far as I know,
there has been no truly convincing explanation for this apparently
universal scaling law, so far.

In Newtonian gravity a local matter-distribution proportional to
$1/r^2$ can be described by the so called isothermal sphere (see
for instance \cite[p. 370]{Peacock/book}), whose spherically
symmetric matter-distribution is given by:

\begin{equation}
\rho = \frac{\sigma_v^2}{2 \pi r^2}
\end{equation}

$\sigma_v^2$ is the constant velocity-dispersion of the objects
within the sphere, which is interpreted as "temperature" (thus the
term isothermal). It appears, that for $\sigma_v^2 \ll 1$ the
isothermal sphere is a useful approximation for the relation
between the matter-density and the velocity dispersion of certain
galaxies. Clearly the Newtonian approach will fail for
relativistic velocity-dispersions. Quite interestingly, the
matter-distribution of the holographic solution can be - formally
- described by $\sigma_v = 1/2$, indicating that the Newtonian
approach cannot be expected to be a reliable approximation for
$\sigma > 0.1$.

\subsection{\label{sec:frames}Some remarks about the frames of the asymptotic and the co-moving observer}

Most of the discussion about the properties of the holostar has
been in the frame of the asymptotic observer, who is at rest in
the ($t, r, \theta, \varphi$)-coordinate system.

If the holostar is to serve as a model for an expanding universe,
one must interpret the phenomena from the frame of the co-moving
observer, who moves nearly geodesically on an almost radial
trajectory through the low density outer regions of the holostar.

The frames of the co-moving and the asymptotic observer are
related by a Lorentz boost in the radial direction. Due to the
small tidal acceleration in the holostar's outer regions, the
extension of the local Lorentz frames can be fairly large. The
proper acceleration - due to exterior forces, such as the pressure
- experienced by the co-moving observer will be very low in the
regions which have a density comparable to the density that is
observed in the universe at the present time.

The holostar's interior space-time is boost-invariant in the
radial direction, i.e. the stress-energy tensor is unaffected by a
radial boost. The co-moving observer moves nearly radially for $r
\gg r_i$. His radial $\gamma$-factor grows as the square root of
his radial coordinate value, whereas his tangential velocity goes
rapidly to zero with $1/r^{3/2}$. The radially boosted co-moving
observer therefore will see exactly the same total stress-energy
tensor - and thus the same total energy density and total
principal pressures - as the observer at rest in the ($t, r,
\theta, \varphi$)-coordinate system. The above statement, however,
only refers to the total energy density and pressures. It is not a
priori clear if the individual contributions to the mass-energy
and pressures, for example from massive particles and photons,
have the same $r$-dependence as the total energy density.

\subsubsection{On the number- and energy-densities of massive particles}

Let us first consider the case of massive particles. The observer
at rest in the coordinate system measures an energy density $\rho$
of the massive particles, which is proportional to $1/r^2$. A
factor of $1/r^{5/2}$ comes from the number density given in
equation (\ref{eq:n:massive}), a factor of $r^{1/2}$ from the
special relativistic $\gamma$-factor that must be applied to the
rest mass of the particles according to equation (\ref{eq:E(r):massive}).

From a naive perspective (neglecting the effects of the pressure)
it appears as if the co-moving observer and the observer at rest
in the ($t, r, \theta, \varphi$)-coordinate system disagree on how
the energy and number densities of massive particles develop with
$r$. Because of the highly relativistic motion of the co-moving
observer, the observer at rest in the ($t, r, \theta,
\varphi$)-coordinate system will find that the proper volume of
any observer co-moving with the massive particles is
Lorentz-contracted in the radial direction. Therefore the
co-moving observer will measure a larger proper volume, enlarged
by the radial $\gamma$-factor, which is proportional to
$\sqrt{r}$. If we denote the volume in the frame of the co-moving
observer by an overline, we find $\overline{V} \propto \gamma \,
r^{5/2} \propto r^3$. As long as the massive particles aren't
created or destroyed, the number-density of the massive particles
in the co-moving frame therefore must scale as $1/r^3$.
Furthermore, for the co-moving observer the neighboring massive
particles are at rest to a very good approximation.\footnote{The
tidal acceleration is negligible; the extension of the local
Lorentz frame of the observer is nearly equal to the local Hubble
length, i.e. of the order the current "radius" of the universe,
$r$.} Therefore the mass-energy density in the co-moving frame
should be nothing else than the (presumably) constant rest-mass of
the particles multiplied by their number-density. From this it
follows, that the mass-energy density of the massive particles
should scale as $1/r^3$ as well.

This naive conclusion, however, is false. It does not take into
account the energy change in the co-moving volume due to the
radial pressure. Any radial expansion in the co-moving frame
affects the internal energy. Due to Lorentz-elongation in the
co-moving volume the radial thickness $\overline{l_r}$ of the
expanding shell develops proportional to $r$ in the co-moving
frame. From this the internal energy-change $\delta E$ in the
shell can be calculated for a small radial displacement $\delta
r$. We find:

\begin{equation} \label{eq:dE:massive:co}
\overline{\delta E} = - \overline{P_r} A \, \overline{\delta l_r}
\propto \frac{\delta r}{2}
\end{equation}

with $\overline{P_r} = P_r = -1 / (8 \pi r^2)$ and $\overline{A} =
4 \pi r^2$. Note that the radial pressure in the co-moving frame
is exactly equal to the radial pressure in the coordinate frame
due to the boost-invariance of the stress-energy tensor in the
radial direction.

From equation (\ref{eq:dE:massive:co}) the radial dependence of
the internal total energy in the co-moving frame follows:

$$\overline{E} \propto r$$

We have already seen that the co-moving volume develops as
$\overline{V} \propto r^3$, so that the energy density of the
massive particles in the co-moving frame, taking the pressure into
account, develops exactly as in the coordinate frame, i.e.
$\overline{\rho} \propto \overline{E} / \overline{V} \propto
1/r^2$.

It is quite obvious that such a dependence is not compatible with
the assumption that both the rest-mass and the number of the
massive particles in the co-moving volume remain constant. Either
the rest-mass of the massive particles must increase during the
expansion or new particles (massive or radiation) have to be
created by the negative pressure. There is no strong experimental
evidence, that the rest mass of the nucleon or the electron have
changed considerably during the evolution of the universe, at
least in last few billon years. If this is the case we cannot
avoid the conclusion that the negative pressure has the effect to
create new particles in the co-moving frame. Particle creation in
an expanding universe is not new. It is one of the basic
assumptions of the steady-state model. Furthermore particle
production via expansion against a negative pressure is a well
known phenomenon from the inflational equation of state. There are
differences. Whereas the isotropic negative pressure of the
inflational phase keeps the energy-density in the expanding
universe constant during the expansion, the energy-density in the
holostar develops as $1/r^2$, because the negative string type
pressure only acts in one of three spatial directions. As a result
the particle creation rate in the holostar is quite low at the
present time: Roughly one neutron per cubic kilometer every 10
years is required.

Note, that we have assumed that the massive particles move
geodesically in this argument. The non-negligible pressure within
the holostar solution might enforce non-geodesic motion. (This is
quite unlikely for large $r$, according to the discussion in
section \ref{sec:massive:acc}.) Would this change the main results
found above, i.e. $\rho \propto 1/r^2$ in both the co-moving and
the coordinate frame? The answer is no, as can be seen by
analyzing the following hypothetical case of non-geodesic motion.
Let us assume that the motion of massive particles takes place in
such a way, that there is no expansion in the radial direction. If
energy-conservation holds globally, it can be shown that in this
case the radial motion is subject to a nearly constant proper
velocity, as viewed by the observer at rest in the coordinate
frame, meaning that the radial $\gamma_r$-factor is constant. The
observers in both frames don't see any pressure-induced particle
production, because there is no expansion against the radial
pressure. Furthermore, in both frames the energy- and
number-densities follow an inverse square law, as the expansion
only takes place in the two tangential directions and both
observers don't see any $r$-dependent Lorentz-contraction due to
the constant $\gamma_r$ factor.\footnote{\label{fn:Unruh} This
mode of non-geodesic motion for the massive particles has the
following interesting feature: Pure geodesic motion of massive
particles leads to the somewhat undesirable result, that there
would be a non-zero - albeit quite small - mass-energy density
outside the membrane, which isn't compatible with the holostar
equations (one might try to modify the equations, though).
Geodesic motion is characterized by large values of $\gamma$. It
can be shown, that for the $\gamma$-values in question the massive
particle will move out as far as $3 r_h/2$, i.e. to the angular
momentum barrier of the photons, before they reverse their motion.
In contrast, low $\gamma_r$ particles will reverse their motion at
a position less that a Planck length outside the membrane, thus
smearing the membrane out over roughly a Planck length, which is
quite what one would expect from a quantum perspective.}

The above analysis indicates that, independent whether the motion
is geodesic or not, the energy density of the massive particles in
the co-moving frame should be proportional to $1/r^2$, exactly as
in the coordinate frame. The conclusion is, that a radial boost
from the coordinate to the co-moving frame not only leaves the
total energy density unaffected, but also the respective energy
densities of the different particle species.\footnote{This
argument is not 100 \% water tight. Assuming a constant rest mass
of the geodesically expanding particles, additional particles must
be produced by the expansion against the negative pressure. If
these particles are different from the (massive) particle species
that catalyze their production, there might be a redistribution of
mass-energy between the different species during the expansion.}
With the assumption that the rest mass of the massive particles is
constant the number densities must also evolve according to
$1/r^2$.

One can see this already from the holostar's interior
energy-density:

\begin{equation} \label{eq:n:massive:S:2}
n_m = \frac{\rho}{m} = \frac{1}{8 \pi m r^2}
\end{equation}

The above equation is nearly exact, when the matter-content of the
universe is dominated by one single massive species (the nucleon).

\subsubsection{On the number- and energy-densities of zero rest mass particles}

Let us now discuss the number- and energy densities of the zero
rest-mass particles in the two frames. This problem is closely
related to the question whether the co-moving observer experiences
a different radiation temperature than the observer at rest in the
($t, r, \theta, \varphi$)-coordinate system. The standard argument
for the red-shift of radiation in an expanding Robertson-Walker
universe is, that the wavelength of the radiation is stretched
proportional to the expansion.\footnote{A more sophisticated
derivation is based on the existence of a Killing vector field in
a Robertson-Walker universe (see for example
\cite[p.101-104]{Wald/GR}). The derivation makes use of the well
known fact, that the scalar product of the photon wave-vector with
the Killing-vector is constant for geodesic motion of photons.
This argument, however, requires the geodesic equations of motion,
and therefore is only water-tight for a dust universe without
significant pressure. Another argument is based on the energy loss
due to the expansion of radiation against its own radiation
pressure. With $P = \rho/3$ this argument also gives the $T
\propto 1/r$ dependence in the Standard Cosmological Models.} This
gives the known $T \propto 1/r$-dependence for the radiation
temperature in the standard cosmological models. If this argument
is applied to the holostar, one naively expects that the $T
\propto 1 / \sqrt{r}$-law (which has been shown to hold in the
coordinate frame) is transformed to a $\overline{T} \propto
1/r$-law in the frame of the co-moving observer, due to
Lorentz-contraction (or rather elongation) of the photon
wavelength.

But a $\overline{T} \propto 1/r$-law wouldn't be consistent for a
small holostar, where - due to the high temperature - the energy
density is expected to be dominated by radiation in true thermal
equilibrium. The energy density of thermalized radiation is
proportional to $T^4$. If the only contributor to the total energy
density is radiation, the radiation energy density will transform
exactly as the total energy density in a radial boost. However,
the total energy density is radially boost-invariant. Therefore in
the radiation dominated era the energy density of the radiation in
any radially boosted frame should scale as $1/r^2$. This however
implies the $\overline{T} \propto 1/\sqrt{r}$-dependence, at least
if the radiation is in thermal equilibrium in the boosted
frame.\footnote{If the holostar were truly static in the high
temperature regime, i.e. geodesic and pressure induced
acceleration cancel exactly, there is no problem. There would be
no directed motion and therefore $\overline{T} = T \propto
1/\sqrt{r}$ trivially.}

In a radially boost-invariant space-time one would expect on more
general grounds, that it is - in principle - impossible to
determine the radial velocity of the motion, at least by direct
measurements performed by the observer co-moving with the
matter-flow. A $\overline{T} \propto 1/r$-law would imply a
radiation temperature incompatible with the energy-density of the
radiation, which would allow the co-moving observer to determine
his radial velocity. This can be seen as follows:

According to equation (\ref{eq:numberdensity}) the number density
of photons in the coordinate frame scales as $n \propto
1/r^{3/2}$. Naively one would assume, that the number-density of
photons in the co-moving frame is given by $\overline{n_\gamma} =
n_\gamma / \gamma \propto 1/r^2$, due to Lorentz elongation of the
co-moving volume with respect to the observer at rest. However, at
a closer look one has to take into account the measurement
process. Photons cannot be counted just by putting them in a box
and then taking out each individual particle. As photons always
move with the local speed of light, such a procedure, which would
be good for massive particles, doesn't work. The right way to
count photons is to place a small ideal (spherical) absorber
somewhere in the space-time, count all the hits per proper time
interval and relate the obtained number to the volume (or surface
area) of the absorber. But this procedure requires, that the
time-delay due to the highly relativistic motion of the co-moving
observer has to be taken into account. The co-moving observer will
count many more photons in a given standard interval of his proper
time, than the observer at rest would count in the same interval.
The time-dilation introduces another $\gamma$-factor that exactly
cancels the $\gamma$-factor from the Lorentz contraction of the
proper volume. Therefore we arrive at the remarkable result, that
the number-density of photons should be the same for both
observers, i.e. independent of a radial boost.\footnote{This
result could also have been obtained by calculating the
pressure-induced energy-change in the co-moving frame. Quite
interestingly this change is zero for photons, because the proper
radial extension of a geodesically moving shell of photons remains
constant in the co-moving frame. Therefore the energy-density of
the photons in the co-moving frame evolves inverse proportional to
the proper surface area of the shell, i.e. $\overline{\rho_\gamma}
\propto 1/r^2$. Assuming that the photon number in the shell
remains constant in the co-moving frame (which is equivalent to
assuming geodesic motion or a thermal spectrum) we then find
$\overline{n_\gamma} \propto 1/r^{3/2}$.}

If we had $\overline{T} \propto 1/r$ in the co-moving frame, we
would get $\overline{\rho_\gamma} \propto \overline{n_\gamma}
\overline{T} \propto 1/r^{5/2}$, which implies
$\overline{\rho_\gamma} \propto \overline{T}^{5/2}$ in the
co-moving frame. However, in thermal equilibrium
$\overline{\rho_\gamma} \propto \overline{T}^4$. Even if the
argument given above for the number-density of photons were
incorrect, i.e. only the volume were Lorentz contracted and time
dilation would play no role, we would have $\overline{\rho_\gamma}
\propto \overline{n_\gamma} \overline{T} \propto 1/r^3$, implying
$\overline{\rho_\gamma} \propto \overline{T}^3$ in the co-moving
frame, which wouldn't work either.

Therefore it seems reasonable to postulate, that a radial boost
from the $(t, r, \theta, \varphi)$-coordinate system to the system
of the co-moving observer should not only leave the total energy
density unaffected, but also other physically important
characteristics such as the thermodynamic state of the system,
i.e. whether or not the radiation can be characterized as thermal.
This then implies that the thermodynamic relation between
energy-density and temperature for an ultra-relativistic gas
should be valid in the radially boosted frame of the co-moving
observer, i.e. $\overline{\rho} \propto \overline{T}^4$, which
requires the $\overline{T} \propto 1/\sqrt{r}$ law.

The reader may object, that any radial boost will produce a large
anisotropy in the radiation temperature, as measured in the
boosted frame. I.e. the radiation will be blue-shifted for photons
travelling opposite to the motion of the co-moving observer and
red-shifted for photons travelling in the same direction, due to
"normal" Doppler shift. However, this point of view implicitly
assumes, that the radiation has originated from matter {\em at
rest} in the coordinate system. If the radiation that the observer
in the co-moving frame sees has originated from the massive
particles in his {\em own} frame\footnote{The - nearly isotropic -
CMB radiation that we can see in the observable universe has
arisen from the so called last scattering surface, i.e. was
emitted (and re-emitted) from the ionized matter at roughly
$z\approx 1000$.}, the radiation will be nearly isotropic: The
observer in the co-moving frame experiences an isotropic
Hubble-expansion of the massive particles within the current
Hubble-radius. Ignoring tidal effects (which are second order) and
assuming that special relativity remains correct even for
$\gamma$-factors as high as occur in the holostar space-time, it
does not matter whether the isotropically expanding volume is at
rest or moves relative to a - preferred or not preferred boosted
frame - with extremely high $\gamma$-factors.

If the radiation is isotropic in the co-moving frame, one might
assume, that it is highly anisotropic in the coordinate frame. But
it is quite unlikely that a - hypothetical - observer in the
coordinate frame will be able to see the radiation that has
originated in the co-moving frame: This radiation has an energy
roughly equal to the Planck-energy and thus an extremely low
cross-sectional area. What the observer at rest in the coordinate
frame will see, however, is Unruh radiation associated with the
pressure-induced acceleration. It will be shown in section
\ref{sec:Unruh} that the Unruh-temperature associated with this
acceleration scales with $1 / \sqrt{r}$, just as required.
Unruh-radiation is always isotropic. If this is the correct
interpretation, the only type of radiation that an observer in the
coordinate frame can see - effectively - is Unruh radiation with a
temperature proportional to the radiation temperature in the
co-moving frame. Self-consistency requires, that both temperatures
are equal.

Another - not too convincing argument - for the isotropy of the
radiation in the co-moving frame can be devised, if one takes the
negative radial pressure into account. The radially boosted
observer will find that the volume that lies in front of him is
Lorentz-contracted in the radial direction. The photons coming
from the front side therefore come from a radially "squeezed"
volume. But any volume contraction in the radial direction will
reduce the energy of the photons because of the negative radial
pressure. The blue-shift of the photons due to the (kinematical)
Doppler-shift will be exactly compensated by the red-shift
originating from the pressure-induced Lorentz contraction. A
similar effect occurs for the photons coming from the rear, i.e.
moving in the same direction as the observer.

Although perhaps some new insights are required in order to
resolve the problem of relating the observations in the different
frames in a satisfactory fashion, nature appears to have taken a
definite point of view: If we live in a large holostar, we clearly
are in the position of the co-moving observer. Except for a small
dipole anisotropy, which can be explained by the small relative
motion of our local group with respect to the isotropic
Hubble-flow (or by tidal effects), the CMBR is isotropic.
Furthermore, in Planck-units the CMBR-temperature is $T_{CMBR}
\simeq 2 \cdot 10^{-32} T_{Pl}$, whereas the radius of the
observable universe is roughly $r \simeq 9 \cdot 10^{60} r_{Pl}$
and the mass-density $\rho \simeq 5 \cdot 10^{-124} \rho_{Pl}$.
These simple figures strongly suggest, that the $T \propto
1/\sqrt{r}$ and $\rho \propto T^4$ laws are realized in our
universe in the system of the co-moving observer.

\subsection{\label{sec:nucleosynthesis}On the baryon to photon ratio and nucleosynthesis}

The discussion of the previous section has paved the way to
address the problem of nucleosynthesis in the holostar universe. A
quite remarkable by-product of the discussion in this section is a
surprisingly simple explanation for the baryon to photon ratio in
the universe.

\subsubsection{The assumption of constant rest mass}

In the following discussion I will assume, that the rest-mass of
the electron and nucleon are constant, i.e. independent of
temperature or radial position. This assumption will be relaxed in
the subsequent section.

We have seen in the previous section that the number density of
massive particles $\overline{n_m}$ and photons
$\overline{n_\gamma}$ in the co-moving frame develop as:

\begin{equation}
\overline{n_m} \propto \frac{1}{r^2} \propto T^4
\end{equation}

and

\begin{equation}
\overline{n_\gamma} \propto \frac{1}{r^{\frac{3}{2}}} \propto T^3
\end{equation}

As a consequence the ratio of the energy-densities per proper
volume of massive particles to photons remains constant in the
co-moving frame in the holostar universe\footnote{Note that
$\rho_\gamma \propto \rho_m \propto 1/r^2$ has already been
established in the coordinate frame, according to the analysis of
sections \ref{sec:motion:shell:rad} and
\ref{sec:motion:shell:mass}.}, i.e:

\begin{equation} \label{eq:rho:T}
\overline{\rho_m} \propto \overline{\rho_\gamma} \propto
\frac{1}{r^2} \propto T^4
\end{equation}

Comparing the energy density of electrons $\overline{\rho_e}$ and
photons $\overline{\rho_\gamma}$ at the present time, we find that
they are almost equal. In fact, assuming the chemical potential of
the photons to be zero, the respective energy densities turn out
as

\begin{equation}
\overline{\rho_\gamma} = 8.99 \cdot 10^{-128} \, \rho_{Pl}
\end{equation}

and

\begin{equation}
\overline{\rho_{e}} = 2.23 \cdot 10^{-127} \, \rho_{Pl} = 2.52
\overline{\rho_\gamma}
\end{equation}

if we assume an electrically uncharged universe, a proton to
nucleon ratio of $7/8$ and a universe consisting predominantly out
of nucleons (no significant dark matter component). $\rho_{Pl}=
m_{Pl}/r_{Pl}^3$ is the Planck-density.

In the context of the standard cosmological model this fact
appears as a very lucky coincidence, which happens just at the
current age of the universe and won't last long: The energy
density of radiation and matter evolve differently in the standard
cosmological model. Basically $\rho_m \propto T^3$, whereas
$\rho_\gamma \propto T^4$, so that the energy density of the
radiation falls off with $T$ compared to the energy density of the
massive particles.\footnote{The standard cosmological model
assumes that the number ratio of baryons to photons $\eta$ remains
constant in the expanding universe. With the recent WMAP data this
ratio is now estimated (at the time of nucleosynthesis) as $\eta
\simeq 6.5 \cdot 10^{-10}$. The postulate $\eta = const$ is quite
different to the evolution of the different particle species in
the holostar, where rather the energy- and entropy densities, and
not the number-densities of the fundamental particle species
remain constant during the evolution.}

The particular value of the baryon to photon ratio $\eta$ is a
free parameter in the standard cosmological model. The most recent
experimental determination of $\eta$ via primordial
nucleosynthesis is given by \cite{WMAP/cosmologicalParameters} as
$\eta \approx 6.5 \cdot 10^{-10}$ (at the time of
nucleosynthesis). Unfortunately we still lack an established
theoretical framework by which this value could be calculated or
even roughly estimated from first principles. In the holostar the
value of $\eta$ is linked to the nearly constant energy densities
of the different fundamental particle species. Whenever we know
what the ratio of the energy densities should be, we can estimate
the present value of $\eta$.

Can the ratio between the energy densities of photons and
electrons in the holostar universe be predicted by first
principles? In order to answer this question let us turn our
clocks backward until the radiation temperature in the holostar
registers somewhere above the electron rest mass, but well below
the mass of the muon or pi-meson (for example $T \approx 1-10 \,
MeV$). The rather fast electromagnetic reactions at the high
temperatures and densities ensure that the energy is distributed
nearly equally among the relativistic degrees of freedom of
electrons/positrons and photons, respectively.\footnote{I never
found the basic assumption of the standard cosmological model
utterly convincing, according to which the mutual annihilation of
baryons and anti-baryons into photons (and other
ultra-relativistic particles) produced an enormous imbalance
between the numbers of photons and baryons of order $10^9$, which
is postulated to have remained nearly constant from the time of
baryogenesis ($T \approx 1 \, GeV$) to the time of nucleosynthesis
($T \approx 0.1 \, MeV$) and forever thereon.

It is clear, that such a scenario is almost inescapable if the
cosmological principle were a law of nature (and not just a
convenient assumption). According to the cosmological principle -
and the Friedman-Robertson-Walker models that follow from this
principle - the universe looks the same for every observer at the
same (universal) cosmological time. It is not difficult to verify
that in such a universe the particle numbers of the different
stable species in any volume co-moving with the expansion is
conserved, at least if particle-changing interactions are
negligible. If one extrapolates the high photon-to-baryon ratio on
the order of $10^9$ today back to the time of baryogenesis, using
the above mentioned constraint on relative and total
particle-numbers, one cannot evade the conclusion, that a nearly
equilibrium energy-distribution shortly before baryogenesis
(nearly equal numbers of baryons, anti-baryons, photons and other
relativistic particles) was transformed to an enormous imbalance
shortly after baryogenesis (energy-density of the relic baryons
roughly a factor of $10^9$ less than the energy-densities of the
other particle species). Yet such a violent departure from
equilibrium in a very short time interval is hard to believe.

In the holostar space-time we don't have the constraints on the
relative particle numbers as in an FRW-type universe, as the
particles move through a static - inhomogeneous - space-time and
can "go anywhere they wan't", as long as they obey the local
equations of motion. Furthermore, in the holostar-universe it is
not the ratio of the particle-numbers of the different species
that remain constant during the expansion, but rather their
energy- and entropy densities. As the energy-densities of
electrons and photons are nearly equal in our universe today, the
holostar model of the universe predicts roughly equal numbers of
electrons/baryons and photons at the time of nucleosynthesis. This
opens up the possibility of a smooth quasi-equilibrium
phase-transition, whenever the temperature falls below the
mass-threshold of a particular particle species and virtually all
of the anti-matter is annihilated. Such a quasi-equilibrium phase
transition is characterized by the property, that the
energy-densities of all particles species participating in the
energy-transfer from the annihilated species to the remaining
species are still very close to their equilibrium values shortly
after the phase-transition, i.e. nearly equal. In order to be
compatible with the second law of thermodynamics, however, such a
quasi-equilibrium phase transition requires a profound
matter-antimatter asymmetry shortly before the phase-transition
sets in. But such a profound matter-antimatter asymmetry is
precisely what is predicted for an ultra-relativistic gas in the
holostar space-time in thermodynamic equilibrium. According to
\cite{petri/thermo} ultra-relativistic fermions in the curved
holostar space-time must acquire a non-zero chemical potential
roughly equal to the radiation temperature, otherwise the
thermodynamic constraint equations don't allow a solution (see
also the discussion in section \ref{sec:Gibbs=0}). The non-zero
chemical potential of the fermions induces a natural asymmetry
between ultra-relativistic fermions and anti-fermions. It can be
shown by simple microscopic statistical thermodynamics, that in
any spherically symmetric space-time with an energy-density $\rho
= 1 / (8 \pi r^2)$ (i.e. equal to the holostar energy-density) at
least $\approx 11.5$  - or more - fermions per anti-fermion are
required in thermal equilibrium at ultra-high temperatures. Any
phase-transition in such a space-time proceeds quite "smoothly",
as the number of anti-particles at high temperatures is always
less than 8 \% of the total particle number of a given species.
Mutual annihilation of particles and anti-particles therefore
reduces the overall energy-density in the given species by
maximally 16 \%. For a more detailed discussion with accurate
numerical figures see \cite{petri/thermo}.

The standard scenario of big-bang baryogenesis (with a primordial
matter-antimatter asymmetry on the order of $10^{-9}$) stands and
falls with the assumption of a homogenous universe with a
universal cosmological time. But there is no physical law that the
universe {\em must} be constructed according to the cosmological
principle. The cosmological principle requires us to accept, that
the universe has a completely different structure than all the
other self-gravitating objects we know, such as black hole type
objects (or elementary particles). Those objects have a center,
whereas a FRW-type universe has none. From a philosophical point
of view it appears much more satisfactory to assume that the
universe is built according to the same plan as any other large
self-gravitating object. The holostar model of the universe points
to the very real possibility of a universe that is constructed
hierarchically out of its most basic building blocks, which appear
to be strings and membranes at the most fundamental level,
followed by compound objects, ranging from elementary particles to
black hole type holostars on subsequent hierarchy levels. Even the
universe itself might be nothing else than a further step in the
hierarchy: a very large - possibly infinitely extended - holostar,
that contains all of its smaller sub-structures.} When the
temperature falls below the threshold defined by the
electron-mass, the last (few!) remaining positrons are quickly
dispatched, so that the number of photons should be nearly equal
to the number of the relic electrons. The same will be true for
the respective energy densities, so that as a very rough estimate
we can assume, that the energy densities of photons and electrons
are equal at decoupling.\footnote{In the simple analysis here I
neglect the difference in energy-densities of photons and
electrons due to the fact, that photons are bosons whereas
electrons are fermions. Likewise the non-negligible effects of any
non-zero chemical potentials of photons and electrons are
neglected.}

At temperatures below nucleosynthesis, i.e. $T < 0.1 \, MeV$, the
energy densities of photons and electrons in the holostar evolve
nearly independently. Even when the reactions that have maintained
thermal equilibrium before have ceased, the energy-densities of
the two particles species will be maintained at a nearly constant
ratio, whose exact value is determined by the physics at the
time-period where electrons and photons have decoupled
chemically.\footnote{It is a characteristic property of the
holostar space-time, that the energy-densities of the different
fundamental particle species evolve with a constant ratio: At high
temperature the nearly equal energy densities are maintained by
the reactions establishing thermal equilibrium. At low
temperatures the negative radial pressure ensures, that the
massive and massless particles maintain a constant energy ratio
during geodesic motion. The motion of the massive particle not
even has to be geodesic. The negative radial pressure will affect
any motion of particles such, that the energy-density evolves
proportional to $1/r^2$.} An exact determination of this ratio is
beyond the scope of this paper. Quite likely chemical potentials
will play an important role (see \cite{petri/thermo}). However, it
is quite encouraging that the very rough estimate $\rho_e \approx
\rho_\gamma$ at the time of chemical decoupling is within a factor
of three of the experimental value $\rho_e \approx 2.5 \,
\rho_\gamma$ determined at the present time. In fact, $\rho_e
\simeq 2.3 -2.5 \, \rho_\gamma$ is the theoretical prediction for
the ratio of the energy-densities of electrons to photons at high
temperatures in thermal equilibrium in the holostar space-time,
when the non-zero chemical potential of the electrons is taken
into account \cite{petri/thermo}.

The discussion above allows us to make a rough estimate for the
value of $\eta$ at the present time. Assuming that photons and
electrons decoupled chemically at $T = m_e/3$ and assuming equal
energy-densities of electrons and photons at this time we find:

\begin{equation} \label{eq:eta}
\eta_{today} \approx \frac{3 \, T_{CMBR}}{m_e} = 1.38 \cdot
10^{-9}
\end{equation}

This very rough estimate is only a factor of 2 higher
than the WMAP-result $\eta_{WMAP} \approx 6.5 \cdot 10^{-10}$.

We can also compare the value in equation (\ref{eq:eta}) to the
maximum possible value of the baryon to photon ratio in the
universe today. Under the assumption that there is no significant
dark matter-component $\eta_{max}$ can be estimated from the
cosmological number densities of photons and
electrons.\footnote{The photon number density is determined by the
Planck formula, assuming the chemical potential of the photons to
be zero. The maximum value for the electron number density is
determined from the total matter-density of the universe according
to WMAP, assuming no significant dark matter component, a proton
to nucleon ratio of 7/8 and an uncharged universe with no
significant antimatter.} We find $\eta_{max} \approx 3.14 \cdot
10^{-9}$, which is a factor of two higher than the estimate of
equation (\ref{eq:eta}).

We are now in the position to discuss nucleosynthesis in the
holostar universe. The primordial synthesis of the light elements
proceeds somewhat differently as in the standard cosmological
model. The two key parameters governing the respective reaction
rates, the number density of the nucleons (baryons) $n_b$ at the
temperature of nucleosynthesis and the (Hubble) expansion rate,
turn out to be significantly different in both models.

In the standard cosmological model the number-density of the
nucleons depends on the cube of the temperature, i.e. $n_b \propto
1/R^3 \propto T^3$, whereas according to the discussion above the
co-moving observer in the holostar finds that the number density
of non-relativistic particles scales as the fourth power of the
temperature, $\overline{n_b} \propto 1/r^2 \propto T^4$. Therefore
the number-density of the nucleons at the temperature of
nucleosynthesis will be roughly nine orders of magnitude higher
than in the standard model.

This won't have a large effect on the ratio of primordial Helium-4
to Hydrogen. The numerical value of the He/H ratio is mainly due
to the n/p ratio in thermal equilibrium at $T \approx 0.8 \, MeV$,
i.e. the temperature where the weak reactions are shut
off.\footnote{Neutron decay during the time where the temperature
drops to roughly $0.1 MeV$, i.e. the temperature where deuterons
are produced in significant numbers (and quickly end up in the
more stable $He^4$), also plays a non-negligible role in the
standard cosmological model. Note that the shut-off temperature of
the weak interactions, which is $0.8 MeV$ in the standard model,
depends on the Hubble rate, and thus will be (slightly) different
in the holostar.} However, the higher number-density will have a
considerable effect on the amount of Deuterium, Helium-3 and
Lithium-7 produced in the first few seconds of the universe.

The relative abundance of these elements depends quite sensitively
on the nuclear reaction rates, which are proportional to the
number-densities of the reacting species. In order to accurately
calculate the relative number densities of all the other elements,
excluding H and He-4, one has to consider the dynamic process in
which the nuclear reactions compete against the cosmic expansion.
Eventually the nuclear reaction rates will fall below the
expansion rate, ending the period of nucleosynthesis. Therefore
the second key parameter in primordial nucleosynthesis, besides
the nucleon number-densities, is the cosmic expansion rate, which
is proportional to the Hubble-value.

In the matter dominated era the standard model predicts $H \propto
1/t \propto 1/R^{3/2} \propto T^{3/2}$, whereas in the radiation
dominated era $H \propto T^2$. In the radiation dominated era the
dependence of the expansion rate on the temperature is equal in
both models. Taking the different dependencies in the
matter-dominated era into account, one can relate the
Hubble-constant $H_{h}$ in the holostar universe and the
Hubble-constant $H_s$ in the standard model at the time of
nucleosynthesis. We find:

\begin{equation}
H_h \approx \sqrt{\frac{T_{eq}}{T_{CMBR}}} H_s = \sqrt{z_{eq}} H_s \simeq 60 H_s
\end{equation}

where $z_{eq} \approx 3450$ is the red-shift at which the standard
model becomes radiation dominated, according to the recent WMAP
findings. The above result is quite consistent with the fact, that
the "age" of the universe at $T = 0.1 \, MeV$  is roughly $\tau
\approx 7 s$ in the holostar, whereas in the standard cosmological
model we find $t \approx 200 s$ at the same temperature.

It should be evident, that due to the differences in the two key
parameters nucleosynthesis in the holostar universe will proceed
quite differently from the standard model. Without actually
solving the rate equations it is difficult to make quantitative
predictions. In general the higher number density of nucleons in
the holostar will lead to a more effective conversion of Deuterons
to the stable Helium-4 nucleus, reducing the relative abundance of
the few Deuterium nuclei that survive their conversion to He-4. On
the other hand, the significantly larger expansion rate enhances
the chance, that the less stable nuclei such as Deuterium and He-3
will survive through the much shorter time period of primordial
nucleosynthesis. The nuclear reactions are shut-off faster,
providing some amount of "compensation" for the faster reaction
rates due to the higher number densities. Yet it would be quite a
coincidence, if nucleosynthesis in the holostar would lead to the
same abundances of Deuterium, Helium-3 and Lithium-7 as in the
standard cosmological model. Whether the holostar model of the
universe can explain the experimentally determined abundances of
the light elements in a satisfactory fashion, therefore must be
considered as an open question.

\subsubsection{A hypothesis of an r-dependent rest mass.}

It might turn out that more radical proposals are required to make
the holostar solution compatible with the observational results,
such as the relative abundances of the light elements.

In this section I will explore the possibility, that the rest-mass
of the massive particles might vary with radial distance parameter
$r$ (or equivalently, with temperature, mean energy, curvature
radius etc.) in the holostar solution. Although there is not much
conclusive observational evidence for such a variation, the
dualities of string theory suggest the possibility of a power-law
relation between mass- and the distance scales. Some implications
of such a variation in the context of an FRW-model have been
discussed \cite{Buchalter/2004}. Therefore let us assume that

\begin{equation}
m = m_0 \left( \frac{r_0}{r} \right)^n
\end{equation}

Both the nucleon and the electron have masses lying in the range
between $m_e \approx 4.2 \cdot 10^{-23}$ and $m_p \approx 7.7
\cdot 10^{-20}$ of the Planck mass. The current radius of the
universe is $r \approx 10^{61}$ in units of the Planck length.
This observation suggest a dependence $m \propto (1/r)^{1/3}$.
Note that this dependence is quite compatible with the current
values of the cross-sectional areas $\sigma_T$
(Thomson-cross-section) of the electron and $\sigma_B$ of the
nucleon ($\sigma_B \approx 1.5 \cdot 10^{40}$ for the nucleon and
$\sigma_T \approx 2.5 \cdot 10^{41}$ for the electron in Planck
units). In the low energy-limit the cross sections of the
elementary particles are roughly proportional to $\hbar^2/m^2$.
All in all we find that the ratio of cross-sectional area of a
particle with respect to its mass $\sigma / m$, which already
turned out to be an important quantity in section
\ref{sec:massive:acc}, is roughly equal to the current radius of
the universe and inverse proportional to the cube of the particle
mass in Planck units, i.e. $\sigma / m \propto 1/m^3 \propto r$.
Furthermore, an assumed variation $m \propto r^{-1/3}$ is
sufficiently small, so that it might have gone unnoticed.

Under the assumption of a matter-dominated universe the
number-density of the massive particles is given by:

$$n_m = \frac{\rho}{m} = \frac{1}{8 \pi r^2 m(r)}$$

Under the assumption of a radiation dominated universe it is easy
to show, using the Hawking-entropy-formula, that the
number-density in the radiation is given by:

$$n_r = \left( \frac{r_0}{r} \right)^\frac{3}{2} \frac{1}{2 \sigma_r r_0 \hbar}$$

$\sigma_r$ is the entropy per particle of the radiation. The above
formula remains correct even after the universe has ceased to be
radiation dominated, if one multiplies the number-density with the
factor denoting the proportion of the energy-density of the
radiation to the total energy-density. For photons this factor is
roughly 5000, assuming $\Omega_m \approx 0.26$ according to WMAP.

The ratio of the number-densities is given by:

\begin{equation}
\frac{n_r}{n_m} = \frac{4 \pi m_0 r_0}{\sigma_r \hbar}
\left(\frac{r}{r_0} \right)^\frac{1}{6} = \frac{m}{\sigma_r T}
\end{equation}

Note, that this relation is {\em independent} of the functional
dependence of $m(r)$. Therefore the ratio of the number-densities
of photons to massive particles will always be proportional to the
ratio of the rest mass of a particle to the radiation temperature,
no matter how complicated $m$ depends on $r$.

However, the temperature at which the radiation temperature
becomes equal to the electron mass (or equivalently, the
number-density of the photons becomes nearly equal to the
number-density of the electrons), will be much higher in the case
$m \propto 1/r^{1/3}$ than in the case $m=const$. It is easy to
calculate the energy at which this happens, by recalling that:

\begin{equation} \label{eq:m:r}
m \propto 1/r^{1/3} \propto T^{2/3}
\end{equation}

so that

\begin{equation} \label{eq:m:m/T}
T \propto \left(\frac{T}{m}\right)^3
\end{equation}

In the matter-dominated era the temperature depends on the cube of
the ratio $T/m$. This ratio is quite low at the present time. With
$m_e / T_{CMBR} = 2.18 \cdot 10^9$ and using equation
(\ref{eq:m:m/T}) we can extrapolate the radiation temperature from
today's value up to the time when the ratio $m_e/T$ is unity.

\begin{equation}
T_{eq} = \left( \frac{m_e}{T_{CMBR}} \right)^3 T_{CMBR} = 2.8
\cdot 10^{28} K
\end{equation}

This extrapolation is quite robust, because there are no unknown
particle-species in the relevant temperature range $T_{CMBR} < T <
m_e$.

A temperature of $3 \cdot 10^{28} K$ corresponds to a mean
particle energy $E = 7.8 \cdot 10^{15} \, GeV = 6.4 \cdot 10^{-4}
E_{Pl}$, which is quite close to the $GUT$-energy. Therefore the
assumption of an $r$-dependent mass with $m^3 \propto 1/r$ leads
to the quite spectacular scenario, that the unification
energy-scale turns out to be nothing else than the scale at which
the electron becomes relativistic in the thermal environment of
the holostar space-time.

Equations (\ref{eq:m:r}, \ref{eq:m:m/T}) allow us to express the
relevant quantities, such as Hubble-constant, number-densities,
energy-densities etc. in terms of $r$, $T$ or the ratio $m/T$. The
ratio $m/T$ is the truly important quantity, because for most
physical processes not the absolute value of $T$ is important, but
rather the ratios $m_e/T$, $m_b/T$ etc. . We find:

\begin{equation}
m \propto \frac{1}{r^{1/3}} \propto T^{2/3} \propto
\left(\frac{T}{m}\right)^2
\end{equation}

\begin{equation}
H \propto \frac{1}{r} \propto T^{2} \propto
\left(\frac{T}{m}\right)^6
\end{equation}

\begin{equation}
n_\gamma \propto \frac{1}{r^{3/2}} \propto T^{3} \propto
\left(\frac{T}{m}\right)^9
\end{equation}

\begin{equation}
n_m \propto \frac{1}{r^{5/3}} \propto T^{10/3} \propto
\left(\frac{T}{m}\right)^{10}
\end{equation}

\begin{equation}
s \propto \frac{1}{r^{3/2}}\propto T^{3} \propto
\left(\frac{T}{m}\right)^9
\end{equation}

\begin{equation}
\rho \propto \frac{1}{r^2} \propto T^{4} \propto
\left(\frac{T}{m}\right)^{12}
\end{equation}

These relations show, that when $T/m$ is unity, i.e. a factor of
roughly $10^9$ higher than today, all physically relevant
quantities such as energy-densities, number-densities, Hubble-rate
etc. are very close to the values of an FRW-type universe at the
GUT-energy scale. Furthermore, if we except that the ratio $m/T$
is the truly relevant quantity to which the local length scale $l
\propto m/T$ must be referenced, the number-densities of the
particles expressed in terms of $l$ evolve as if the "full"
underlying space-time consisted out of 9 (or 10) spatial
dimensions: The number-density of the {\em massless} particles
evolves as $n_\gamma \propto 1/l^9$, whereas the number-density of
the {\em massive} particles evolves as $n_m \propto 1/l^{10}$. It
is quite remarkable and maybe not a coincidence that the
fundamental space-time dimensions of string theory are $9+1$, and
that the fundamental theory ($M$-theory) is believed to be
formulated in $10+1$ dimensions. If the different dependencies of
$n_m$ and $n_\gamma$ on $l$ in holostar space truly have their
fundamental origin in string theory, one expects that $10+1$
dimensional $M$-theory will be the full theory, incorporating
massless and massive strings, whereas its $9+1$ dimensional
offsprings rather refer to the massless sectors of the full
theory.

However, these are speculations that rely on the validity of the
purely phenomenological assumption that $m \propto 1/r^{1/3}$. At
the time this paper is written there appears to be no hard
theoretical evidence for such an assumption. An - albeit shaky -
argument might be this: $m \propto 1/r^{1/3}$ is equivalent to
$\sigma / m \propto r$, due to the dependence of the
cross-sectional area $\sigma$ on $1/m^2$. According to equation
(\ref{eq:aP}) the pressure-induced acceleration in the co-moving
frame is given by:

$$a_P = \frac{\sigma}{m} P_r = -\frac{\sigma}{m} \frac{1}{8 \pi r^2}$$

This acceleration is negative and tends to slow any massive object
compared to its trajectory calculated under the assumption of a
pressure free space-time. It is intriguing to relate the
pressure-induced acceleration $a_P$ to the "cosmological"
acceleration, given by the isotropic Hubble-expansion.\footnote{In
natural units the Hubble-constant has the dimension of an
acceleration: [$d^2s/dt^2$] = [$1/t$]} In the holostar-space time
we have $H \propto 1/r$. By setting $a_P = H$ we find:

$$\frac{\sigma}{m} \propto \frac{1}{m^3} \propto 8 \pi r$$

This is just the desired dependence

It should be possible to notice an acceleration $a_P = H$, if it
is real. Assuming a Hubble-constant of $H = 72$ km/s / MPc the
acceleration turns out as $a_P \approx 6.9 \cdot 10^{-10} m/s^2$
The anomalous acceleration of the Pioneer space-crafts has just
the right magnitude: $a_P \approx 8 \cdot 10^{-10}$. See
\cite{Pioneer} for a discussion of the experimental determination
of $a_P$ and possible explanations.

\subsection{\label{sec:Unruh}The Unruh temperature and a determination of the parameter $r_i$}

\subsubsection{Local geodesic acceleration and Unruh-temperature}

A massive particle at rest in the ($r, \theta, \varphi, t$)
coordinate system is subject to a geodesic acceleration given by
equation (\ref{eq:g}). With this acceleration the following
Unruh-temperature can be associated

\begin{equation} \label{eq:TUnruh}
T_{U} = \frac{g \hbar}{2 \pi} = \frac{\hbar}{4 \pi r}
\sqrt{\frac{r_0}{r}} =  T_\gamma \frac{r_0}{r}
\end{equation}

where $T_\gamma$ is the local radiation temperature given by
equation (\ref{eq:Tlocal}).

At $r \approx r_0$ the Unruh temperature is comparable to the
radiation temperature, which is quite close to the
Planck-temperature. Therefore particles with masses up to nearly
Planck-mass can be created out of vacuum in the vicinity of a
roughly Planck-size region close to the center of the holostar.

Note that the geodesic acceleration $g$ for a stationary observer
falls off with $1/r^{3/2}$, whereas the radiation temperature
scales with $1/r^{1/2}$. Therefore the Unruh-temperature due to
{\em geodesic} acceleration will be negligible with respect to the
radiation temperature, whenever $r \gg r_0$.

If we multiply the Unruh-temperature with the local radiation
temperature of equation (\ref{eq:Tlocal}) we find the following
interesting relation, which doesn't include the - unknown -
parameter $r_0$:

\begin{equation} \label{eq:TU:Trad}
T_{U} T_\gamma = \frac{\hbar^2}{16 \pi^2 r^2} = \frac{\hbar^2}{2
\pi} \rho
\end{equation}

\subsubsection{The Unruh-temperature at the membrane}

The Unruh temperature, as derived in equation (\ref{eq:TUnruh}),
only applies to an observer {\em at rest} in the holostar
space-time. There are only two conceivable positions in the
holostar, where a particle can remain at rest: The membrane, which
constitutes the global minimum of the effective potential, and the
central position $r \approx r_0$, where the effective potential
has a local maximum.\footnote{An asymptotic observer at spatial
infinity will also be "at rest", but spatial infinity is not a
position that any real observer can occupy.}

Let us first discuss Unruh-radiation at the membrane. As the
membrane constitutes the boundary to the exterior space-time, we
have to consider the geodesic acceleration induced by the exterior
space-time in addition to the geodesic acceleration in the
interior. Except for a Planck size black hole the exterior
geodesic acceleration is much larger than the interior
acceleration (at the position at the membrane), so that the
Unruh-temperature at the membrane will be dominated by
Unruh-radiation tied to the exterior acceleration. The exterior
geodesic acceleration is nothing else than that of a black hole
outside of the event-horizon:

$$g_{ext} = \frac{r_+}{2 r^2} \sqrt{A(r)}$$

which amounts to

$$g_{ext} = \frac{r_+}{r_0} g(r_h) = \frac{1}{2 \sqrt{r_0 r_h}} \left(1-\frac{r_0}{r_h}\right) \simeq  \frac{1}{2 \sqrt{r_0 r_h}} $$

at the position of the membrane $r_h$. With $g(r_h)$ I have
denoted the interior geodesic acceleration at the membrane.

The Unruh-temperature at the membrane is $T = \hbar g_{ext} / (2
\pi) \simeq \hbar / (4 \pi \sqrt{r_0 r_h})$. Except for the factor
$1- r_0/r_h$ the radiation temperature and the Unruh-temperature
at the membrane are equal. The difference between the two
temperatures is practically indistinguishable for large holostars.
As the surface red-shift of the holostar is $z = \sqrt{r_h/r_0}$,
both temperatures measured at infinity are equal to the
Hawking-temperature of a black hole with the same gravitating
mass. This is quite a satisfactory and self-consistent result.

\subsubsection{The Unruh temperature at the center and a determination of $r_i$}

Let us now turn to the holostar's central position, which is
defined as the position, where a particle can be considered to be
nearly at rest (in unstable equilibrium). In anticipation of the
result, I denote this position by $r_i$. At $r_i$ self-consistency
requires, that the Unruh-temperature should be equal to the
radiation temperature. It does not make sense to attribute two
different temperatures to the same physical situation. The
difficult question is how to properly define this "rest" position.
The concept of point-like particles doesn't make sense in the
holostar space-time, so that $r_i=0$ is meaningless. Such an exact
value would also be in conflict with the uncertainty principle. In
section \ref{sec:uncertainty} is has been shown, that the
uncertainty principle forbids that any particle in the holostar
space-time can be localized better than $r \approx r_0$.
Furthermore, the smallest possible "radius" of a particle is
$r_0/2$ (see the discussion in section \ref{sec:uncertainty}).
According to the arguments given in section
\ref{sec:energy:conditions} the "region" $r < r_0/2$ should not be
considered to be a genuine part of the space-time, so that the
smallest "distance" of any real particle from the center is $r
\approx r_0$. From this perspective one expects, that the central
"rest" position in the holostar space-time should be the position
closest to the center that can be occupied by one single (real)
particle, leading to $r_i \approx r_0$.

A better definition explicitly takes into account that the Unruh-
and radiation temperature must be equal at the radial position
where a (real) particle/observer is {\em at rest}. As the
equilibrium at the center is unstable, any particle/observer
nearly at rest at the center must eventually start to move out. At
late times the particle's motion will be nearly geodesical. As
geodesic motion at late times is nearly radial, the equations of
geodesic motion are governed by just one parameter, $r_i$, which
is defined as the radial position, where the motion started out
from rest.\footnote{This parameter determines the local
$\gamma$-factor of the motion via equation (\ref{eq:gamma}), the
locally measured Hubble-constant via equation (\ref{eq:Hubble})
etc.} Therefore it makes imminent sense to identify the "rest"
position, where Unruh and radiation temperature are equal, with
$r_i$. With this identification equation (\ref{eq:TUnruh})
immediately allows us to make the prediction:

\begin{equation} \label{eq:ri=r0}
r_i = r_0
\end{equation}

\subsubsection{Does the geodesically moving observer see an Unruh temperature?}

So far we only analyzed the Unruh temperature for an observer at
rest in the coordinate system. Will a geodesically moving observer
at an arbitrary position within the holostar also experience an
Unruh-temperature, in addition to the local radiation temperature
produced by the co-moving matter?

The answer is trivial: A geodesically moving observer is by
definition unaccelerated in his own frame of reference, so such an
observer will not experience an Unruh temperature. However, pure
geodesic motion is not possible within a holostar, due to the
pressure. Even in the low-density regions of a holostar where the
motion can be considered by all practical purposes to be geodesic,
the pressure provides a small deceleration. It is quite easy to
calculate this deceleration in the co-moving frame. We find:

\begin{equation} \label{eq:ap:coord}
a_P = \frac{\sigma}{m} P = \frac{\sigma}{m} \frac{1}{8 \pi r^2}
\end{equation}

This deceleration is quite low and virtually unnoticeable for
radial positions comparable to the current radius of the universe.
For example, if we take $\sigma = \hbar$ and $m = m_e$ we find
$a_P \simeq 6.3 \cdot 10^{-50} \,  m/s^2$. Even when the
cross-sectional area $\sigma$ is taken to be equal to the
electro-magnetic cross section of an electron, $\sigma \approx
\hbar / m_e^2$, we find that the pressure-induced deceleration in
the frame of the co-moving observer is of the order $10^{-10}  \,
m/s^2$.\footnote{Quite interestingly this is roughly the magnitude
of the anomalous Pioneer acceleration.}

Furthermore, in order to attain an Unruh-temperature comparable to
the temperature of the CMBR, the proper acceleration in the
co-moving frame would have to be enormous: $a = 2 \pi T_U / \hbar
\approx 10^{20} m/s^2$ for $T_U \simeq 2.73 K$. Therefore Unruh
radiation does not play any significant role for the co-moving
observer (except at $r \simeq r_i = r_0$).

\subsubsection{Unruh temperature due to the pressure-induced acceleration}

Let us now turn back to a stationary observer, i.e. an observer at
rest in the coordinate system. This observer is subject to two
different sources of acceleration: the geodesic acceleration and
the pressure-induced acceleration. The {\em geodesic} acceleration
falls off with $1/r^{3/2}$, so that the Unruh-temperature due to
the geodesic acceleration is negligible with respect to the
radiation temperature. However, for an observer at rest in the
holostar space-time the Unruh-temperature induced by the pressure
is not negligible. It is proportional to $1/\sqrt{r}$, i.e.
proportional to the radiation temperature. In order to see this,
let us go back to the co-moving  moving observer, who moves highly
relativistically with respect to a stationary observer on a nearly
geodesic path. Although the co-moving observer cannot move exactly
geodesically with zero proper acceleration, due to the pressure,
he is singled out from other reference frames by the requirement,
that the pressure-induced acceleration will be smallest in his
frame. In order to calculate the pressure-induced acceleration in
any other frame, such as the frame of the stationary observer, we
only have to transform the proper acceleration measured by the
co-moving observer to the other frame. This is an easy exercise in
special relativity.

If we denote the pressure-induced acceleration in the stationary
frame with a tilde, we get $\widetilde{a_P} = \gamma^3 a_P$,
because the deceleration $a_P$ lies in the direction of motion,
i.e. is parallel to the (radial) boost. $\gamma$ is given by
equation (\ref{eq:gamma}). Therefore the acceleration (or rather
deceleration) due to the pressure in the frame of the {\em stationary}
observer turns out as:

\begin{equation}
\widetilde{a_P} = \frac{\sigma}{m} \frac{1}{8 \pi r^2}
\left(\frac{r}{r_i}\right)^{\frac{3}{2}} = \frac{\sigma}{2 m r_i}
\frac{1}{4 \pi \sqrt{r r_i}}
\end{equation}

This acceleration can be associated with an Unruh temperature,
which the stationary observer should be able to measure in
principle.

\begin{equation}
\widetilde{T_U}(r) = \frac{\hbar \widetilde{a_P}}{2 \pi} =
\frac{\sigma}{4 \pi m r_i} \sqrt{\frac{r_0}{r_i}} T_\gamma(r)
\end{equation}

We find that the pressure-induced Unruh-temperature in
the frame of the stationary observer is proportional to the local
radiation temperature $T_\gamma$, and therefore much higher than
the Unruh-temperature produced by geodesic acceleration as given
in equation (\ref{eq:TUnruh}).

In the above equation we have taken $r_i$ to be arbitrary. This
will allow us to determine $r_i$ from a slightly different
perspective. For $r_i \gg r_0$ the Unruh-temperature in the
stationary frame would be much lower than the radiation
temperature experienced by the co-moving observer. The ratio of
the Unruh-temperature to the local radiation temperature is nearly
constant and given by:

\begin{equation} \label{eq:TU/T}
\frac{\widetilde{T_U(r)}}{T_\gamma(r)} = \frac{\sigma}{\hbar}
\frac{T_i}{m} \frac{r_0}{r_i}
\end{equation}

$T_i = T_\gamma(r_i)$ is the local radiation temperature at the
turning point of the motion of the geodesically moving observer.

It seems paradoxical, that the Unruh-temperature of the stationary
observer should depend on the - presumably arbitrary - position
$r_i$, from which a geodesically moving observer started out his
motion. The solution to this apparent paradox is, that there is
only one possible position $r_i = const$ in the holostar
space-time, where a real observer can start his motion from rest.

It is possible to determine the value of $r_i$ from equation
(\ref{eq:TU/T}). Again it does not make sense to attribute
different temperatures to an observer at rest at $r_i$. Setting
the Unruh temperature and the radiation temperature at $r_i$ equal
we get

\begin{equation} \label{eq:sigma:m}
\frac{r_i}{r_0} = \frac{\sigma}{\hbar} \frac{T_i}{m} \simeq 1
\end{equation}

The last equality follows from the discussion in
\ref{sec:massive:acc} and the results obtained in
\cite{petri/charge}, according to which $\sigma / \hbar \approx
\sigma_S = m / T_i$, so that we get the already known result $r_0
\simeq r_i$. Note that this argument is independent of the
argument leading to equation (\ref{eq:ri=r0}). There the
Unruh-temperature due to the {\em geodesic} acceleration was set
equal to the radiation temperature. Here we demand that the
Unruh-temperature due to the {\em pressure-induced} acceleration
should be equal to the radiation temperature throughout the whole
space-time.\footnote{Note, that $r_i$ is defined as the position,
where a particle is at rest. This requires that the
net-acceleration vanishes. As the pressure-induced acceleration
and the geodesic acceleration have opposite signs, zero
net-acceleration is obtained, whenever the pressure-induced Unruh
temperature and the Unruh temperature due to geodesic acceleration
are equal. However, this does not mean, that the stationary
observer does not feel any acceleration. Any one who has ever set
his feet on the earth knows, that being stationary in a static
gravitational field does not mean that one does not feel its
effects. Therefore zero net-acceleration should not confuse us to
assume, that the Unruh-temperature will be zero.}

On the other hand, we already know that $r_0 = r_i$, so we can use
equation (\ref{eq:sigma:m}) to obtain an estimate for the mass $m$
of a geodesically moving particle starting out from $r_i$:

\begin{equation} \label{eq:m:T0}
m = \frac{\sigma}{\hbar } T_0= \frac{\sigma}{4 \pi r_0}
\end{equation}

where

\begin{equation} \label{eq:T0}
T_0 = T(r_0) = \frac{\hbar}{4 \pi r_0}
\end{equation}

is the temperature at $r_i = r_0$.

\subsubsection{An alternative derivation of $r_i$}

Equations (\ref{eq:T0}, \ref{eq:ri=r0}) can be derived by a very
elegant argument. In section \ref{sec:entropy:density}, equation
(\ref{eq:s:r0}) we found the following expression for the geodesic
acceleration:

\begin{equation} \label{eq:g:gU}
g = 4 r_0^2 \left( \frac{2 \pi T}{\hbar}\right)^3 = 4 r_0^2 g_U^3
\end{equation}

where $g_U$ is the "Unruh-acceleration" associated with the
radiation temperature $T$. In general we cannot identify the
radiation temperature with a proper acceleration via the
Unruh-formula. This is only possible when the particle is
stationary, i.e. at radial position $r=r_i$. But at any stationary
position we can set $g_U = g$ and consequentially $T_U(r_i) =
T_i$. Applying this to equation (\ref{eq:g:gU}) and solving for
$g$ we find:

\begin{equation} \label{eq:g:gU:r0}
g = g_U = \frac{1}{2 r_0}
\end{equation}

from which $T(r_i) = \hbar / (4 \pi r_0)$ and thus $r_i = r_0$
follow immediately.

\subsubsection{Are the acceleration horizon of the stationary observer and the Hubble-horizon of the geodesically moving observer related?}

Naively one would assume, that a stationary observer, situated at
the holostar's center, should be able to "see" the whole holostar
universe from his "elevated" position, at least in principle. On
the other hand, we know from section \ref{sec:Hubble} that a
geodesically moving observer, starting out from the center, will
experience an isotropic Hubble-expansion at late times with $H
\propto 1/r$. As a geodesically moving observer cannot look beyond
his local Hubble-length, it is reasonable to assume that the local
Hubble-constant defines a "true" horizon with $r_H = 1 / H$. The
horizon due to the Hubble-expansion shrinks, when the observer
approaches the holostar's center. At radial position $r_0$ one
would expect a horizon size of order $r_0$. This appears to be in
conflict with the assumption, that a stationary observer at the
center can see everything.

However, this assumption does not take into account the enormous
geodesic acceleration of a stationary observer at the holostar's
center, with which a true acceleration horizon is associated. The
geodesic acceleration at the center is given by equation
(\ref{eq:g:gU:r0}). The associated acceleration horizon is given
by $r_g = \hbar / g = 2 r_0$, i.e. two Planck-lengths.

Therefore, at least at the holostar's center the acceleration
horizon and the Hubble-horizon are comparable to each other. It is
natural to assume, that this will remain to be the case throughout
the whole holostar space-time.

\subsubsection{An argument for $\sigma = \sigma_S \hbar$ at $r_i$}

In \cite{petri/thermo} it was shown, that the energy per
ultra-relativistic particle in the holostar space-time is related
to the radiation temperature via $E = \overline{\sigma_S} T$.
$\overline{\sigma_S}$ is the (average) entropy per particle, which
is slightly larger than $\pi$. The exact value of
$\overline{\sigma_S}$ depends on the thermodynamic model. In the
simple model discussed in \cite{petri/thermo}
$\overline{\sigma_S}$ only depends on the ratio of bosonic to
fermionic degrees of freedom ($\overline{\sigma_S} \simeq 3.3792$,
when there are only fermions at $r_i$, $\overline{\sigma_S} \simeq
3.2299$ when there are equal numbers of fermions and bosons and
$\overline{\sigma_S} \simeq 3.1568$, when there are 8 times as
much bosons than fermions). If we assume that the radiation quanta
at $r_i = r_0$ have just the right energy to create a pair of
massive particles at rest at $r_0$, we find $m = E =
\overline{\sigma_S} T_0$ from which $\sigma = \overline{\sigma_S}
\hbar$ follows. This result was obtained independently in
\cite{petri/charge}. It is the relation that one expects from the
Hawking entropy-area formula for a spherically symmetric black
hole: $S = A / (4 \hbar) = \pi r_h^2 / \hbar = \sigma / \hbar$.

In \cite{petri/thermo} it was demonstrated, that $ r_h^2 \approx
(r_0/2)^2 = (\sigma_S \hbar / \pi) \approx \hbar$ at energies
close to the Planck-energy. If we insert $\sigma \approx \sigma_S
\hbar$ and $r_0 \approx 2 \sqrt{\sigma_S \hbar / \pi}$ into
equation (\ref{eq:m:T0}) for the mass $m$ of the particle, that
starts out moving geodesically from $r_i$, we find the following
value:

$$m \approx \frac{1}{8} \sqrt{\frac{\sigma_S}{\pi}} \sqrt{\hbar} \simeq \frac{m_{Pl}}{8}$$

Remarkably this is quite close to the preon-mass estimate in
section \ref{sec:preon:2}.

\subsection{A relation between the string tension and length}

It has been pointed out in the introduction, that the holostar is
a string solution. The equation of state of the total interior
matter is identical to the equation of state of a radial
collection of densely packed classical strings.

The main purpose of this paper was to show, that despite the
overall string equation of state, part of the interior matter can
be interpreted in terms of ordinary particles.

A discussion which focusses predominantly on the the string-nature
of the interior matter can be found in \cite{petri/string}. It is
not the purpose of this paper, to reiterate results which have
been published elsewhere. However, in light of a recently
published result in string theory, which has gained some
considerable coverage in the media, I feel obliged to point out
how these results, derived purely in the context of string-theory,
relate to the properties of the - purely classical - holostar
solution.

In \cite{Mathur/2004} it has been shown, that by so called
"fractionation" the string tension will become inverse
proportional to the string length, which leads to the prediction,
that the strings can occupy a fairly large volume and "black
holes" should be filled with low-energy strings. The radius of
such a stringy "fuzzball" has been estimated in the context of
string theory. It turned out to be nearly equal to the
Schwarzschild radius of a black hole.

This result is essentially identical to the predictions that can
be obtained from the classical holostar solution. The
string-tension at the radial coordinate position of the membrane
$r_h$ is given by

\begin{equation}
\mu = \frac{1}{8 \pi r_h^2}
\end{equation}

Assuming the string extends from the holostar's center to the
membrane, its "length" $R$, determined as the travel time of a
photon along the string length measured by an asymptotic observer
at infinity, is given by:

\begin{equation}
R = \int_0^{r_h}{\frac{r}{r_0} dr} = \frac{r_h^2}{2 r_0}
\end{equation}

We immediately see that the string tension is inverse proportional
to its length. The constant of proportionality is

\begin{equation} \label{eq:R:mu}
R \mu = \frac{1}{16 \pi r_0} \approx \frac{1}{32 \pi r_{Pl}}
\end{equation}

If one takes into account, that not all strings attached to the
membrane extend to the holostar's center one can calculate a mean
string-length, which is a factor of 2 smaller than the maximum
length:

\begin{equation}
\overline{R} = \frac{r_h^2}{4 r_0}
\end{equation}

However, one has to take into account that according to string
theory the strings cannot just "end" in the interior holostar
space-time. Either they must form a closed loop, or both
string-ends must attach to a D-brane. The only D-brane in the
classical holostar solution is the boundary membrane at $r = r_h$.
Disregarding any closed (interior) loops, any string with one end
attached to the holostar's boundary membrane must have its other
end attached to the membrane as well. This doubles the mean length
of the string, so that equation (\ref{eq:R:mu}) should still be
the correct result.

\section{The holostar as a unified description for the fundamental building blocks of nature?}

As has been demonstrated in the previous chapter, the holostar has
properties which are in many respects similar to the properties of
black holes and the universe.

It is quite remarkable, that the fairly simple model of the
holostar appears to have the potential to explain the properties
of objects, that so far have been treated as distinct building
blocks of nature. Black holes and the universe didn't appear to
have much in common, although both are believed to be
self-gravitating systems. The holostar solution points at a deeper
connection between self gravitating systems of any size. The
holostar solution might even be of some relevance for elementary
particles.

This chapter is dedicated to a first preliminary discussion of the
question, whether the holostar might be able to serve as an
alternative - possibly unified - description for black holes, the
universe and elementary particles, at a semi-classical level.

\subsection{The holostar as alternative to black holes?}

The holostar's properties are very similar to the properties
attributed to black holes. It has an entropy and temperature
proportional, if not equal to the Hawking entropy and temperature.
Its exterior space-time metric is equal to that of a black hole,
disregarding the Planck coordinate distance between the membrane
and the gravitational radius. Therefore, as viewed from an
exterior observer, the holostar is virtually indistinguishable
from a black hole.

With respect to the interior space-time, i.e. the space-time
within the event horizon (or within the membrane), there are
profound differences:

A black hole has no interior matter, except - presumably - a point
mass $M$ at the position of its central singularity.\footnote{A
rotating Kerr-black hole has a ring-singularity.} The number and
the nature of the interior particles of a black hole, or rather
the particles that have gone into the black hole, is undefined.
Any concentric sphere within the event horizon is a trapped
surface. The holostar has a continuous interior matter
distribution, no singularity at its center, no trapped surfaces
and - as will be shown in \cite{petri/thermo} - a definite number
of interior particles proportional to its boundary area. The
absence of trapped surfaces and of an accompanying singularity is
a desirable feature in its own right. Furthermore, the holostar
appears to be the most radical fulfillment of the holographic
principle: The number of the holostar's interior particles is
exactly proportional to its proper surface area. Many researchers
believe the holographic principle to be one of the basic
ingredients from which a possible future universal theory of
quantum gravitation can be formed.

A black hole has an event horizon. The unique properties of the
event horizon, i.e. its constant surface gravity and its
(classically) non-decreasing area, determine the thermodynamic
properties of a black hole, i.e. its Hawking temperature and
entropy. The holostar's thermodynamic properties are determined by
its interior matter configuration. Therefore the holostar solution
"needs" no event horizon. In fact, it possesses no event horizon,
because the metric coefficients $B$ and $1/A$ never become zero in
the whole space-time. Note, however, that for a large holostar the
minimum value of $B$, the metric coefficient with respect to the
exterior time coordinate $t$, becomes arbitrarily close to zero.
$B$ reaches it's lowest value at the boundary $r_h = r_+ + r_0$:

\begin{equation}
B_{min} = B(r_h) = \frac{r_0}{r_h} = \frac{1}{1 + \frac{r_+}
{r_0}} \simeq \frac{1}{1 + \frac{M} {M_{Pl}}}
\end{equation}

For a holostar with a gravitational radius of $n$ Planck lengths,
i.e. $r_+ = n r_0$, one gets $B_{min} = 1/(n+1)$ and $A_{max} =
n+1$. A holostar with the mass of the sun has $n \approx 2 \cdot
10^{38}$ and therefore $B_{min} \approx 10^{-38}$.

Instead of an event horizon the holostar possesses a two
dimensional membrane of tangential pressure, who's properties are
exactly equal to the properties of the event horizon of a black
hole as expressed by the membrane paradigm \cite{Price/Thorne/mem,
Thorne/mem}. Therefore the action of the holostar on the exterior
space-time is indistinguishable from that of a same-sized black
hole.

The absence of an event horizon is a desirable feature of the
holostar. If there is no event horizon, there is no information
paradox. Unitary evolution of states is possible throughout the
whole space-time.

From the viewpoint of an exterior observer the holostar appears as
a viable alternative to a black hole. It is in most respects
similar to a black hole but doesn't appear to be plagued with the
undesirable consequences of an event horizon (information paradox)
and of a central singularity (breakdown of causality and/or
predictability).

Furthermore the holostar's interior is truly Machian. What matters
are the relative positions and motions of the interior particles
with respect to the whole object. The holostar's spherical
membrane serves as the common reference for all events, not the
ill-defined notion of empty asymptotic Minkowski space. In fact,
the active gravitational mass-density in the holostar space-time
is non-zero only at the membrane, so that the membrane can be
regarded as the "true" source of the holostar's gravitational
field.

Both the holostar and a black hole are genuine solutions to the
field equation of general relativity. From the viewpoint of the
theory both solutions have a good chance to be physically realized
in the real world. At the time this paper is written the holostar
solution looks like an attractive alternative to the black hole
solutions. However, only future research - theoretical and
experimental - will be able to answer the question, what
alternative, if any, will provide a better description of the
phenomena of the real world. Some more evidence in favor of the
holostar solution will be presented in \cite{petri/thermo,
petri/charge, petri/string}.

\subsection{The holostar as a self-consistent model for the
universe?}

A large holostar has some of the properties, which can be found in
the universe in its actual state today: At any interior position
there is an isotropic radiation background with a definite
temperature proportional to $1/\sqrt{r}$. Geodesic motion of
photons preserves the Planck-distribution. Massive particles
acquire a radially directed outward motion, during which the
matter-density decreases over proper time. Likewise the
temperature of the background radiation decreases. The temperature
and mass density within the holostar are related by the following
formula, $\rho = 2^5 \pi^3 r_0^2 T^4$, which yields results
consistent with the observations, when the mass density and
microwave-background temperature of the universe are used as input
and $r_0$ is set to roughly twice the Planck-length.

The theoretical description of the universe's evolution -
according to the equations of motion in the holostar space-time -
is in many aspects similar to the standard cosmological model. In
the standard model the energy density is related to the
cosmological time as $\rho \propto 1/t^2$. This relation is valid
in the matter dominated as well as in the radiation dominated
epoch. It is also valid in the holostar universe, if $t$ is
interpreted as the proper time of the co-moving observer. The
standard cosmological model furthermore has the following
dependence between the total energy density and the radiation
temperature: $\rho \propto T^4$. In the holostar universe the same
relation is valid.

On the other hand, the holostar has some properties, which might
not be compatible with our universe: When massive particles and
photons are completely decoupled, the ratio of the number-density
of photons with respect to the number density of massive particles
is predicted to increase over time in the holostar-universe ($n_m
\propto 1/r^2$, whereas $n_\gamma \propto 1/r^{3/2}$). The
standard cosmological models assumes that this ratio remains
nearly constant up to high redshift. Bigbang nucleosynthesis in
the standard cosmological models wouldn't be compatible with the
observations, if the photon to baryon ratio at high redshift would
be very different from today.

Yet the holostar appears to have the potential to explain some
phenomena, which are unexplained in the standard cosmological
models:

The standard model has the $\Omega$- or flatness-problem. Why is
$\Omega$ today so close to 1? If $\Omega$ is not exactly 1, it
must have been close to 1 in the Planck-era to an accuracy of
better than $10^{-60}$, i.e. the ratio of one Planck length to the
radius of the observable universe today. Such a finetuning is
highly improbable. One would expect $\Omega = 1$, exactly.
However, in the standard cosmological models $\Omega$ is a free
parameter. There is no necessity to set it equal to one. The
holostar "solves" this problem in that there is no free parameter.
The total energy density is completely fixed. Any other total
energy density would result in a very different metric, for which
most of the results presented in this paper, as well as in
\cite{petri/thermo, petri/charge}, would not hold.

The standard model has the cosmological constant problem: The
recent supernova-experiments \cite{Barris/Tonry,
Perlmutter/Supernovae_Lambda0.7, Perlmutter/Schmidt, Riess,
Tonry/Schmidt} indicate that the universe is accelerating
today.\footnote{Note, that this statement is somewhat
model-dependent. The observations indicate, that the universe has
recently undergone a transition from deceleration (at $z>0.6$) to
acceleration (at $z < 0.6$). However, the scatter in the data is
very high, so that it is difficult to achieve quantitative results
without setting priors on the theoretical model, such as assuming
a true cosmological constant or putting restrictions on the
equation of state, such as $P+\rho \geq 0$. Therefore other
possibilities which are compatible with the data, such as nearly
unaccelerated expansion in the recent past, cannot be ruled out
with high statistical confidence level.} In the standard
cosmological model such an acceleration is "explained" by a
positive cosmological constant, $\Lambda > 0$. However, the
natural value of the cosmological constant would be equal to the
Planck-density, whereas the cosmological constant today is roughly
a factor of $10^{124}$ smaller than its "natural" value. Such a
huge discrepancy suggests, that the cosmological constant should
be exactly zero. The supernova red-shift surveys, however, appear
to demand a cosmological constant which amounts to roughly $0.7$
of the critical mass-density today. Why does $\Lambda$ have this
particular value? Furthermore, $\Lambda$ and the mass-density
scale differently with $\tau$ (or $r$). Why then do we just happen
to live in an epoch where both values are so close?\footnote{Some
cosmological models therefore assume a time-varying cosmological
constant, which however is quite difficult to incorporate into the
the context of general relativity.} The holostar solution provides
a solution to the problem of an accelerating universe without need
for a cosmological constant. The holostar solution in fact
requires a cosmological constant which is exactly zero, which in
consequence can be interpreted as an indication that supersymmetry
might well be an exact symmetry of nature.

The standard model has the so called horizon problem. This problem
can be traced to the fact, that in the standard cosmological
models the scale factor of the universe varies as $r \propto
t^{2/3}$ in the matter-dominated era and as $r \propto t^{1/2}$ in
the radiation dominated era, whereas the Hubble-length varies
proportional to $r_H \propto t$. If the scale factor varies slower
than the Hubble length, the region that can be seen by an observer
today will have originated from causally unconnected regions in
the past. For example, at red-shift $z_{eq} \approx 1000$, i.e.
when matter and radiation have decoupled according to the standard
model, the radius of the observable universe consisted of roughly
$d_{e} / d_{c} \approx \sqrt{z_{eq}} \approx 30$ causally
disconnected radial patches, or $30^3$ regions. $d_e$ is the
distance scale of the expansion, $d_c$ of the causally connected
regions. The horizon problem becomes even worse in the radiation
dominated era.\footnote{In the radiation dominated era one finds
$d_{e} / d_{c} \approx z$.} The horizon problem makes it difficult
to explain why the CMBR is isotropic to such high degree.
Inflation is a possible solution to this problem. However, it is
far from clear how and why inflation started or ended.
Furthermore, the inflational scenarios need quite a bit of
finetuning and there appears to be no theoretical framework that
could effectively restrain the different scenarios and/or their
parameters. The holostar space-time solves the horizon problem in
a quite elegant way. The Hubble length $r$ is always proportional
to the scale-factor in the frame of the co-moving observer, as $r
\propto \tau$ and $H \propto 1/r$, therefore everything that is
visible to the co-moving observer today, was causally connected to
him in the past.

Related to the horizon problem is the problem of the scale factor.
With the $T \propto 1/r$-law, in the standard cosmological models
the scale factor is $r \approx 10^{30} r_{Pl}$ at the
Planck-temperature. Why would nature choose just this number?
Inflation can solve this problem, albeit not in a truly convincing
way. The holostar universe with $T \propto 1 / \sqrt{r}$ and $r
\propto \tau$ elegantly gets rid of this problem. The scale factor
evolves proportional (and within the errors equal!) to the
Hubble-length. At the Planck-temperature, both the scale-factor
and the Hubble-length are nearly equal to the Planck-length.

Inflation was originally introduced in order to explain the so
called monopole-problem. The standard cosmological model without
inflation predicts far too many monopoles. Inflation, if it sets
in at the right time, thins out the number of monopoles to a
number consistent with the observations. It might be, that the
holostar universe has an alternative solution to the problem of
monopoles or other heavy particles: Heavy particles will acquire
geodesic motion very early in the holostar, i.e. at small
$r$-coordinate value. But pure geodesic motion in the holostar
universe tends to thin out the non-relativistic particles with
respect to the still relativistic particles.

Inflation poses another cosmic coincidence problem: According to
the recent $WMAP$-data $H t = 1$ to a very high degree of
accuracy. $H t = 1$ is the relation expected for permanently
unaccelerated expansion. But the Standard Cosmological Model
assumes, that the universe first accelerated (in the inflationary
phase), then decelerated (in the matter- and radiation dominated
phase) and then started accelerating again (in the "dark energy"
dominated phase). All these cosmological era's are characterized
by $H t \neq 1$. Why do we happen to live just in the time, where
the strange sequence of acceleration, deceleration and
acceleration produces a value $H t = 1$, i.e. exactly the result
expected from constantly undecelerated expansion? And why did the
"switch" from deceleration to acceleration appear to have happened
just in our rather recent past?\footnote{In fact, as has been
shown in section \ref{sec:dL:z} a $\chi^2$-test of the recent
supernova yields nearly identical results for the holostar model
(permanently constant expansion) and the standard $\Lambda
CDM$-model with $\Omega_\Lambda \simeq 0.76$ and $\Omega_m \simeq
0.24$. (The holostar-model is slightly preferred, according to the
$chi^2$-test, but not with an confidence-level, that any decent
researcher would tolerate as acceptable.) Therefore is impossible
- with the present observational data - to decide between the two
models, at least not if one uses accepted statistical methods.}

If we actually live in a large holostar, we should be able to
determine our current radial position $r$ by the formulae given in
the previous sections. In principal it safest to determine $r$ by
the total matter-density via $\rho = 1 / (8 \pi r^2)$, as no
unknown parameters enter into this determination. Using the recent
WMAP-data we find:

$$r = \frac{1}{\sqrt{8 \pi \rho}} = 9.129 \cdot 10^{60} \, r_{Pl} = 1.560 \cdot 10^{10} ly$$

This is quite close to the value $r \approx 1.5 \cdot 10^{10} ly
\approx 8.8 \cdot 10^{60} r_{Pl}$, which was thought to be the
radius of the observable universe for a rather long period of
time.

However, the (total) matter density is difficult to determine
experimentally. Although other estimates of the total mass-energy
density, for example by examining the rotation curves of galaxies
and clusters, all lie within the same range, the accuracy of the
measurements (including systematic errors) is most likely not
better than 5 \%.

A much better accuracy can be achieved through the precise
measurements of the microwave background radiation, whose
temperature is known to roughly $0.1\%$. The determination of $r$
by the $CMBR$-temperature, however, requires the measurement or
theoretical determination of the "fundamental area" $r_0^2$. Its
value can be determined experimentally from both the
matter-density and the radiation temperature, or theoretically as
suggested in \cite{petri/charge}. If we use the theoretical value
($r_0^2 \simeq 2 \sqrt{3} \hbar$) at low energies we find:

$$r = \frac{\hbar^2}{16 \pi^2 T^2 r_0} = 9.180 \cdot 10^{60} \, r_{Pl} = 1.569 \cdot 10^{10} ly$$

In a large holostar the momentary value of the $r$-coordinate can
also be calculated from the local Hubble-value. In order to do
this, the starting point of the motion $r_i$ and the fundamental
length parameter $r_0$, or rather their ratio, have to be known.
Theoretically one would expect $r_i \simeq r_0$ (see for example
section \ref{sec:Unruh}), whereas experiments point to a somewhat
lower value. If we use $r_i/r_0 \approx 1$ and $H = 71 (km/s) /
Mpc$ we find:

$$r = \sqrt{\frac{r_0}{r_i}}\frac{1}{H} = 8.062 \cdot 10^{60} \, r_{Pl} = 1.378 \cdot 10^{10} ly$$

Very remarkably, all three different methods for the determination
of $r$ give quite consistent results, which agree by 15\%. A
rather large deviation comes from the Hubble-value, which is not
quite unexpected taking into account the difficulty of matching
the different cosmological length scales. Note however, that the
rather good agreement depends on the assumption $r_0 \simeq 1.87
\, r_{Pl}$ and $r_i/r_0 \simeq 1$. These assumptions, especially
the second, could well prove wrong by a substantial factor.

From a theoretical point of view $r_i = r_0$ is interesting
because it allows us to interpret the microwave background
radiation in terms of Unruh radiation. If $r_i=r_0$ turns out to
be the right ansatz, the Hubble constant can be predicted. Its
value should be:

\begin{equation}
H = \frac{1}{r} = 62.36 \, \frac{km/s}{Mpc}
\end{equation}

This value is well within the measurement errors, which are mostly
due to the calibration of the "intermediate" ladder of the
cosmological distance scale via the Cepheid variables.

Note that the absolute measurements of the Hubble-constant via
gravitational lensing, the Sunyaev-Zeldovich effect, the
Baade-Wesslink method and radioactive decay models consistently
yield results in the range $60 \pm 10$ (km/s)/Mpc, as remarked by
\cite[p. 144-145]{Peacock/book}.

Without strong theoretical support for a definite value of $r_0$
the matter-density seems to be the best way to determine $r$. If
the fundamental length parameter $r_0$ can be pinned down
theoretically, such as suggested in \cite{petri/charge}, the
extremely precise measurements of the microwave background
temperature will give the best estimate for $r$.

The simplicity of the holostar solution and the fact that it has
no adjustable parameters\footnote{Possibly an overall charge Q or
angular momentum could be included. $r_0$ is not regarded as an
adjustable parameter: The analysis here, as well as in
\cite{petri/thermo, petri/charge} strongly suggest, $r_0^2 \simeq
4 \sqrt{3/4} r_{Pl}$ at low energies. As long as the theoretical
"prediction" $r_i = r_0$ hasn't been confirmed independently $r_i$
should be regarded as an adjustable parameter. It is encouraging,
however, that $r_i/r_0$, as determined from the relative growth of
the density perturbations after radiation and matter became
kinematically decoupled, is quite close to the value determined
from the measurements of the Hubble-constant.} makes the holostar
solution quite attractive as an alternate model for the universe,
a model that can be quite easily verified or falsified.

At the time this paper is written it not clear, whether the
holostar will provide a serious alternative to the standard
cosmological model. It has the potential to solve many of the
known problems of the standard cosmological model. But the
standard cosmological model has been very successful so far. One
of its great achievements is the remarkably accurate explanation
for the distribution of the light elements, produced by bigbang
nucleosynthesis. Although the holostar universe is in many
respects similar to the standard cosmological models, it is
doubtable that the synthesis of the light elements will proceed in
exactly the same way as in the standard model. It would be quite a
coincidence if the holostar could give a similarly accurate
agreement between theory and observation.

Furthermore, it is not altogether clear how the "true" motion of
the massive particles within the cosmic fluid will turn out.
Therefore some results from the simple analysis in this paper
might have to be revised. There also is the problem of relating
the observations in the frames of the co-moving observer and the
observer at rest, which was discussed briefly in section
\ref{sec:frames}, but which quite certainly requires further more
sophisticated consideration.

On the other hand, it is difficult to believe that the remarkably
consistent determination of $r$ by three independent methods is
just a numerical coincidence with no deeper physical meaning.

The holostar-model of the universe also allows a prediction for
the baryon-to photon number ratio, which comes close (within a
factor of 2-3) to the actually observed value. It explains the
evolution of the density-perturbation fairly well and it might
even have an answer for the low quadrupole and octupole moments
found in the CMBR-power spectrum.

Therefore the question whether the holostar can serve as an
alternative description of the universe should be regarded as
open, hopefully decidable by research in the near future. Even if
the holostar doesn't qualify as an accurate description of the
universe, its particularly simple metric should make it a valuable
theoretical laboratory to study the so far only poorly understood
physical effects that arise in a universe with significant
(anisotropic) pressure.

\subsection{The holostar as a simple model for a self
gravitating particle with zero gravitational mass ?}

Another unexpected feature of the holographic solution is, that
one can choose $r_+ = 2 M = 0$ and still get a genuine solution
with "structure", i.e. an interior "source region" with non-zero
interior matter-distribution bounded by a membrane of roughly
Planck-area ($r_+=0$ implies $r_h = r_0 \neq 0$). Note that the
interior "source region" $r < r_0$ should not be considered as
physically accessible for measurements. Neither should the
interior matter distribution be considered as a meaningful
description of the "interior structure". In fact, the holostar
solution with $r_h = r_0$ and $M = 0$ should be regarded to have
no physically relevant interior sub-structure. The only physically
relevant quantity is its finite boundary area.

The $r_+ = 0$ solution therefore might serve as a very simple, in
fact the simplest possible model for a particle of Planck-size. It
describes a (spin-0) particle with a gravitating mass that is
classically zero, as can be seen from the properties of the metric
outside the "source region" (i.e. $A=B=1$):

Although it quite unlikely that an extrapolation from the
experimentally verified regime of general relativity right down to
the Planck scale can be trusted quantitatively, the holographic
solution hints at the possibility, that an elementary particle
might be - at least approximately - describable as a
self-gravitating system.

There are remarkable similarities between the properties of black
holes and elementary particles. As has already been noted by
Carter \cite{Carter/electronAsBh} the Kerr-Newmann solution has a
gyromagnetic ratio of 2, i.e. its g-factor is equal to that of the
electron (disregarding radiative corrections). The (three)
quadrupole moments of the Kerr-Newman solution are $2/3$ and
$-1/3$ in respective units, an interesting analogue to the
fractional charges of the three quarks confined within the
nucleon?!

It is not altogether folly to identify elementary particles with
highly gravitating systems. For energies approaching the
Planck-scale gravity becomes comparable to the other forces.
Therefore an elementary particle will quite certainly become a
strongly gravitating system at high energies. However, the
identification of an elementary particle with a solution to the
classical field equations of general relativity at the low end of
the energy scale has so far met serious obstacles. It is difficult
to explain why the masses of the light elementary particles are
about 20 orders of magnitude smaller than the Planck-mass.

The common expectation, that the Planck-mass will be the minimum
mass of a compact self gravitating object can be traced to the
fact, that the only "natural" unit of mass which can be
constructed from the three fundamental constants of nature $c, \,
G$ and $\hbar$ is the Planck mass. Thus it seems evident, that any
"fundamental" theory of nature which incorporates the three
constants, must have fundamental particles of roughly
Planck-mass.\footnote{Supersymmetry suggests another possibility:
Although the fundamental mass-scale in any supersymmetric theory
is the Planck-mass, the near zero masses of the elementary
particles are explained from a very precise cancellation of
positive and negative contributions of bosons and fermions, due to
the (hidden) supersymmetry of nature.} This quite certainly would
be true, if mass always remains a "fundamental" parameter of
nature, even at the most extreme energy scales.

However, there might be a slight misconception about the
significance of mass in high gravitational fields. Without doubt,
"mass" is a fundamental quantity both in Newton's theory of
gravitation and in quantum field theories such as QED or QCD.
These theories have very heavily influenced the conception of mass
as a fundamental quantity of physics. In a geometric theory, such
as Einstein's theory of general relativity, mass comes not as a
natural property of a space-time-manifold. From a general
relativistic viewpoint mass isn't a geometric property at all and
enters into the equations of general relativity in a rather crude
way, as stated by Einstein himself.\footnote{Einstein once
described his theory as a building, whose one side (the left,
geometric side of the field equations) is "constructed of marble",
whereas the right side (matter-side) is "constructed from low
grade wood".} Furthermore, mass has dimensions of length. From the
viewpoint of loop quantum gravity length - in contrast to area (or
volume) - is a concept which is difficult to define.\footnote{See
for example \cite{Thiemann/length}} Therefore it is questionable
if mass will remain a fundamental parameter\footnote{"fundamental"
is used in the current context as "optimally adapted" for the
description of the phenomena.} in situations of high gravity,
where the geometry cannot be considered static but becomes dynamic
itself. In these situations it seems more "natural" to describe
the interacting (geometric) objects not by their mass, but rather
by the area of their boundaries. Mass appears rather as a
perturbation. Surfaces as basis for the "natural" description of
systems with high gravitational fields are not only motivated by
recent research results, such as the holographic principle
\cite{Hooft/hol, Susskind/hol}, the study of light-sheets
\cite{Bousso/lightsheets} and isolated horizons
\cite{Ashtekar/IsolatedHorizons, Ashtekar/IsolatedHorizons2}, or
M-theory, but also by the simple fact, that in natural units
$c=G=1$ area has dimension of action (or angular momentum), from
which we know that it is quantized in units of $\hbar / 2$.

Therefore the "elementary" holostar with zero gravitational mass,
but non-zero boundary area might serve as a very crude classical
approximation for the most simple elementary particle with zero
angular momentum and charge. Unfortunately none of the known
particles can be identified with the "elementary" uncharged and
spherically symmetric holostar-solution with $r_h = r_0$. Even if
a spin-zero uncharged and zero mass particle
exists\footnote{According to loop quantum gravity, a spin-network
state with a single spin-0 link has zero area, and thus zero
entropy. There are reasons to believe, that a spin-network state
with a single spin can be identified with an elementary particle
(see \cite{petri/charge}). However, a zero area, zero-entropy
particle is quite inconceivable. Therefore it is likely, that a
fundamental (i.e. not composite) particle with zero spin and
charge doesn't exist. Any uncharged spin-zero particle therefore
should be composite. For all the particles that are known so far
this is actually the case.}, it is unlikely that more than its
cross-sectional area can be "predicted" by the holostar solution.
Still it will be interesting to compare the properties of a
charged and/or rotating holostar to the properties of the known
elementary particles in order to see how far the equations of
classical general relativity allow us to go. Some encouraging
results, which point to a deep connection between the holostar
solution and loop quantum gravity spin-network states, are
reported in \cite{petri/charge}.

\section{Discussion}

The holostar solution appears as an attractive alternative for a
black hole. It has a surface temperature, which - measured at
spatial infinity - is proportional to the Hawking temperature. Its
total number of interior relativistic particles is proportional to its proper
surface area, which can be interpreted as evidence that the
Hawking entropy is of microscopic-statistical origin and the
holographic principle is valid in classical general relativity for
self gravitating objects of arbitrary size. In contrast to a black
hole, the holostar has no event horizon, no trapped surfaces and
no central singularity, so there is no information paradox and no
breakdown of predictability.

The interior matter-state of a holostar consists of string-type
matter, which extends to the holostar's boundary membrane,
situated roughly 2 Planck coordinate lengths outside the
holostar's gravitational radius. The string tension at the
boundary is given by $\mu = 1 / (8 \pi r^2)$, which is inverse
proportional to the mean string length $R = r^2 / ( 2 r_0)$. This
result agrees with a very recent result in string theory
\cite{Mathur/2004}.

Furthermore, the holostar solution has some potential to serve as
an alternative model for a universe with anisotropic negative
pressure, without need for a cosmological constant. It also admits
microscopic self-gravitating objects with a surface area of
roughly the Planck-area and zero gravitating mass.

The remarkable properties of the holostar solution and its high
degree of self-consistency should make it an object of
considerable interest for future research.

A field of research which presents itself immediately is the the
generalization of the holostar solution to the rotating and / or
charged case. The derivation of the charged holostar solution is
straight forward and is discussed in \cite{petri/charge}. A
generalization to the case of a rotating body, including spin (and
charge), will be an interesting topic of future research.

An accurate description of the thermodynamic properties of the
holostar solution is of considerable interest. In
\cite{petri/thermo} the entropy/area law and the temperature-law
for the interior holostar matter state are put on a sound
foundation. The thermodynamic analysis allows one to relate the
Hawking temperature (at spatial infinity) to the local temperature
and energy-density in the interior holostar space-time. Using this
relation the Hawking prediction is verified to an accuracy of
roughly 1\%.

Another valuable field of research will be the examination,
whether the holostar solution can serve as an alternative model
for the universe. The holostar solution appears to have the
potential to solve many problems of the standard cosmological
models, such as the horizon-problem, the cosmological constant
problem, the problem of structure formation from the small density
perturbations in the CMBR, etc. . It provides an explanation for
the numerical value of the baryon to photon ratio $\eta$. Even
some of the more recent observational results, such as the missing
angular correlation of the CMBR-fluctuations at angular
separations larger than $60^\circ$, are explainable in the context
of the holostar space-time. On the other hand, it is far from
clear whether the holostar solution will ever be able to explain
the observed abundances of the light elements, especially
Deuterium and Lithium, in a convincing fashion, such as the
Standard Cosmological Model can. Nucleosynthesis in a holostar
universe will be a demanding challenge. If the holostar can pass
this test, it should open up a new field of interesting research
in cosmology and particle physics. Quite likely chemical
potentials and supersymmetry will play an important role in the
holostar universe, at least at temperatures at the GUT-scale.

Lastly, the properties of the membrane, how it is formed, how it
is composed and how it manages to maintain its two-dimensional
structure will be an interesting research topic, if the holostar
turns out to be a realistic alternative for a black hole. The
study of anisotropic string type matter states in high
gravitational fields should provide fruitful as well.


\end{document}